\documentclass[twocolumn]{aastex631}

\usepackage{multirow}
\usepackage{comment}   

\shorttitle{AASTeX v6.3.1 Sample article}
\shortauthors{Duggal et al.}
\graphicspath{{./}{figures/}}

\begin{document}

\title{Optical- $\&$ UV-Continuum Morphologies of Compact Radio Source Hosts}


\correspondingauthor{Chetna Duggal}
\email{duggalc@myumanitoba.ca}

\author[0000-0001-7781-246X]{C. Duggal}
\affiliation{Department of Physics and Astronomy, University of Manitoba, Winnipeg, MB R3T 2N2, Canada} 

\author[0000-0001-6421-054X]{C. P. O'Dea}
\affiliation{Department of Physics and Astronomy, University of Manitoba, Winnipeg, MB R3T 2N2, Canada}

\author[0000-0002-4735-8224]{S. A. Baum}
\affiliation{Department of Physics and Astronomy, University of Manitoba, Winnipeg, MB R3T 2N2, Canada}

\author[0000-0002-0690-8824]{A.\ Labiano}
\affiliation{Telespazio UK for the European Space Agency, ESAC, Camino Bajo del Castillo s/n, 28692 Villanueva de la Ca\~nada, Spain.}

\author[0000-0002-2951-3278]{C. Tadhunter}
\affiliation{Department of Physics $\&$ Astronomy, University of Sheffield, University of Sheffield, Sheffield S3 7RH, UK}

\author[0000-0002-1516-0336]{D. M. Worrall}
\affiliation{H.H. Wills Physics Laboratory, University of Bristol, Tyndall Ave, Bristol BS8 1TL, UK}

\author[0000-0002-9482-6844]{R. Morganti}
\affiliation{Kapteyn Astronomical Institute, University of Groningen, 9700 AB Groningen, The Netherlands}
\affiliation{ASTRON, the Netherlands Institute for Radio Astronomy, Postbus 2, NL-7990 AA Dwingeloo, The Netherlands}

\author[0000-0002-5445-5401]{G. R. Tremblay}
\affiliation{Harvard-Smithsonian Center for Astrophysics, 60 Garden Street, Cambridge, MA 02138, USA}

\author[0000-0003-0589-5969]{D. Dicken}
\affiliation{UK Astronomy Technology Centre, Royal Observatory Edinburgh, Blackford Hill, Edinburgh EH9 3HJ, UK}


\begin{abstract}

We present the first systematic search for UV signatures from radio source-driven AGN feedback in Compact Steep Spectrum (CSS) radio galaxies. Owing to their characteristic sub-galactic jets ($1-20$ kpc projected linear sizes), CSS hosts are excellent laboratories for probing galaxy scale feedback via jet-triggered star formation. The sample consists of 7 powerful CSS  galaxies, and 2 galaxies host to radio sources $>$20 kpc as control, at low to intermediate redshifts ($z<$ 0.6). Our new \textit{HST} images show extended UV continuum emission in 6/7 CSS galaxies; with 5 CSS hosts exhibiting UV knots  co-spatial and aligned along the radio-jet axis. Young ($\lesssim$ 10 Myr), massive ($\gtrsim$ 5 M$_\odot$) stellar populations are likely to be the dominant source of the blue excess emission in radio galaxies at these redshifts. Hence, the radio-aligned UV regions could be attributed to jet-induced starbursts. Lower near-UV SFRs compared to other indicators suggests low scattered AGN light contribution to the observed UV. Dust attenuation of UV emission appears unlikely from high internal extinction correction estimates in most sources. Comparison with evolutionary synthesis models shows that our observations are consistent with recent ($\sim$1$-$8 Myr old) star forming activity likely triggered by current or an earlier episode of radio emission, or by a confined radio source that has frustrated growth due to a dense environment. While follow-up spectroscopic and polarized light observations are needed to constrain the activity-related components in the observed UV, the detection of jet-induced star formation is a confirmation of an important prediction of the jet feedback paradigm.


\end{abstract}





\section{Introduction} \label{sec:intro}




Galaxies hosting powerful radio-luminous active galactic nuclei (AGN) are known to be profoundly affected by the extreme energy output of their central supermassive blackhole (SMBH) engines (as reviewed by \citealt{2012NewAR..56...93A, 2012ARA&A..50..455F, 2017NatAs...1E.165H, 2017FrASS...4...42M, 2018NatAs...2..179C, 2023Galax..11...21H}). 
From a theoretical standpoint, quenching of star formation by energy feedback from AGN is needed to halt galaxy growth, to produce the observed galaxy luminosity function (e.g., \citealt{2003ApJ...599...38B, 2006MNRAS.365...11C}) as well as to explain the strong correlations that exist between SMBH masses and the mass, luminosity and stellar velocity dispersions of galaxy bulges (e.g., \citealt{2012ARA&A..50..455F, 2015ARA&A..53..115K}). Observationally, there is a wealth of evidence that SMBH activity affects the interstellar medium (ISM) of the host galaxy. 
Powerful radio galaxies at high redshifts ($z\geq0.6$) exhibit emission-line regions co-spatial with radio emission and optical/UV continuum elongated and aligned along the radio source direction (e.g., \citealt{1993ARA&A..31..639M, 1996MNRAS.280L...9B, 1997MNRAS.292..758B, 1999AJ....117..677B, 2000MNRAS.311...23B, 2005MNRAS.359.1393I}), providing strong argument in favour of AGN feedback. On kiloparsec (kpc) scales, outflows from the AGN influence formation of new stars via regulating mechanisms that can extinguish star forming activity (``negative" feedback), thereby directly impacting host galaxy evolution. 
In high accretion rate quasars, radiation-driven winds and outflows heat the surrounding gas, leading to suppression of cooling and star formation (e.g., \citealt{2012A&A...537L...8C, 2015ARA&A..53..115K, 2015Natur.519..436T}). In the low-power accretion regime, radio-jet emission dominates feedback; where the kinetic energy of jet plasma drives the expulsion and/or heating of ambient gas from the galaxy core (e.g., \citealt{2008A&A...491..407N, 2012NJPh...14e5023M}). But in addition to the inhibitive mechanisms, AGN emission is also expected to boost star formation ("positive" feedback) in the host. Theoretical models and simulation studies have predicted shock-driven enhancement of starburst activity in the vicinity of the radio jets (\citealt{1989MNRAS.239P...1R, 1989ApJ...345L..21B, 2009MNRAS.396...61T, 2012MNRAS.425..438G, 2014ApJ...796..113D, 2017ApJ...844...37D, 2017ApJ...850..171F}) and radiative outflows (\citealt{2005ApJ...635L.121K, 2012MNRAS.427.2998I, 2013ApJ...772..112S, 2017ApJ...844...37D}). Observational evidence of this effect on sub-galactic scale has been limited in regard to both quasar-mode (\citealt{2015A&A...582A..63C, 2015ApJ...799...82C, 2016A&A...591A..28C}) and radio-mode (\citealt{2015A&A...574A..34S, 2015A&A...574A..89S, 2020MNRAS.499.4940Z}) processes, but has been growing with the advent of integral-field studies of active galaxies at higher redshifts, where strong impacts of feedback are expected. The emerging picture from recent works suggests that AGN jets and outflows might be playing a double role$-$ providing negative feedback as a fundamental large-scale mechanism that shapes the growth of SMBH and the host, with the starburst-enhancing positive feedback acting locally and/or in ``episodes'' that occur at short timescales of a few Myr (e.g., \citealt{2015ApJ...799...82C, 2018MNRAS.479.5544M, 2023MNRAS.519.3338T}). However, suppression of new star formation, while inferred from thermal and kinematic properties of the ISM gas, is not straightforward to observe without comparative analysis with a control sample. Feedback-induced star forming regions, on the other hand, are expected to show direct association with the jet emission. Hence, the detection of the theoretically predicted star formation along the radio source would, in fact, serve as a confirmation of the AGN feedback paradigm.
\newline

The main objective of this work is a search for positive feedback signatures in hosts of compact, sub-galactic scale radio sources, likely representing an early stage in FRI/FRII 
radio source expansion. Compact Steep-Spectrum (CSS) sources, GHz-Peaked Spectrum (GPS) sources and High-Frequency Peakers (HFP) $-$ collectively called Peaked Spectrum (PS) sources, along with Compact Symmetric Objects (CSOs) populate the young (or possibly, short-lived) radio source category (see \citealt{2021A&ARv..29....3O} for a review). These small yet powerful radio sources present excellent laboratories for the radio-mode feedback, since the interaction of the expanding radio source with the surrounding ISM is likely to be most vigorous in this infancy phase. As the galaxy-sized nuclear jets push through dense ambient medium, they drive a powerful bow shock at velocities of $\sim$10$^3$ km s$^{-1}$ through the surrounding gas clouds (\citealt{1997ApJ...485..112B, 2002AJ....123.2333O}). The gas clouds in the vicinity of the radio source are caused to collapse, triggering the formation of new stars along the jet axis (\citealt{1989ApJ...345L..21B, 2011ApJ...728...29W, 2012MNRAS.425..438G}). On the other hand, shock-heating excites gas in the more extended ambient clouds, causing them to accelerate outwards, thereby suppressing compression and formation of stars. Compact radio galaxies show a bimodal distribution in rates of star formation; while some appear to be passive and non-star forming, others have moderate star formation rates of $\sim$ few to few tens of M$_\odot$ per year (e.g., \citealt{2011A&A...528A.110F, 2021A&ARv..29....3O, 2023arXiv230410538G}).

Characterized by projected radio sizes of 1 to 20 kpc and steep ($\alpha \geq$ 0.5; flux density S $\propto \nu^{-\alpha}$) radio spectra, CSS sources are our main objects of interest in this study. Their radio size being an order of magnitude larger than PS/CSOs, CSS sources are the only compact radio sources currently resolvable at the scale of the jets. A compelling evidence for jet-ISM interaction in CSS galaxies has been the detection of strong spatial association of extended emission-line regions (EELR) with radio structure (\citealt{1997ApJS..110..191D, 1999ApJ...526...27D, 2000AJ....120.2284A, 2008ApJS..175..423P}). 
Gas kinematics in the emission-line regions is consistent with shocks (\citealt{1994ApJS...91..491G, 2008MNRAS.387..639H, 2013ApJ...772..138S, 2013MNRAS.435.1350R}) combined with AGN photoionization (\citealt{2005A&A...436..493L, 2009AN....330..226H, 2013ApJ...772..138S, 2016MNRAS.455.2242R}) as the main excitation mechanism. Thus, the radio-EELR alignment in CSS sources strongly suggests jet-driven feedback to the host ISM.

\begin{deluxetable*}{lcc@{\hskip 6mm}ccclccc}[ht!]
\tablecaption{The target sample \label{tab:first}}
\tabletypesize{\footnotesize}
\tablehead{
\colhead{Source} & 
\colhead{$\alpha$} & \colhead{$\delta$} & \colhead{\textit{z}} & \colhead{Angular scale} & \colhead{Radio size} &
\colhead{$LS$} & \colhead{\textit{P}$_\textup{1.4 GHz}$} & \colhead{Sample} & \colhead{Spectral type}\\
 & & & & \colhead{(kpc/ $''$)}  & \colhead{( $''$ )} &
\colhead{(kpc)} & \colhead{(10$^{27}$ \textup{W Hz$^{-1}$})} & & \\
\colhead{(1)} &\colhead{(2)} & \colhead{(3)}& \colhead{(4)}& \colhead{(5)}  & \colhead{(6)} & \colhead{(7)} & \colhead{(8)} & \colhead{(9)} & \colhead{(10)}
}
\startdata
0258+35 & 03 01 42.40 & +35 12 21.00 & 0.017 & 0.346 & 3.8 & 1.32 & 0.001 & G05 & NLRG$^1$ \\
1014+392$^\dagger$ & 10 17 14.20 & +39 01 23.00 & 0.536 & 6.400 & 6.1 & 39.03$^\dagger$ &  1.607 & F01 & NLRG$^2$ \\
1025+390 & 10 28 44.30 & +38 44 36.70 & 0.361 & 5.079 & 3.2 & 16.28 & 0.296 & F01 & NLRG$^3$ \\
1037+30 & 10 40 29.96 & +29 57 57.99 & 0.091 & 1.699 & 3.3 & 5.63 & 0.008 & G05 & NLRG$^3$ \\
1128+455 & 11 31 38.89 & +45 14 51.15 & 0.404 & 5.453 & 0.9 & 4.91 & 1.201 & F01 & BLRG$^3$ \\
1201+394 & 12 04 06.86 & +39 12 18.17 & 0.445 & 5.777 & 2.1 & 12.14 & 0.356 & F01 & NLRG$^4$ \\
1203+645 & 12 06 24.70 & +64 13 36.80 & 0.371 & 5.169 & 1.4 & 7.25 & 1.781 & O98 & BLRG$^3$ \\
1221$-$423 & 12 23 43.30 & $-$42 35 38.00 & 0.171 & 2.923 & 1.5 & 4.40 & 0.205 & B06 & NLRG$^5$ \\
1445+410$^\dagger$ & 14 47 12.76 & +40 47 45.00 & 0.195 & 3.249 & 8.1 & 26.41$^\dagger$ & 0.046 & F01 & NLRG$^4$ \\
\enddata
\tablecomments{(1) Target galaxy ($^\dagger$control sample),
(2) R.A. (J2000), (3) Dec. (J2000), (4) Redshift, (5) Angular scale at target redshift, (6) Angular size of the radio source (separation between the outermost component peaks, taken from source reference in Column 9), (7) Projected linear size of radio source (calculated from Columns 5 $\&$ 6), (8) 1.4 GHz radio luminosity (W Hz$^{-1}$), (9) Source sample references: G05 
\citep{2005A\string&A...441...89G} = low power CSS; F01 \citep{2001A\string&A...369..380F} = moderate power CSS; O98 \citep{1998PASP..110..493O} = \cite{1997A\string&A...325..943S} plus 
\cite{1990A\string&A...231..333F} = powerful CSS sources; B06 \citep{2006AJ....131..100B} = southern 3C equivalent, (10) NLRG/BLRG = Narrow-line/Broad-line radio galaxy; optical spectrum references: 1. \cite{1995ApJS...98..477H}, 2. \cite{2006MNRAS.369.1566G}, 3. \cite{2020MNRAS.491...92L}, 4. SDSS/DR12 spectral catalog, 5. \cite{2005MNRAS.356..515J}.}

\end{deluxetable*}


\begin{figure*}
\epsscale{1.0}
\plottwo{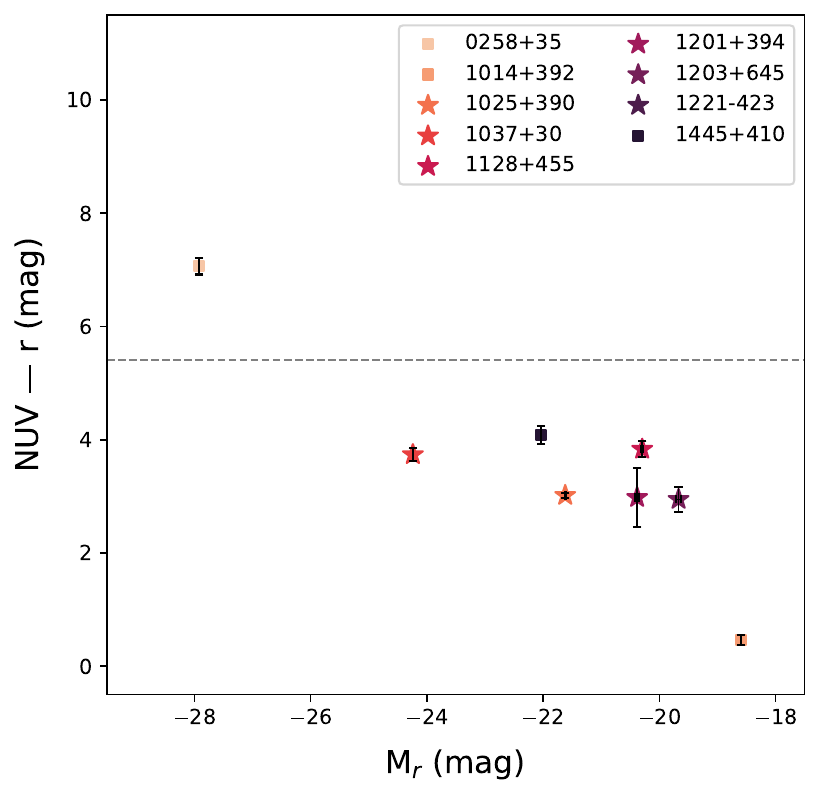}{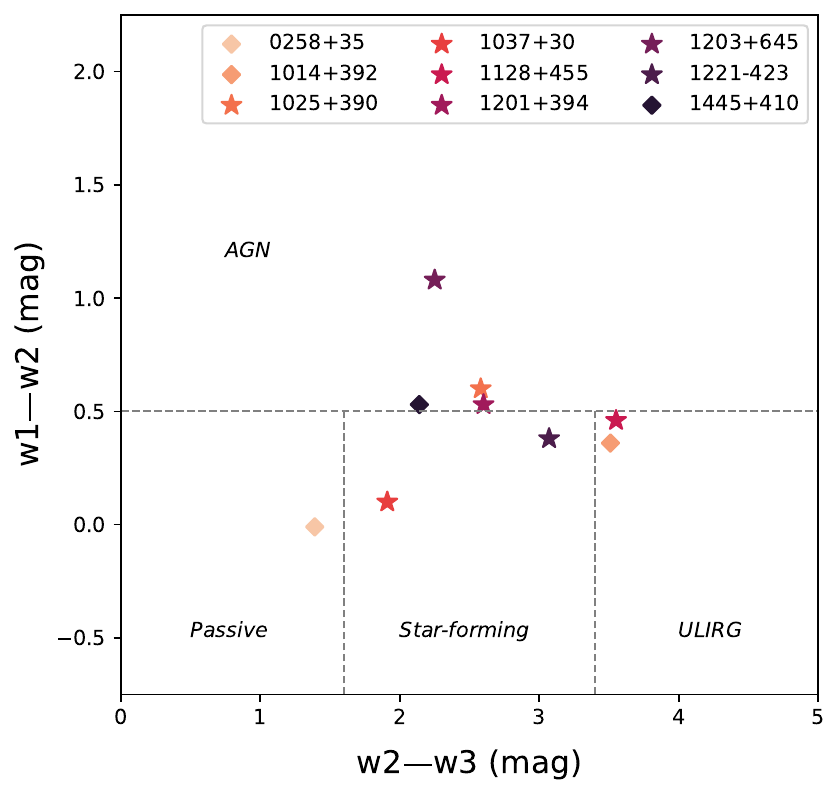}
\caption{(\emph{Left panel}) [NUV$-r$] vs. M$_r$ color-magnitude distribution for our sample. The horizontal solid line represents the \cite{2007ApJS..173..619K} threshold for recent star formation: the galaxies below [NUV$-r$] = 5.5 are likely to show young stellar population. \textit{GALEX}-NUV magnitudes and \textit{SDSS/r} magnitudes are used here ($r$-band photometry is not available for the Southern galaxy 1221-423). "$\star$" indicates the galaxies with clumpy, extended star-forming regions in the \textit{HST} images. (\emph{Right panel}) \textit{WISE} color-color plot for our sample of compact radio source host galaxies. Nearly all our sources lie around starburst IR colors. The dotted lines show classification criteria by \cite{2016MNRAS.462.2631M} as follows: $1.6<$ [W2$-$W3] $<3.4$ and [W1$-$W2] $<0.5$ are star-forming galaxies; [W2$-$W3] $<1.6$ and [W1$-$W2] $<0.5$ are passive galaxies; while galaxies with [W1$-$W2] $>0.5$ are AGN-dominated; the region with [W2$-$W3] $>3.4$ and [W1$-$W2] $<0.5$ belongs to (Ultra) Luminous Infrared Galaxies, the most actively star-forming objects in the universe. \label{fig:wise}  \label{fig:nuv-r}}
\end{figure*}


We seek to trace star formation triggered by the radio source in the shocked ISM$-$ a testament to radio-mode feedback operating on galaxy scales$-$ in compact, young CSS radio galaxies. 
As the youngest stellar populations emit the bulk of their energy in the rest-frame UV ($<$0.3  $\mu$m) band, observations at UV wavelengths
are ideal for investigating star formation in galaxies over timescales of $\sim$100 Myr; typically the O- and B-type stars with maximum main sequence lifetimes of $\sim$10 Myr and 100 Myr, respectively, that are brighter in UV than at other wavelengths. We carried out the first systematic UV imaging study focused on CSS sources, in search of radio-aligned UV light from jet-induced starbursts. Following a pilot snapshot observation with the \textit{HST} which detected extended UV light aligned with the radio source in the CSS host galaxy 3C 303.1 \citep{2008A&A...477..491L}, we broadened our search with a larger sample. A brief description of the motivation and preliminary findings were also published in \cite{2021AN....342.1087D}. In this paper, we present detailed analysis of the \textit{Hubble Space Telescope (HST)} observations which is organized as follows.
Sections \ref{sec:target} and \ref{sec:obsred} contain the details of the sample, the observed and archival data, and image processing. Extinction correction, photometric measurements and morphological decomposition methods are described in Section \ref{sec:analysis}. The main results are discussed in \ref{sec:results}.
This section also presents a quantitative analysis of observed star formation, in conjunction with population synthesis modelling. Section \ref{sec:compiled} discusses previous research on each of the individual target galaxy in the context of this work. The conclusions are summarized in Section \ref{sec:concl}. Throughout this work, a flat $\Lambda$CDM cosmology with H$_0$ = 69.6 km s$^{-1}$ Mpc$^{-1}$, $\Omega_M$ = 0.3 and $\Omega_{vac}$ = 0.7 is assumed.

\section{Target Selection} \label{sec:target}

Our sample consists of 9 radio galaxies$-$ 7 CSS host galaxies, along with 2 galaxies hosting $>$20 kpc radio sources acting as the control sample. There was no previous evidence for jet-induced star formation in any of the 9 target sources. The sample details are listed in Table \ref{tab:first}. The target galaxies are drawn from well-defined compact radio source samples of \cite{1997A&A...325..943S}, \cite{1990A&A...231..333F, 2001A&A...369..380F}, \cite{2005A&A...441...89G} and \cite{2006AJ....131..100B}, that are spread over a range in radio power (1.4 GHz luminosities given in Table \ref{tab:first}). The targets have been chosen to represent compact radio galaxies at low and intermediate redshift range $z\lesssim0.6$, in order to eliminate strong effects due to cosmic evolution. Projected linear radio size between $\sim$1$-8''$ forms another constraint for source selection, to enable resolution of optical/UV continuum at the scale of the radio source. 
\newline

Although the UV properties of the radio galaxies were not part of the sample criteria so as to avoid any selection bias, archival photometry shows blue [NUV$-r$] colors for most of the sample (see Figure \ref{fig:nuv-r}, left panel), typical of galaxies that have experienced star formation (SF) in the last $\lesssim$1 Gyr. [NUV$-r$ vs. M$_r$] is an excellent diagnostic for recent SF (\citealt{2006Natur.442..888S, 2007ApJS..173..619K}) in early-types, and has been extensively used as SF indicator in radio galaxies (e.g., \citealt{2008A&A...489..989B}) and Brightest Cluster Galaxies (BCGs) (e.g., \citealt{2009MNRAS.395..462P}). Substantial direct AGN contribution is not expected in the majority of sample hosted in NLRGs. 
On the \textit{WISE}\footnote{Wide-field Infrared Survey Explorer telescope \citep{2010AJ....140.1868W}} color-color plot (Figure \ref{fig:nuv-r}, right panel), most of the compact radio source hosts lie in the starburst-dominated region, consistent with our expectation of low AGN contamination in optical/UV colors.

\section{Observations $\&$ data reduction} \label{sec:obsred}

\begin{figure*}
\epsscale{1.1}
\plotone{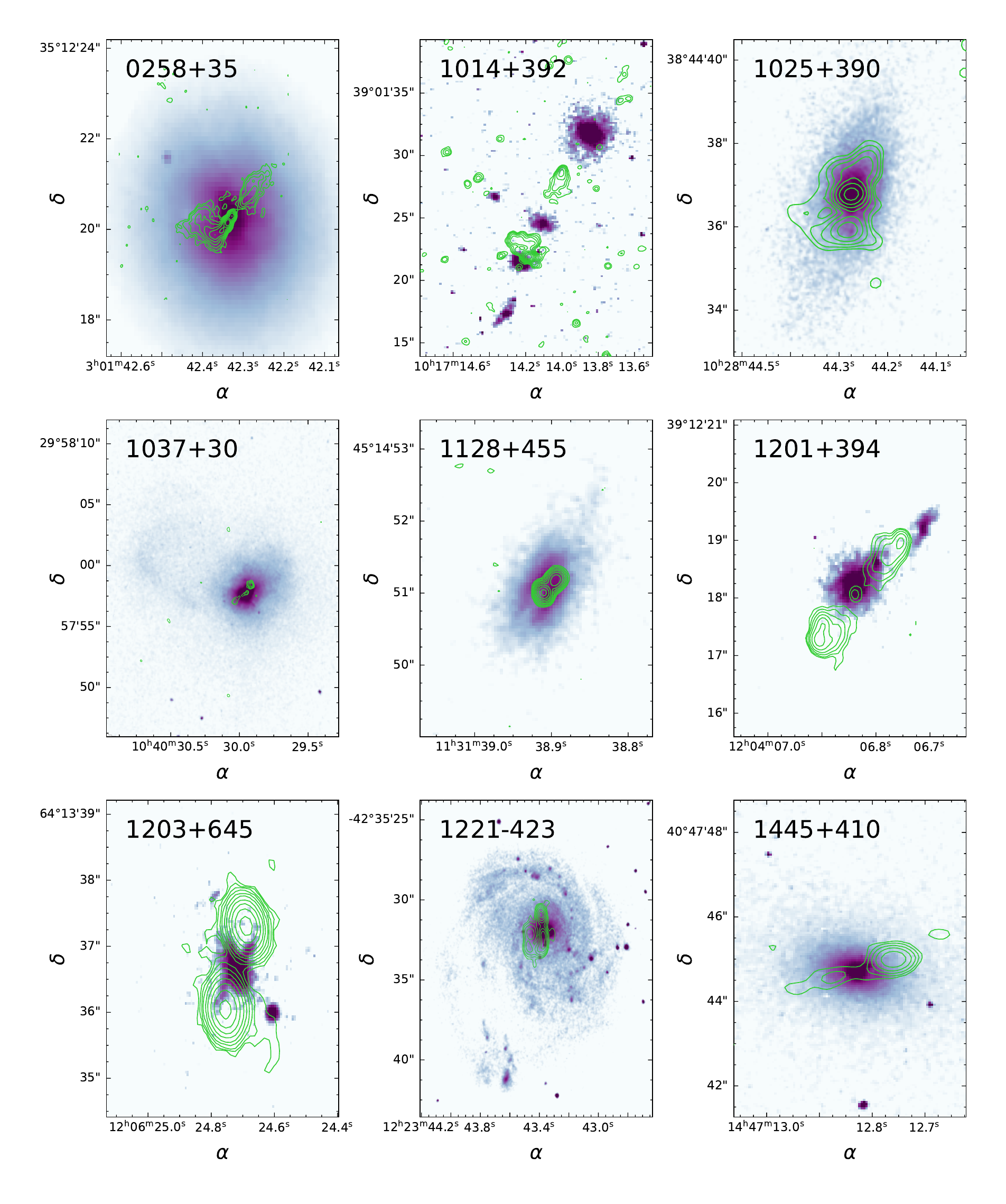}
\caption{\textit{HST} optical continuum band morphologies, overlaid with radio contours for the 9 radio galaxies in our sample (5 GHz: 1014+392, 1025+390, 1203+645, 1445+410; 8 GHz: 1037+30, 1201+394; 15 GHz: 1128+455; 22 GHz: 0258+35; 24 GHz: 1221$-$423). The contours are plotted at intervals defined by 2$^i \times 3\sigma$ mJy beam$^{-1}$ (where i$ = -$1, 1, 2, 3, ..., 10). All panels are rotated such that North is pointed up and East is left. The \textit{HST} vs. \textit{VLA} image registration is tied to the radio core positions in all sources. (In case of 1128+455 and 1203+645, the core positions are approximate. A possible core position for 1203+645 identified in 5 GHz map by \cite{1998MNRAS.299..467L} would shift the overlaid contours 0.3$''$ towards the SE.) \label{fig:opradio}}
\end{figure*}

\subsection{Imaging with HST}

High-resolution ($\sim$0.05$''$/pix) imaging was obtained for the 9 radio galaxies with the UVIS channel on \textit{HST}'s Wide Field Camera 3 (WFC3) in optical ($6000-8500$ \AA) and near-UV  ($2000-3500$ \AA) bands, in Cycle 25 GO program 15245 (PI: C. O'Dea) over 14 orbits. The filters and exposure details are summarized in Table \ref{tab:obs}. Filter selection was based on the need to capture the line-free continua. This allows our data to be free of contamination from emission line regions. We utilised narrow-band NUV continuum (between rest-frame [CIII]$\lambda$1909 and MgII$\lambda$2798) for sensitivity to emission from hot, massive (O- and B-type) stars, and medium-band optical continuum (between [OIII]$\lambda$ 5007 and [NII]$\lambda$ 6548) to determine the optical colors.

\begin{deluxetable}{lcccl}     
\tablecaption{Observation details \label{tab:obs}}
\tabletypesize{\footnotesize}
\tablehead{
\colhead{Source} & & 
\colhead{HST filter} & \colhead{Pivot} & \colhead{Exposure} \\
 & & & \colhead{wavelength} & \colhead{time} \\
 & & & \colhead{(\AA)} & \colhead{(sec)}
}
\startdata
\multirow{2}{*}{0258+35} & & F621M & 6219 & 1 x 700  \\
 & & F225W & 2359 & 1 x 1650 \\
\multirow{2}{*}{1014+392} & & F845M & 8436 & 2 x 700  \\
 & & F336W & 3355 & 2 x 1650 \\
\multirow{2}{*}{1025+390} & & F763M & 7612 & 2 x 700  \\
 & & F336W & 3355 & 2 x 1650 \\
\multirow{2}{*}{1037+30} & & F621M & 6219 & 1 x 476  \\
 & & F225W & 2359 & 1 x 1770 \\
\multirow{2}{*}{1128+455} & & F763M & 7612 & 2 x 700  \\
 & & F336W & 3355 & 2 x 1740 \\
\multirow{2}{*}{1201+394} & & F845M & 8436 & 2 x 700  \\
 & & F336W & 3355 & 2 x 1650 \\
\multirow{2}{*}{1203+645} & & F763M & 7612 & 2 x 700  \\
 & & F336W & 3355 & 2 x 1860 \\
\multirow{2}{*}{1221$-$423} & & F689M & 6876 & 1 x 700  \\
 & & F275W & 2704 & 1 x 1680 \\
\multirow{2}{*}{1445+410} & & F689M & 6876 & 1 x 700  \\
 & & F275W & 2704 & 1 x 1680 \\
\enddata
\end{deluxetable}

The imaging data were pre-calibrated through \textit{HST}’s standard \textit{calwf3} pipeline (includes bias $\&$ dark current subtraction, flat fielding and charge transfer efficiency corrections).  
Post-pipeline processing was done using the  \textsc{drizzlepac}\footnote{\url{https://www.stsci.edu/scientific-community/software/drizzlepac.html}} software \citep{2021drzp}. The \textit{tweakreg} task performs astrometric alignment of the individual exposures followed by bad-pixel/cosmic-ray rejection, geometric distortion correction and dithering to produce the final drizzled images with the task  \textit{astrodrizzle}. The drizzling process involves mapping the input image pixels onto pixels in the subsampled output image, taking into account the shifts and rotations between individual exposures. To avoid convolving the image with the large pixel ``footprint” of the instrument, \textit{astrodrizzle} allows the user to shrink the pixel before it is mapped into the output image by selecting the ``drop size" via \textit{pixfrac} parameter. A smaller drop size results in higher resolution and lower correlated noise, but tends to reduce sensitivity to low-surface brightness features. On the other hand, higher values would compromise on resolution. Hence, \textit{pixfrac} selection depends on science goals and is decided through visual inspection of the output image.

\subsection{Archival data} 

In this study, we use archival imaging and/or photometric magnitudes in near-UV, optical, infrared and radio bands.
NUV band (1771$-$2831 \AA) photometry with the \textit{Galaxy Evolution Survey (GALEX}; \cite{2005ApJ...619L...1M}) for the sample was collected from the MAST/\textit{GALEX} archive. Stacked image cutouts from the \textit{Panoramic Survey Telescope and Rapid Response System (PanSTARRS}; \cite{2016arXiv161205560C}) in $g,r,i,z,y$ bands were obtained from the PS1 catalog (only available for 8/9 galaxies because of \textit{PanSTARRS}' Southern declination limit of $-30^\circ$. We sourced pipeline-processed radio maps (5 GHz, 8 GHz and 15 GHz bands; A-config.) for 8 radio sources from the NRAO \textit{Very Large Array (VLA)} Image Archive; though in some cases, raw $uv$-data were obtained from the \textit{VLA} Data Archive and re-reduced. For 1221$-$423, 12 mm radio image from the \textit{Australia Telescope Compact Array (ATCA)} has been used (provided by \citealt{2010MNRAS.407..721J}; private communication). Infrared imaging and photometry data were obtained from the NASA/IPAC Infrared Science Archive, in the \textit{WISE} \citep{2010AJ....140.1868W} bands— W1 (3.4$\mu$m), W2 (4.3$\mu$m), W3 (12$\mu$m) and W4 (22 $\mu$m); as well as the K$-$band of \textit{Two Micron All Sky Survey (2MASS}; \cite{2006AJ....131.1163S}) available for 6/9 galaxies.

\section{Analysis} \label{sec:analysis}

\begin{deluxetable*}{lcccccccc}[ht!]    
\tablecaption{Extinction corrections for optical/UV photometry \label{tab:extcorr}}
\tabletypesize{\footnotesize}
\tablehead{
\colhead{Source} & & Band & \multicolumn{2}{c}{Galactic Extinction} & & \multicolumn{3}{c}{Internal Extinction} \\
\tableline
 & & & \colhead{E(B-V)$_{Gal}$} & \colhead{A($\lambda$)$_{Gal}$} 
  & &  
  \colhead{H$_{\alpha}/H_{\beta}$} & \colhead{E(B-V)$_{H_{\alpha}/H_{\beta}}$} & \colhead{A($\lambda$)$_{H_{\alpha}/H_{\beta}}$} \\
 & & & \colhead{(mags)} & \colhead{(mags)} & & \colhead{} & \colhead{(mags)} & \colhead{(mags)}
}
\startdata
\multirow{2}{*}{0258+35} & & F621M & \multirow{2}{*}{0.157} & 0.42 & & \multirow{2}{*}{5.54$^a$} & \multirow{2}{*}{0.67} & 1.80 \\
 & & F225W & & 1.30 & & & & 5.50 \\
\multirow{2}{*}{1014+392} & & F845M & \multirow{2}{*}{0.012} & 0.02 & & \multirow{2}{*}{\nodata} & \multirow{2}{*}{\nodata} & \multirow{2}{*}{\nodata} \\
 & & F336W & & 0.06 & & & & \\
\multirow{2}{*}{1025+390} & & F763M & \multirow{2}{*}{0.009} & 0.02 & & \multirow{2}{*}{4.73$^a$} & \multirow{2}{*}{0.51} & 1.03 \\
 & & F336W & & 0.05 & & & & 2.56 \\
\multirow{2}{*}{1037+30} & & F621M & \multirow{2}{*}{0.016} & 0.04 & & \multirow{2}{*}{8.82$^a$} & \multirow{2}{*}{1.13} & 3.06 \\
 & & F225W & & 0.13 & & & & 9.36 \\
\multirow{2}{*}{1128+455} & & F763M & \multirow{2}{*}{0.015} & 0.03 & & \multirow{2}{*}{33.87$^b$} & \multirow{2}{*}{2.46} & 5.05 \\
 & & F336W & & 0.08 & & & & 12.59 \\
\multirow{2}{*}{1201+394} & & F845M & \multirow{2}{*}{0.022} & 0.04 & & \multirow{2}{*}{\nodata} & \multirow{2}{*}{\nodata} & \multirow{2}{*}{\nodata} \\
 & & F336W & & 0.11 & & & & \\
\multirow{2}{*}{1203+645} & & F763M & \multirow{2}{*}{0.015} & 0.03 & & \multirow{2}{*}{17.86$^b$} & \multirow{2}{*}{1.84} & 3.74 \\
 & & F336W & & 0.07 & & & & 9.33 \\
\multirow{2}{*}{1221$-$423} & & F689M & \multirow{2}{*}{0.085} & 0.20 & & \multirow{2}{*}{6.36$^c$} & \multirow{2}{*}{0.80} & 1.92 \\
 & & F275W & & 0.54 & & & & 5.06 \\
\multirow{2}{*}{1445+410} & & F689M & \multirow{2}{*}{0.013} & 0.03 & & \multirow{2}{*}{8.22$^a$} & \multirow{2}{*}{1.06} & 2.53 \\
 & & F275W & & 0.08 & & & & 6.68 \\
\enddata
\tablecomments{A($\lambda$) corrections computed at the pivot wavelengths of UVIS filters (Table \ref{tab:obs}). Galactic foreground extinction estimated using E(B-V) from the NASA/IRSA Milky Way reddening map \citep{2011ApJ...737..103S}. Internal extinction is calculated from H$\alpha$ and H$\beta$ fluxes sourced from: (a) SDSS/DR12 catalog (central 3$''$ aperture), (b) \cite{2020MNRAS.491...92L} (SDSS/DR12 3$''$ aperture), (c) \cite{2010MNRAS.407..721J} (central 2.7$''$ aperture).}
\end{deluxetable*}

\begin{deluxetable*}{L|CCCCCCCCC}
\tablecaption{Optical to mid-infrared photometry of the 9 radio galaxies \label{tab:phot}}
\tablehead{
\colhead{Band} & \colhead{\textbf{0258+35}} & \colhead{\textbf{1014+392$^\dagger$}} & \colhead{\textbf{1025+390}} & \colhead{\textbf{1037+30}} & \colhead{\textbf{1128+455}} & \colhead{\textbf{1201+394}} & \colhead{\textbf{1203+645}} & \colhead{\textbf{1221-423}} & \colhead{\textbf{1445+410$^\dagger$}}
}
\tabletypesize{\footnotesize}
\startdata
\\
V & 12.32\pm0.00  & 20.64\pm0.24 & 18.27\pm0.07 & 16.05\pm0.02 & 19.54\pm0.13 & 19.68\pm0.16 & 20.33\pm0.18 & 17.75\pm0.06 & 18.47\pm0.07 \\
U & \nodata    &   \nodata    & 22.66\pm0.39 & 19.85\pm0.22 & 22.91\pm0.67 & 24.57\pm1.06 & 22.74\pm0.81 & 19.45\pm0.43 &   \nodata    \\
NUV & 18.63\pm0.15 & 21.75\pm0.03 & 21.28\pm0.04 & 19.36\pm0.11 & 23.41\pm0.14 & 22.46\pm0.52 & 23.15\pm0.22 & 19.66\pm0.14 & 21.87\pm0.16 \\
FUV &   \nodata    &   \nodata    & 21.91\pm0.06 & 20.66\pm0.28 &   \nodata    &   \nodata    &   \nodata    & 21.14\pm0.4  &   \nodata    \\
W1     &  8.37\pm0.00  & 14.70\pm0.03  & 13.96\pm0.03 & 12.45\pm0.02 & 15.15\pm0.04 & 15.04\pm0.03 & 13.51\pm0.02 & 11.98\pm0.02 & 14.38\pm0.03 \\
W2     &  8.38\pm0.00  & 14.34\pm0.05 & 13.36\pm0.03 & 12.35\pm0.01 & 14.69\pm0.06 & 14.51\pm0.05 & 12.43\pm0.02 & 11.60\pm0.01  & 13.85\pm0.03 \\
W3     &  6.99\pm0.00  & 10.83\pm0.13 & 10.78\pm0.10  & 10.44\pm0.00  & 11.14\pm0.14 & 11.91\pm0.27 & 10.18\pm0.05 &  8.53\pm0.00  & 11.71\pm0.14 \\
W4     &  4.96\pm0.00  & 8.69\pm0.46  &  8.22\pm0.00  &  7.34\pm0.00  & 8.01\pm0.24  &  8.78\pm0.00  & 7.62\pm0.13  &  6.76\pm0.00  &  8.76\pm0.00  \\
g & 12.88\pm0.01 & 22.05\pm0.05 & 19.51\pm0.01 & 16.01\pm0.01 & 21.03\pm0.01 & 21.39\pm0.01 & 23.34\pm0.01 &   \nodata    & 19.17\pm0.01 \\
r     & 11.59\pm0.01 & 20.30\pm0.00  & 17.97\pm0.01 & 15.21\pm0.01 & 19.34\pm0.01 & 19.64\pm0.01 & 20.10\pm0.01  &   \nodata    & 17.73\pm0.01 \\
i     & 11.22\pm0.01 & 19.64\pm0.01 & 17.22\pm0.00 & 14.83\pm0.00 & 18.79\pm0.01 & 18.61\pm0.01 & 19.51\pm0.01 &   \nodata    & 17.00\pm0.01  \\
z     & 10.62\pm0.01 & 19.41\pm0.01 & 16.74\pm0.01 & 14.32\pm0.00 & 18.24\pm0.01 & 18.05\pm0.01 & 18.12\pm0.01 &   \nodata    & 16.58\pm0.01 \\
y     & 9.45\pm0.01  & 17.53\pm0.01 & 15.86\pm0.01 & 13.21\pm0.01 & 16.99\pm0.01 & 16.63\pm0.01 & 20.96\pm0.01 &   \nodata    & 15.44\pm0.01 \\
\\
\enddata
\tablecomments{ $^\dagger$ denotes the control sample. All magnitudes are in the AB system. Bandpasses: \textit{V}— \textit{HST}/UVIS optical channel ($6000-8500$ \AA);  \textit{U}— \textit{HST}/UVIS ultraviolet channel $2000-3500$ \AA); 
\textit{NUV}— \textit{GALEX} near-ultraviolet ($1750-2800$ \AA); \textit{FUV}— \textit{GALEX} far-ultraviolet ($1350-1750$ \AA); \textit{W1, W2, W3, W4}— \textit{WISE} 3.4 $\mu$m, 4.3 $\mu$m, 12 $\mu$m and 22 $\mu$m bands; \textit{g, r, i, z, y}— \textit{PanSTARRS1} bands. The measured and catalog-obtained photometric magnitudes are corrected only for Galactic extinction (not for internal dust reddening; see Sec. \ref{subsec:photometry}).} 
\end{deluxetable*} 

\subsection{1D profiles $\&$ photometry}
\label{subsec:photometry}

Surface photometry is measured by fitting elliptical isophotes to extended sources to derive radial profiles, i.e., the variation of intensity and ellipticity with radius \citep{1987MNRAS.226..747J, 1999BaltA...8..535M, 2016ARA&A..54..597C}. This is an extensively-used method of measuring photometry for radio galaxies in the literature (e.g., \citealt{2000A&AS..143..369G, 2016ApJ...818..182V}).

Figure \ref{fig:opradio} presents the optical band $HST$ images of the target sample. We performed the isophotal analysis with the \textsc{ellipse}\footnote{\url{http://stsdas.stsci.edu/documents/SUG/UG_33.html}} routine using Python-based\footnote{\url{https://iraf-community.github.io/pyraf.html}} IRAF \citep{1993ASPC...52..173T}. Unrelated neighbouring galaxies and stars were masked out prior to the fitting. 
The isophotal profiles (surface brightness, ellipticity and position angle vs. radial distance) extracted from the optical and UV images, respectively, are presented in the Appendix. All of the host galaxies in the sample exhibit isophotal distortions indicating complex  structure.  
\newline

Photometry was computed based on the 1$\sigma$ isophote (i.e., the isophote with intensity one standard deviation above the mean of the sky background) selected as the integration aperture. Systematic uncertainty in the computed magnitudes was derived by adding the Poisson noise in source flux and RMS error from the sky background, in quadrature.

\begin{figure*}
\epsscale{1.1}
\plotone{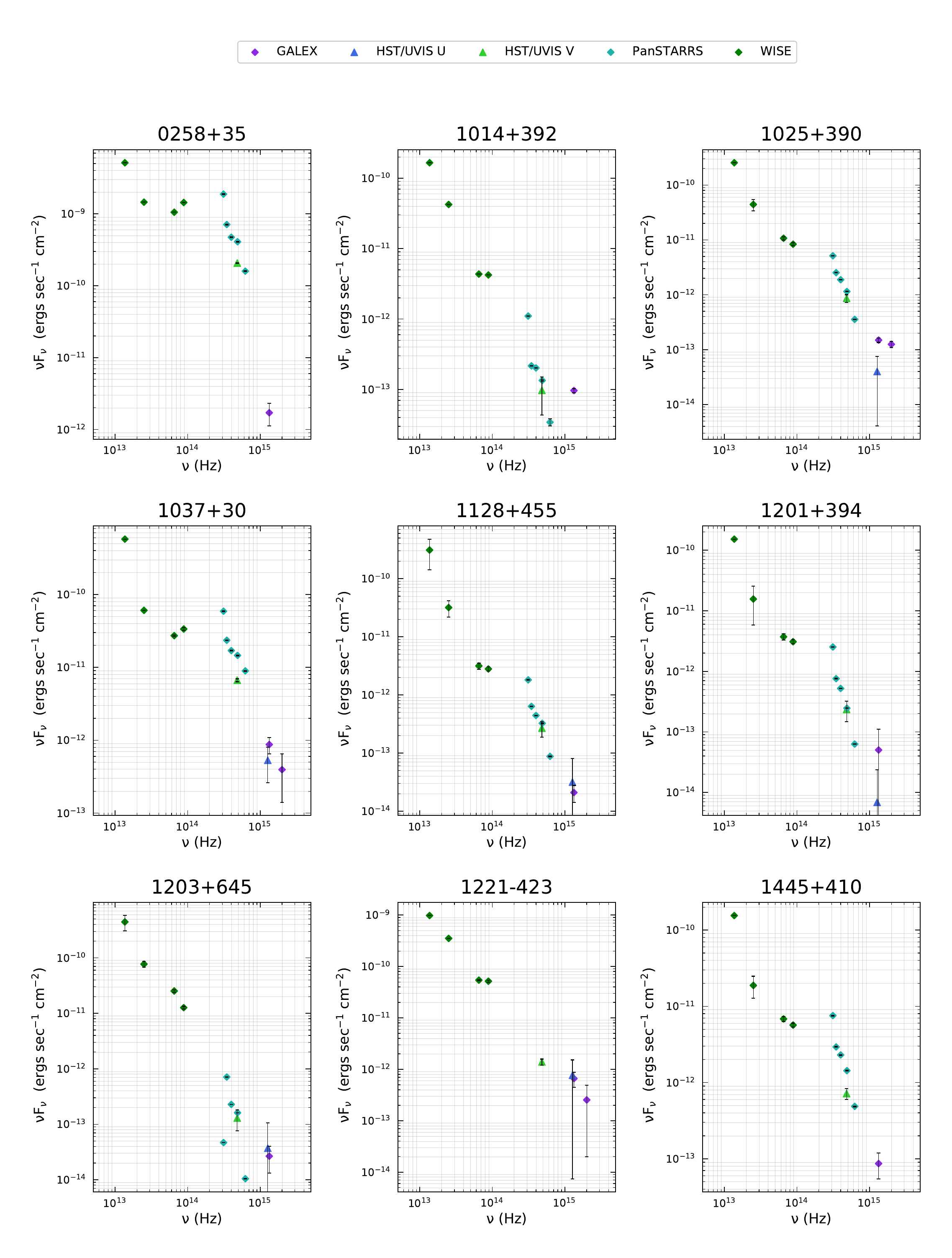}
\caption{Ultraviolet-to-infrared SEDs for the 9 compact radio galaxies in our sample. The data combine our photometric measurements for \textit{HST (UVIS} bands) and \textit{PanSTARRS (g, r, i, z, y)} imaging, with catalogued \textit{WISE (W1, W2, W3, W4)} and \textit{GALEX} NUV and FUV (where available) magnitudes. \label{fig:sed}}
\end{figure*}

The photometric measurements were corrected for Galactic extinction using the scaling relation by \citealt{1989ApJ...345..245C} and the E(B$-$V) color excess sourced from the NASA/IPAC\footnote{\url{https://irsa.ipac.caltech.edu/applications/DUST/}} archive. We also computed internal extinction corrections for the sources with available Balmer decrement ratios. The computed corrections and source references are listed in Table \ref{tab:extcorr}. Since the computed adjustments for internal dust reddening in the UV band turn out to be extremely high, we conclude that there must be very little extinction in the observed UV light, i.e., no significant obscuration of UV light by dust in our line of sight. Consequently, we did not correct the photometry for internal extinction. With regard to archival data, the \textit{GALEX} pipeline includes Galactic extinction correction for the cataloged magnitudes; and due to minimal dust extinction in the infrared bands, corrections for \textit{WISE} bands were ignored. 
The results from our \textit{HST} photometric analysis are given in Table \ref{tab:phot}, while the photometric spectral energy distributions (SEDs) are shown in Figure \ref{fig:sed}. 

\subsection{2D surface brightness modelling} \label{subsubsec:galfit}

Isophote fitting of host morphologies made it clear that there was underlying structure in almost all cases. The ubiquitous cross-shaped residual features in the central region in most of the sample warranted further examination in terms of bulge, disk and possible unresolved nuclear AGN source contributions. To this end, we performed two-dimensional galaxy modelling with \textsc{galfit}\footnote{\url{https://users.obs.carnegiescience.edu/peng/work/galfit/galfit.html}} (\citealt{2002AJ....124..266P, 2010AJ....139.2097P})— an iterative algorithm for structural decomposition of imaging data based on Levenberg-Marquardt minimization. \textsc{galfit} creates a model from a set of user-selected components (analytic functions that describe radial intensity distribution), convolves it with the instrument point spread function (PSF) and matches it to the input object via least-$\chi^2$ fitting. The PSF images for the WFC3 detector were generated with \textit{HST}-specific point spread simulation tool \textsc{tinytim}\footnote{\url{https://www.stsci.edu/hst/instrumentation/focus-and-pointing/focus/tiny-tim-hst-psf-modeling}} (\citealt{2011SPIE.8127E..0JK}). The image input is required to be in counts units (produced from count rates by multiplying exposure time), so \textsc{galfit} can self-construct the sigma map for $\chi^2$ computations.
\newline

\textit{PSF construction}. It is known that WFC3 pixels generally undersample the PSF, i.e., full width at half-maximum (FWHM) of PSF is less than 2 pixels. As this hinders proper convolution within \textsc{galfit}, pixel subsampling is needed for modelling the PSF. The best way for this is to generate a \textsc{tinytim} PSF that is over-sampled compared to data, to be convolved and re-binned within \textsc{galfit} before matching with data. The output \textsc{tinytim} PSF model has pixel dimensions $1/x$ of normal in each direction (for $1< x < 10$ , where x is the PSF sampling factor relative to data). We found subsampling factor of 2 to be the optimum value for creating better-than-Nyquist sampled model PSFs for the majority of source images (in case of 0258+25 and 1201+394, critically sampled \textsc{tinytim} PSF was sufficient).
\newline

\textit{Modelling strategy}. A typical \textsc{galfit} structural component has parameters that control basic source properties— profile, scale length, shape, and orientation. Starting with a single Sersic power law, other components can added to the model as needed to improve the fit, based on examining the residuals. Symmetric patterns, bipolar or quadrupolar features usually imply the requirement for additional components. We initially kept the entire parameter space free to vary, e.g., galaxy centroid position, effective radius, position angle for major axis and sky background. This produced the optimum results in most cases. For some galaxies where multiple Sersic profiles were fit, the model centroids and/or component Sersic indices were kept fixed. The best-fit was selected based on goodness of fit (reduced-$\chi^2$) and through visual inspection of residual images.  For the cases that generated relatively similar residuals or resulted in equivalent $\chi^2$ values, the combination with lower number of components was selected as the best-fitting model.

Stars and background galaxies were masked out prior to fitting. Detached, but possibly related, sources in the vicinity of the target galaxies were also masked to focus on target components. These neighbours were modelled separately and usually fit with single-component profiles. They could likely be cluster companions (e.g., the two sources towards north in case of 1201+394 and the southern source in 1203+645). Results of \textsc{galfit} modelling for the optical and UV band images are detailed in the Appendix. 
We use the \textsc{ellipsect}\footnote{\url{http://github.com/canorve/EllipSect}} (\citealt{2020zndo...4033448A}) software to generate radial surface brightness profiles displaying the constituent components of the best fitting \textsc{galfit} models. The best-fit model parameters are tabulated in Table \ref{tab:galfit}.

\begin{deluxetable*}{lcccccccccccc}
\tablecaption{Best-fit parameters from \textsc{galfit} modelling of host galaxies \label{tab:galfit}}
\tabletypesize{\footnotesize}
\tablehead{
\colhead{Source} & Component & $\Delta\alpha$ & $\Delta\delta$ & N & M$_{total}$ & R$_{eff}$ & ($\alpha, \beta, \gamma$) & R$_b$ & $\mu_{R_b}$ & R$_s$ & $b/a$ & PA \\
& & ($''$) & ($''$) & & (AB mag) & ($''$) & & ($''$) & (mag/$''$) & ($''$) & & (deg)
}
\decimalcolnumbers
\startdata
\multirow{3}{*}{0258+35} & \textit{sc} 
& 0.2 & 0.4 & 4.0 & 12.5 & 70.9 & \nodata & \nodata & \nodata & \nodata & 0.67 & --43 \\
 & \textit{sc} & 0.3 & 0.0 & 4.0 & 13.6 & 13.1 & \nodata & \nodata & \nodata & \nodata & 0.84 & --82 \\
 & \textit{sc} & 0.0 & 0.6 & 4.0 & 13.4 & 44.8 & \nodata & \nodata & \nodata & \nodata & 0.74 & +4 \\
 \\
1014+392 & \textit{sc} & 0.01 & 0.0 & 2.2 & 20.6 & 1.1 & \nodata & \nodata & \nodata & \nodata & 0.73 & +84 \\
\\
\multirow{3}{*}{1025+390} & \textit{sc} & 0.2 & 0.01 & 2.0 & 18.6 & 4.6 & \nodata & \nodata & \nodata & \nodata & 0.43 & --85 \\
 & \textit{sc} & 0.12 & 0.04 & 1.7 & 20.5 & 0.7 & \nodata & \nodata & \nodata & \nodata & 0.78 & --38 \\
 & \textit{ps} & 0.0 & 0.01 & \nodata & 22.0 & \nodata & \nodata & \nodata & \nodata & \nodata & 1.00 & \nodata \\
 \\
\multirow{3}{*}{1037+30} & \textit{sc} & 0.1 & 0.2 & 9.7 & 12.1 & 17.1 & \nodata & \nodata & \nodata & \nodata & 0.51 & +84\\
 & \textit{sc} & 0.04 & 0.02 & 3.4 & 26.4 & 1.3 & \nodata & \nodata & \nodata & \nodata & 0.64 & --41 \\
 & \textit{ps} & 0.05 & 0.05 & \nodata & 20.2 & \nodata & \nodata & \nodata & \nodata & \nodata & 1.00 & \nodata \\
 \\
1128+455 & \textit{sc} & 0.08 & 0.07 & 1.0 & 20.1 & \nodata & \nodata & \nodata & \nodata & 0.6 & 0.58 & --34 \\
\\ 
\multirow{2}{*}{1201+394} & \textit{nu} & 0.1 & 0.1 & \nodata & \nodata & \nodata & (4.1,4.0,0.4) & 0.1 & 20.3 & \nodata & 0.85 & +35 \\
 & \textit{sc} + \textit{m=1} & 0.1 & 0.1 & 1.2 & 21.2 & 0.9 & \nodata & \nodata & \nodata & \nodata & 0.49 & --37 \\
 & & & & (0.3,10.5)$^{**}$ & & & & & & & & \\
 \\
\multirow{2}{*}{1203+645} & \textit{nu} & 0.12 & 0.14 & \nodata & \nodata & \nodata & (0.1,2.2,0.2) & 0.2 & 20.8 & \nodata & 0.60 & --11\\
 & \textit{ps} & 0.06 & 0.06 & \nodata & 25.3 & \nodata & \nodata & \nodata & \nodata & \nodata & 1.00 & \nodata \\
 \\
\multirow{3}{*}{1221$-$423} & \textit{sc} & 0.01 & 0.05 & 0.3 & 19.8 & 0.3 & \nodata & \nodata & \nodata & \nodata & 0.71 & +65\\
 & \textit{sc} & 0.05 & 0.0 & 1.0 & 16.2 & \nodata & \nodata & \nodata & \nodata & 2.1 & 0.86 & +23 \\
 & \textit{sc} & 0.04 & 0.02 & 1.1 & 21.5 & 0.1 & \nodata & \nodata & \nodata & \nodata & 0.49 & +22 \\
 \\
\multirow{3}{*}{1445+410} & \textit{nu} & 0.08 & 0.05 & \nodata & \nodata & \nodata & (1.6,2.2,0.5) & 0.5 & 19.8 & \nodata & 0.67 & --65\\
 & \textit{sc} & 0.1 & 0.25 & 9.99 & 19.4 & 22.9 & \nodata & \nodata & \nodata & \nodata & 0.33 & --57\\
 & \textit{ps} & 0.0 & 0.0 & \nodata & 23.2 & \nodata & \nodata & \nodata & \nodata & \nodata & 1.00 & \nodata \\
 \tableline
 UV band \\
\tableline
1025+390 & \textit{sc} & 0.02 & 0.01 & 1.5 & 20.5 & 0.8 & \nodata & \nodata & \nodata & \nodata & 0.80 & --78 \\
\\
\multirow{2}{*}{1037+30} & \textit{nu} & 0.01 & 0.02 & \nodata & \nodata & \nodata & (1.9,9.6,1.1) & 2.8 & 23.5 & \nodata & 0.32 & --47\\
 & \textit{ps} & 0.05 & 0.05 & \nodata & 20.5 & \nodata& \nodata & \nodata & \nodata & \nodata & 1.00 & \nodata \\
 \\
1128+455 & \textit{sc} & 0.04 & 0.01 & 1.7 & 21.4 & 1.0 & \nodata & \nodata & \nodata & \nodata & 0.54 & --30 \\
1201+394 & \textit{sc} & 0.08 & 0.02 & 2.1 & 22.9 & 0.8 & \nodata & \nodata & \nodata & \nodata & 0.35 & --35 \\
\\
\multirow{3}{*}{1203+645} & \textit{sc} & 0.76 & 0.48 & 1.0 & 24.3 & \nodata & \nodata & \nodata & \nodata & 0.1 & 0.31 & --69\\
 & \textit{sc} & 0.18 & 0.33 & 0.1 & 22.0 & 0.8 & \nodata & \nodata & \nodata & \nodata & 0.39 & --56 \\
 & \textit{sc} & 0.54 & 0.24 & 0.1 & 22.6 & 1.2 & \nodata & \nodata & \nodata & \nodata & 0.18 & --65 \\
1221-423 & \textit{sc} & 0.02 & 0.05 & 2.5 & 19.4 & 1.0 & \nodata & \nodata & \nodata & \nodata & 0.98 & +8 \\
\enddata
\tablecomments{\textsc{galfit} modelling results for the compact radio source host galaxies. (1) Target name, (2) Best-fit model components: $sc=$ Sersic (bulge) model; $m=$ Fourier mode index ($^{**}$Fourier mode amplitude and phase angle relative to the Sersic component axis); $nu=$ Nuker (nuclear bar) model; $pc=$ nuclear point source component, (3) $\&$ (4) RA and declination offset of component centroid from galaxy's optical center (arcsec), (5) Sersic index, (6) integrated magnitude of the Sersic component, (7) The half-light or "effective" radius of the Sersic component (arcsec), (8) Nuker profile indices, (9) effective radius of the Nuker component (arcsec), (10) Surface brightness of the Nuker component, (11) Scale length of exponential disk ($n=1$ Sersic) component (arcsec), (12) Axis ratio, (13) Component position angle (Up=$0^\circ$; Left=$90^\circ$)}
\end{deluxetable*}

\section{Results} \label{sec:results}

\subsection{Continuum morphology} \label{subsec:maps}

The registration offset between \textit{HST} and \textit{VLA} images (up to $0.4''$) was eliminated by aligning the galaxy nucleus in the optical and UV images with the radio nucleus position. The optical continuum maps overlaid with \textit{VLA} radio contours are shown in Figure \ref{fig:opradio}. One-third of the sample exhibits disturbed morphological features e.g., large scale tidal tails and extended filamentary structures— clear signatures of tidal interaction— hinting at possible merger history or ongoing galaxy interactions. Galaxies 1201+394, 1203+645 and 1221$-$423 have well-resolved close companions that may be physically interacting; the former two being part of cluster environments. Two galaxies, 0258+30 and 1128+455, show prominent extended dust features. 
\newline

\begin{figure*}
\epsscale{1.2}
\plotone{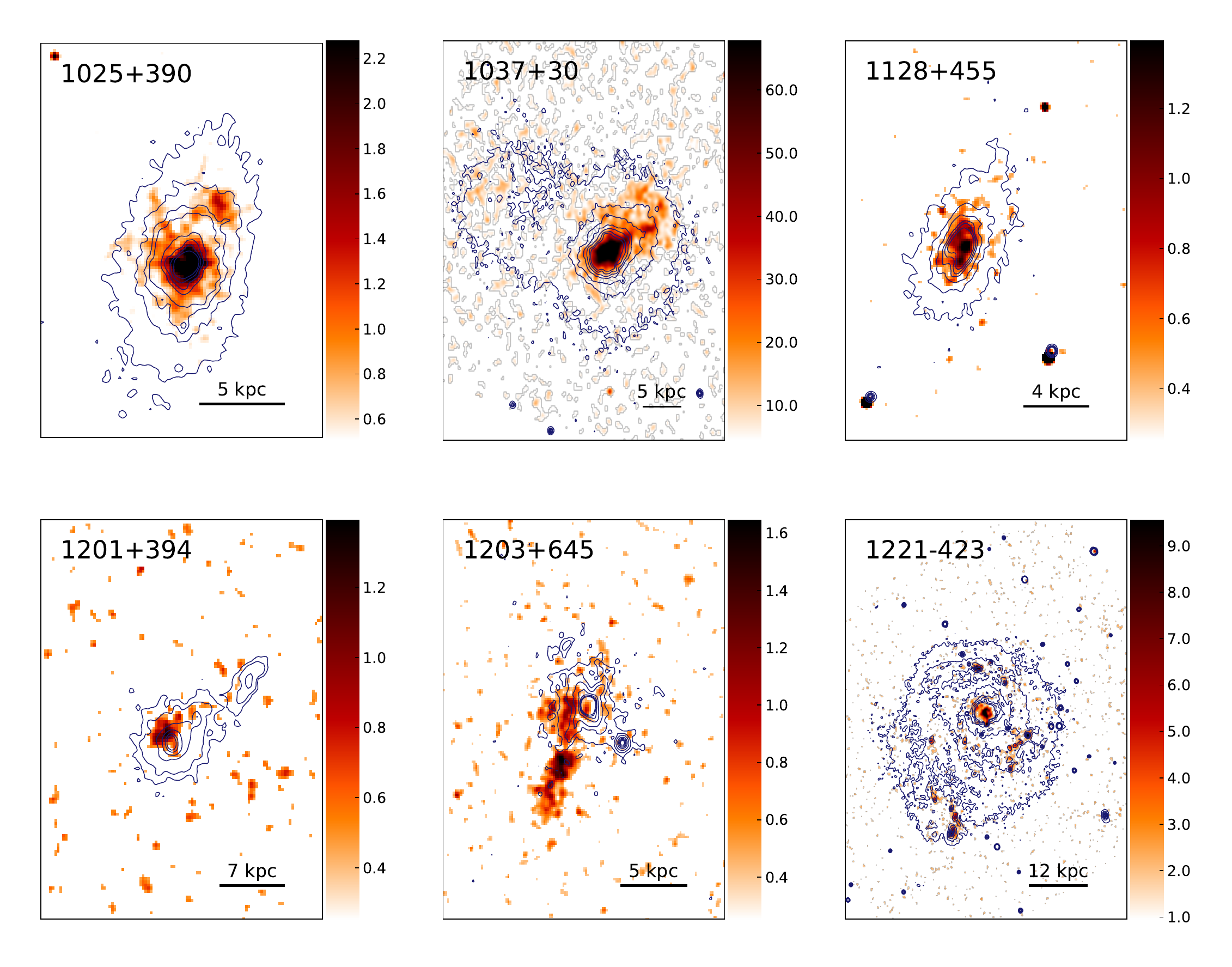}
\caption{\textit{HST} near-UV continuum maps (smoothed with a 1-pixel Gaussian) for 6 CSS radio galaxies overlaid with contours of optical continuum emission. The distribution of the younger stellar populations evident from extended UV emitting regions relative to the general galaxy morphology. All panels are rotated such that North is pointed up and East is left. Flux density on the colorbar is in units of $10^{-20}$ ergs sec$^{-1}$ cm$^{-2}$ \AA$^{-1}$. \label{fig:opuv}}
\end{figure*}

Our near-UV continuum images reveal extended UV emission in 6 out of the 7 CSS host galaxies. No UV emission was detected in the two non-CSS control targets. Figure \ref{fig:opuv} shows the UV band continuum maps relative to the visible morphology for the 6 UV-detected CSS galaxies. These overlays display clear distinction between the dominant old stars and the extended clumps of near-UV emission, revealing the young, massive star populations that possibly give these galaxies their blue [NUV$-r$] colors (Fig. \ref{fig:nuv-r}).

A great deal of research has been done in the past to decode the observed blue/UV excesses in powerful radio galaxies when compared with passive early-type galaxies (\citealt{1984MNRAS.211..833L, 1989ApJ...341..658S}). In addition to the starburst component, several activity-related factors could also contribute to the observed UV continuum: direct light \citep{1995MNRAS.275..703S}, scattered AGN radiation (e.g., \citealt{1992MNRAS.256P..53T, 1993MNRAS.264..421C, 1999AJ....118.1963C, 2007MNRAS.381..611H}), and nebular continuum from AGN-ionized emission-line nebulae (e.g., \citealt{1995MNRAS.273L..29D, 2002MNRAS.330..977T, 2002MNRAS.333..211W, 2007MNRAS.381..611H}). We plan to conduct follow-up observations to test for AGN-related contributions. 

Our sample includes relatively low redshift sources (\emph{z}$\leq$0.6) which are expected to show UV-emitting young stars, as earlier studies have found 30$-$50$\%$ of powerful radio galaxies at low and intermediate redshifts \emph{z}$<0.7$ show young stellar populations make a significant contribution to the UV/optical continua, after taking into account the AGN-related components (\citealt{2001MNRAS.325..636A, 2002MNRAS.330..977T, 2005MNRAS.356..480T, 2002MNRAS.333..211W, 2004MNRAS.347..771W}). 
Further, young and intermediate-age (few Myr to 1 Gyr old) stellar populations have been detected in compact radio galaxies \citep{2008A&A...477..491L, 2009AN....330..226H}. In light of these findings, we suggest that while it is possible that the flux near the galaxy cores in UV-bright hosts (Figure \ref{fig:opuv}) could be AGN-contaminated and possibly also include emission due to circumnuclear starbursts; the more extended regions and UV knots detected farther from the nucleus ($\sim$3-4 kpc in most cases) are likely to be shining with considerable emission from newborn, massive stars.
\newline

\begin{figure*}
\epsscale{1.2}
\plotone{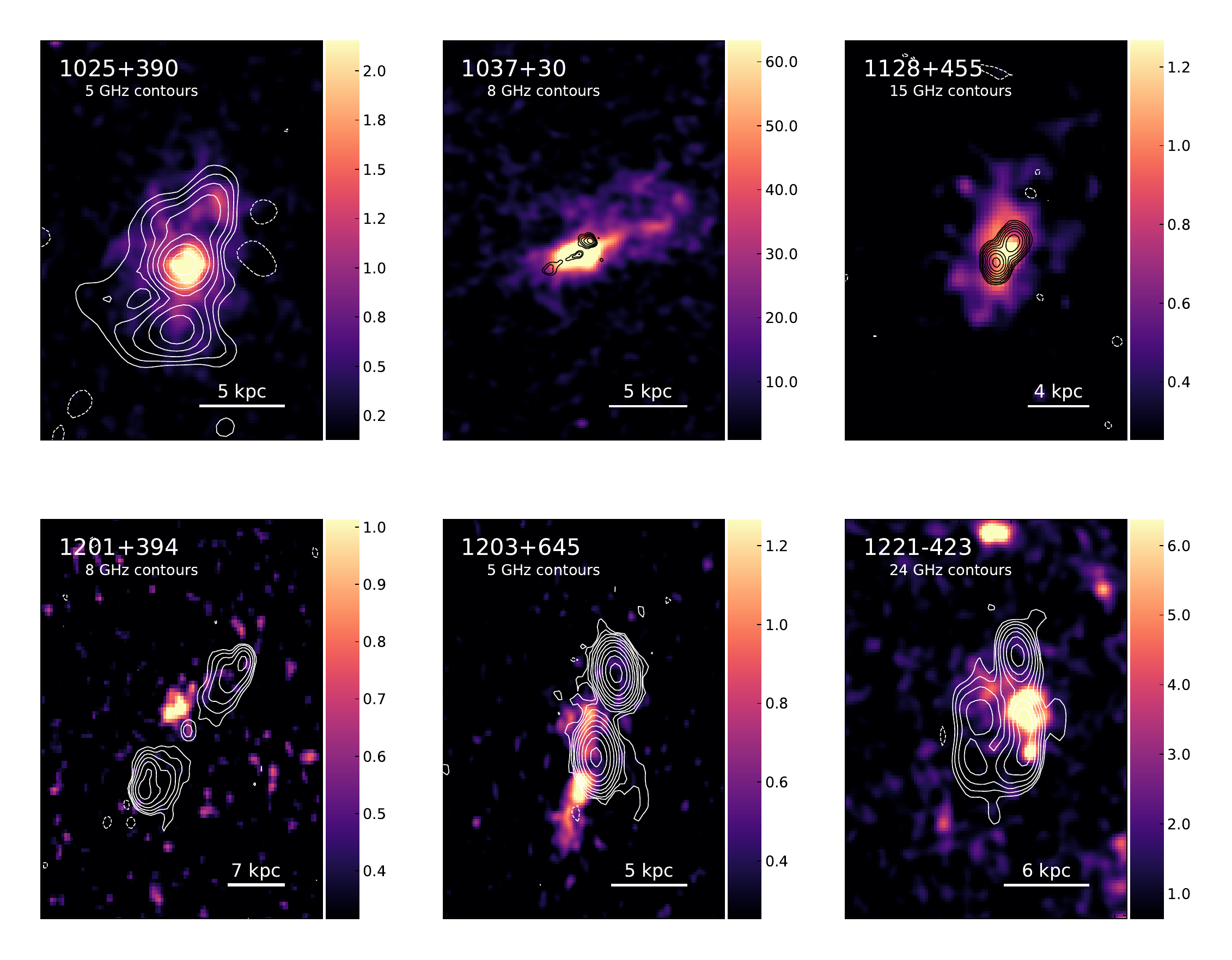}
\caption{\textit{HST} near-UV continuum maps (smoothed with a 1-pixel Gaussian) for 6 CSS radio galaxies overlaid with radio emission contours. The contours are plotted at intervals defined by 2$^i \times 3\sigma$ mJy beam$^{-1}$ (where i$ = -$1, 1, 2, 3, ..., 10). 
The UV-emitting regions show remarkable alignment with radio morphology, strongly suggesting jet-induced shock-triggered starbursts due to the expanding radio source. 
All panels are rotated such that North is pointed up and East is left. Flux density on the color bar is in units of $10^{-20}$ ergs sec$^{-1}$ cm$^{-2}$ \AA$^{-1}$. The \textit{HST} vs. \textit{VLA} image registration is tied to the radio core positions in all sources. (In case of 1128+455 and 1203+645, the core positions are approximate. A possible core position for 1203+645 identified in 5 GHz map by \cite{1998MNRAS.299..467L} would shift the overlaid contours 0.3$''$ towards the SE, further coinciding with the UV tail.) 
\label{fig:uvradio}}
\end{figure*}

Figure \ref{fig:uvradio} presents one of the main results of this paper: comparison of the near-UV extent with radio jet structure. We find remarkable spatial correlation between the size and position angles (PA) of the UV regions and radio lobes. The UV regions are strongly aligned along the jet axis in 5 out of 6 CSS hosts— 1025+390, 1037+30, 1128+455, 1203+645 and 1221$-$423. The radio/UV co-spatiality in these sources strongly suggests jet-driven starbursts, as the jet propagating through a dense ISM shock-triggers star forming activity by compressing the nearby gas clouds. 
The CSS host 1201+394 shows a small UV knot (of $\sim$8 kpc projected linear width) peripheral to the nucleus and the radio core, with slight elongation along the direction of the jet. It is interesting to note that the apparent offset of the young star population means that it is unlikely to be nuclear starburst or AGN-related continuum feature, but might be a result of starburst activity induced at a previous epoch when the radio source was smaller, before the lobes expanded farther outwards to their current size.

In some of these cases, the UV regions extend beyond the apparent radio source influence. Galaxy-wide starbursts are not uncommon in compact radio galaxies, having been found in both CSS and GPS hosts \citep{2007MNRAS.381..611H, 2008MNRAS.387..639H}. The UV emission spreads out to $\sim$3 kpc beyond the radio source in 1128+455, while the tidally-disrupted BCG 1037+30 has distinct star-forming clumps/filaments scattered out to roughly $\sim$8 kpc projected distance. 
Some of these UV regions lie beyond the radio lobes but are aligned along the radio source axis. This could be due to SF triggered by the bow shock$-$ the shock front expands outside the radio lobe and will cause compression of massive clouds in its path
(\citealt{1989ApJ...345L..21B, 2002ApJS..141..337C, 2011ApJ...728...29W}). 
The possible causes for the presence of the more dispersed young stellar populations could be— generic gas infall in local gas-rich environments, SF fuelled by merger events (likely in the hosts that exhibit tidal  features, e.g., 1037+30 and possibly, 1203+645) or persisting starburst activity originally ignited by an earlier cycle of radio emission whose remnant is no longer energetic enough to be observed in the high (GHz) frequency range. We delve into the latter scenario in light of the ages of SF activity in Sec. \ref{subsubsec:ages}. The $\sim$15 kpc tail-like feature extending beyond the galactic continuum in 1203+645 is a peculiar case, discussed in detail in Sec. \ref{sec:compiled}.

\begin{figure*}[h!]
\epsscale{1.1}
\plotone{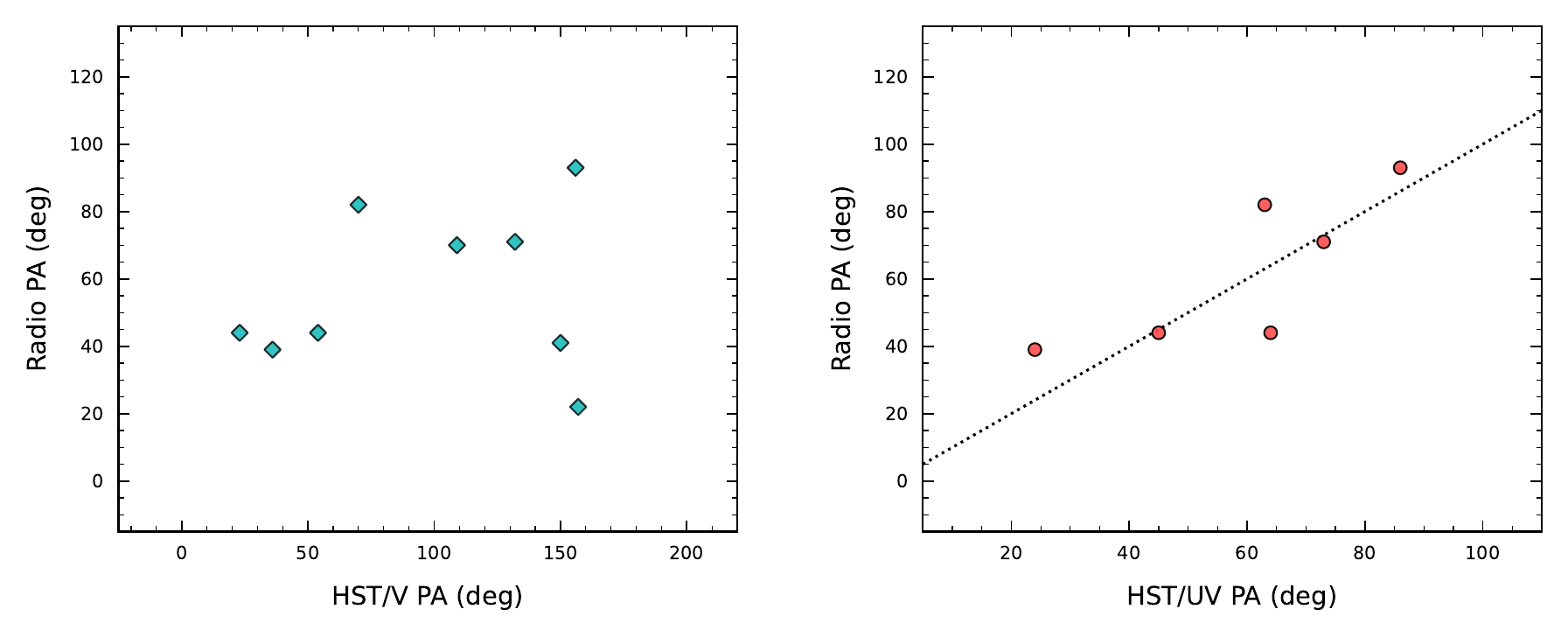}
\caption{A comparison of the position angles of the observed structure of the stellar \emph{(left panel)} and the near-UV continua \emph{(right panel)}, as measured from the $HST$ imaging, with radio source position angles. The general stellar population in the sample does not show spatial relationship with jet direction, while star-forming UV regions are closely correlated with the radio source axis. The dots mark the line with slope unity.\label{fig:pa1}}
\end{figure*}

\begin{figure*}[b!]
\epsscale{1.2}
\plotone{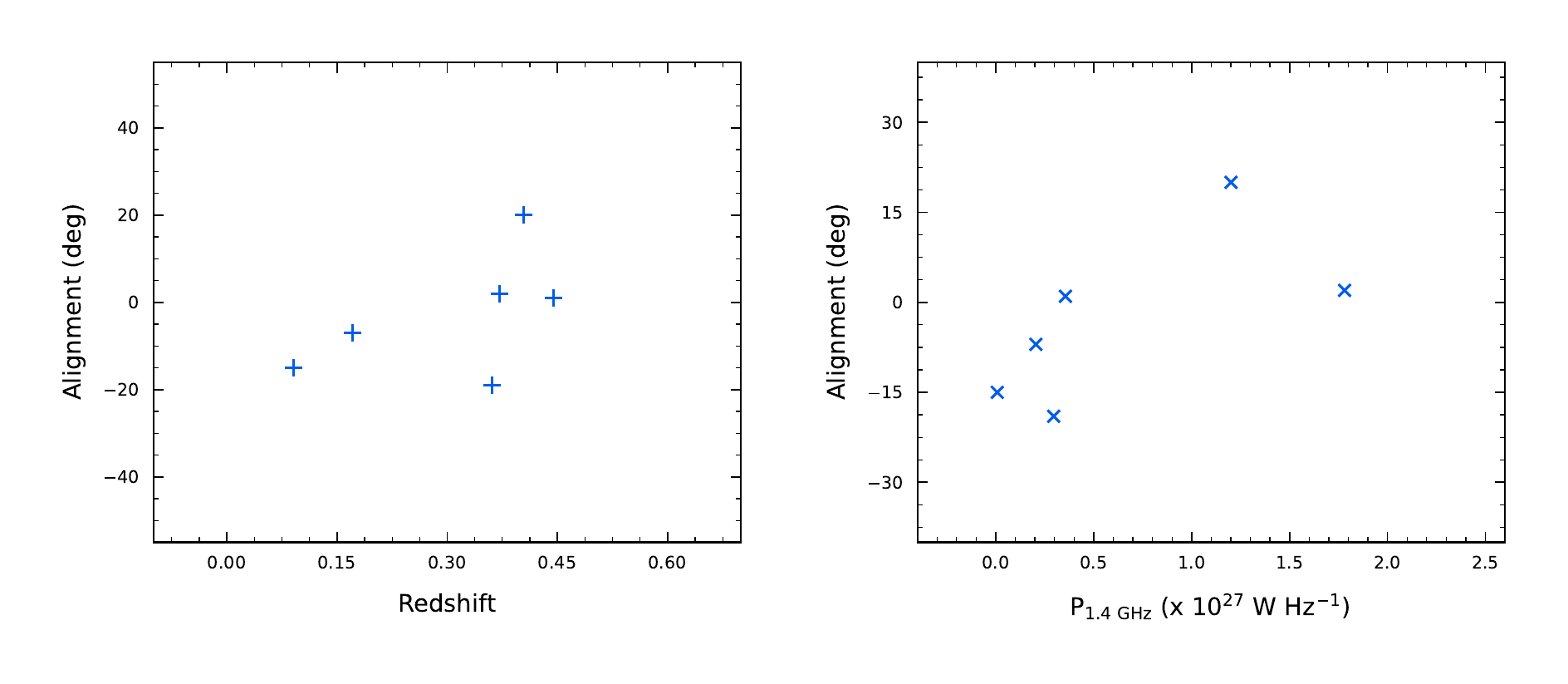}
\caption{Examination of the observed alignment [UV PA $-$ Radio PA] with respect to source redshifts \emph{(left panel)} and 1.4 GHz radio luminosity \emph{(right panel)}. No dependence or bias is apparent in either case. \label{fig:pa2}}
\end{figure*}

\begin{deluxetable}{lcccc}
\tablecaption{Observed position angles \label{tab:pa}}
\tabletypesize{\footnotesize}
\tablehead{
\colhead{Source} &
\colhead{$z$} & \colhead{\emph{HST}(V)} & \colhead{\emph{HST}(UV)} & \colhead{Jet axis} \\
 & & \colhead{(deg)} & \colhead{(deg)} & \colhead{(deg)}
}
\decimalcolnumbers
\startdata
0258+35 &   0.017    &    150    & \nodata &   41    \\
1014+392 &   0.536    &    109    & \nodata &   70    \\
1025+390 &   0.361    &    70     &  63   &   82    \\
1037+30 &   0.091    &    36     &  24   &   39    \\
1128+455 &   0.404    &    54     &  64   &   44    \\
1201+394 &   0.445    &    23     &  45   &   44    \\
1203+645 &   0.371    &    132    &  73   &   71    \\
1221$-$423 &   0.171    &    156    &  86   &   93    \\
1445+410 &   0.195    &    157    & \nodata &   22    \\
\enddata
\tablecomments{Major axis PAs for observed structure in the optical, near-UV and radio bands. The angles are measured from the horizontal in the North-up/East-left image orientation.}
\end{deluxetable}

A comparison of position angles of the observed structure in the optical and UV continua, with the radio source axis, is presented in Table \ref{tab:pa}. The PAs of the stellar continuum and radio emission in the sample do not show any general trend (Figure \ref{fig:pa1}). The direction of the UV emission, however, shows clear correlation with radio source orientation. In Figure \ref{fig:pa2}, we examine the offset between the radio and UV major axes considering source redshifts and 1.4 GHz luminosity. We do not find any dependence of UV/radio PA alignment on radio power or redshift. 
This extends similar results observed by \cite{2008A&A...477..491L}, with a GPS-dominated compact source sample, to CSS sources. 
The redshift vs. radio-UV alignment comparison in Fig. \ref{fig:pa2} also shows agreement with the observed trend in radio-EELR alignment$-$ unlike large-scale radio sources that only show the alignment effect at $z>0.6$, CSS sources can exhibit aligned light at all redshifts (e.g., \citealt{2008ApJS..175..423P, 2022JApA...43...97S}).
\newline

Other than the radio source inducing star formation in the host ISM, jet-cloud alignment in compact radio galaxies may hint at a possible observational selection effect. In a scenario suggested by some studies of powerful compact radio sources (e.g., \citealt{2011MNRAS.412..960T, 2012ApJ...745..172D, 2021AN....342.1200T, 2023arXiv230612636E}), the radio sources might so happen to be expanding into an ISM that is unusually rich in cool gas, e.g., the densest parts of the extended, but asymmetrical merger debris or in the plane of a gas-rich disk (in case of a late-type galaxy host). The subsequent boosting of the radio flux caused by the jet-cloud interactions as the jets expand into the ISM could lead to these compact radio sources being preferentially selected in radio flux-limited samples. The dense, cool gas will then be likely to trace regions of star formation even if not interacting with the radio source. Therefore, the radio-UV alignments might be explained by the jets expanding into the densest parts of large-scale gas structures where star formation is already taking place. This effect could be a factor in some of the gas-rich CSS hosts in our sample.

\subsection{\textsc{galfit} modelling}

Our objectives for modelling the surface brightness profiles of our sample were three-fold: (a) to extract morphological information about the compact source hosts in optical and UV bands-- do they show signs of interactions e.g., asymmetric or irregular features, hidden companions? What are the sizes and position angles of UV-emitting knots relative to the radio source? (b) to detect any hidden nuclear structure like bars or unresolved point source; (c) to disentangle dust features resolved by \textit{HST} in some of these galaxies (e.g., extended dust loops around the core of 0258+35; dense egde-on lane across 1128+455). These would be clearly discernible in the residual images after the visible components of the galaxy have been subtracted.
\newline

Clear evidence of substructure is confirmed in all the sources that show irregular isophotes in optical band. Most galaxies are fit by a combination of Sersic bulges; disk component (an $n=1$ Sersic) is not detected in the majority of sources, consistent with the general finding that CSS sources tend to be hosted in massive ellipticals. On the other hand, 30$\%$ of the sample is found to be without a detectable bulge. These galaxies are best fit by either a pure exponential disk (1128+455), a pure bar (1203+645) or a bar plus an exponential disk (1201+394). Bulgeless galaxies hosting AGN activity are a rare phenomemon (e.g., \citealt{2009ApJ...704..439S, 2009ApJ...690..267D}) and remarkably interesting in light of the well-known BH-bulge scaling relations. However, in our case, this may be an observation bias, as it is possible that bulges in these $z>0.3$ galaxies might be too faint to be detected in our imaging. Deeper, higher-resolution observations would be needed to confirm their bulgeless nature. In the sub-sample of 5 galaxies with detectable bulge, 60$\%$ of them need a Nuker profile (which translates into detection of a nuclear bar structure), either instead of or in conjunction with Sersic components, to properly fit the surface brightness distribution in the core. It is interesting to note that some studies (e.g., \citealt{2000ApJ...529...93K, 2009ASPC..419..402H}) have found correlation between the presence of stellar bars and AGN activity. In addition, our fitting suggests the presence of faint compact source components in the nuclei of about half of the targets-- typically in the sources classified in literature as NLRGs. This is consistent with faint nuclear emission due to obscuration. Another possibility is that some compact radio galaxies may  have a weak nuclear activity from a radiatively-inefficient low-luminosity AGN, as opposed to a bright, quasar-like nucleus.

In UV band, the bright knots are mostly best fit with single-component models (Sersic or Nuker profiles), with index range $0.1<n<3$. The one exception is an extended tail-like feature exhibited by 1203+645 in UV continuum, where two more low-index Sersic components (of flat but sharply truncated intensity curves) are needed for a close fit. The fit for 1025+390 shows a $\sim$3 kpc-sized UV-bright knot in the path of the jet. The presence of such sub-galactic UV clusters is consistent with star-forming regions. The UV band fitting for 1037+30 maintains the nuclear compact source suggested with optical fit; although given the disturbed, irregular inner structure of the galaxy, the point source component may not be reliable.  

The 2D fitting results in the optical and UV for the 9 galaxies are discussed further individually in Section \ref{sec:compiled}.

\subsection{High vs. Low excitation} \label{subsec:excitation}

A key requirement for predicting AGN activity-related UV emission is the estimate of the power of the accretion process. Accretion efficiency correlates with 'excitation type' in radio galaxies (\citealt{2007MNRAS.376.1849H, 2009MNRAS.396.1929H, 2012MNRAS.421.1569B}). High-Excitation radio galaxies (HERGs), i.e., those with strong high excitation-level emission lines in their optical spectrum, have larger total energy output than the weak-lined Low-Excitation radio galaxies (LERGs). These two AGN divisions differ fundamentally based on whether the accretion onto the central SMBH is radiatively efficient or inefficient. HERGs typically have accretion rates between 1-10 per cent of their Eddington rate, whereas LERGs predominately accrete at a rate $<$1 per cent of the Eddington rate. Because of the large radiative output of the strongly accreting HERGs, they are expected to be more capable of producing non-stellar UV continuum by scattering of accretion radiation as well as by ionizing gas in host ISM, compared to the LERG types.

\begin{deluxetable}{lccccc}
\tablecaption{Excitation classification based on emission line strengths \label{tab:excit}}
\tabletypesize{\footnotesize}
\tablehead{
\colhead{Source} & & \colhead{EI} & \colhead{[O II]/[O III]} & \colhead{Type} & \colhead{Reference}\\
\colhead{(1)} & & \colhead{(2)} & \colhead{(3)} & \colhead{(4)} & \colhead{(5)}
}
\startdata
0258+35 &  & 0.27    &  \nodata & LERG &   1   \\
1014+392 &  & \nodata &  \nodata & LERG &   2   \\
1025+390 &  & 0.64  &  \nodata &  LERG   &   3  \\
1037+30 &  &  0.51  &   \nodata &  LERG   &   3  \\
1128+455 &  & 0.89  &  \nodata &  LERG   &   3  \\
1201+394 &  & \nodata  &  15.24  &  LERG   &   4  \\
1203+645 &  & 1.53 &  \nodata  &  HERG   &   3  \\
1221$-$423 &  & \nodata &  1.66  &  LERG   &    4   \\
1445+410 &  & 0.98 &  0.22   &   HERG   &   1,4  \\
\enddata
\tablecomments{(1) Target name (2) Excitation Index (separation = 0.95) (3) [O II]/[O III] line flux ratio (4) Deduced LERG/HERG class (5) References: 1. EI = log([O III]/H$_\beta$)$-\frac{1}{3}$[log([N II]/H$_\alpha$)+log([S II]/H$_\alpha$)+log([O I]/H$_\alpha$)], \cite{2010A\string&A...509A...6B}; 2. Classified by \cite{2013MNRAS.430.3086G}; 3. EI values given by \cite{2020MNRAS.491...92L}; 4. LERG/HERG distinction based on [OII]/[OIII] line ratio ($>1$ corresponds to LERGs), \cite{1997MNRAS.286..241J}. The emission line data were sourced from \cite{2006PhDT........60E} (0258+35), \cite{2010MNRAS.407..721J} (1221$-$423), and the SDSS/DR12 spectral catalog (1201+394 and 1445+410).}
\end{deluxetable}

The HERG/LERG distinctions for our radio galaxy sample are listed in Table \ref{tab:excit}. 
We use two different methods for this classification, based on the availability of nuclear emission-line measurements for the sample. The first approach is the excitation index (EI), defined by \cite{2010A&A...509A...6B}, which combines flux ratios of emission lines$-$ H$\alpha \lambda$6563, H$\beta\lambda$4861, [O I]$\lambda$6300, [O III]$\lambda$5007, [O II]$\lambda$3727, [S II]$\lambda$6716+$\lambda$6731 and [N II]$\lambda$6583 (see Table \ref{tab:excit} caption for EI definition). In this system, the galaxies for which the nuclear spectra show EI$>0.95$ are categorized as HERGs. For the cases where data for the higher (rest) wavelength lines are not available, we switch to the excitation diagnostic by \cite{1997MNRAS.286..241J}, where the sources with an [O II]/[O III] flux ratio $>1$ are classified as LERGs. 
\newline

Out of the 9 radio galaxies in our sample, 7 are consistent with LERG-type emission lines. This includes 5 of the UV-detected CSS galaxies. Due to their low-power accretion, the observed UV continuum in these galaxies is not expected to have significant contribution from scattered AGN radiation; although nebular continuum emission from regions shocked and ionized by the expanding radio lobes (e.g., \citealt{1995MNRAS.273L..29D}) may still have some contribution to the extended UV emission, which demands further investigation. Of the 2 radio sources diagnosed as HERGs, the CSS galaxy 1203+645 exhibits radio-aligned UV light and is likely to have contamination from AGN-related factors in the UV. This is consistent with its BLRG nature and activity-dominated IR colors (Fig. \ref{fig:wise}). The other HERG radio source 1445+410 is a control source, larger than the CSS linear size, from which our imaging did not detect any UV emission.

\subsection{Quantifying star formation} \label{subsec:sf}

\subsubsection{Estimates from observed photometry} \label{subsubsec:sfindic}

The extinction-corrected photometric measurements can be used to estimate rates of star formation (SFR; in units of solar mass per year), utilizing several young star tracers over a range of wavelengths. While SFR indicators in the UV/optical range probe the direct stellar light emerging from galaxies (e.g., \citealt{1998ARA&A..36..189K, 2007ApJS..173..267S, 2013seg..book..419C}), those in the mid/far-IR bands probe the stellar light reprocessed by dust (e.g., \citealt{1998ARA&A..36..189K, 2009ApJ...692..556R, 2017ApJ...850...68C}). Other than observing stellar emission, the ionizing photon rate, as traced by the gas ionized by massive stars, can be used to define SFR tracers; these include recombination lines (primarily H$\alpha$) and forbidden emission lines in metals. It should be noted that the star formation rates derived from the emission-line and mid-IR continuum estimators are likely to represent upper limits due to possible AGN contamination.


We derive star formation rates using multiple tracers to be able to draw better conclusions on stellar populations by comparison of results at different wavelengths. It is important to note that the observed flux-to-SFR calibration with these star formation tracers are prone to systematic uncertainties from the initial mass function (IMF; an empirical function that describes the initial distribution of masses for a population of stars), dust content and metallicity. So, we take the assumed IMF and other model parameters into account to aid our subsequent comparative analysis with synthesised data.
\newline

\textit{UV continuum}: 
The integrated spectrum of galaxies in the UV band is dominated by young stars and the star forming rate scales linearly with luminosity. We use the \cite{1998ARA&A..36..189K} calibration (eq. [\ref{usfr}]) to compute SFRs from the \textit{HST}/UV observations, while the \textit{GALEX}-specific conversion relation by \cite{2007ApJS..173..267S} (eq. [\ref{galexsfr}]) is used for estimating SFRs with archival \textit{GALEX} photometry. Both of these relations are derived using a Salpeter IMF with mass limits 0.1 and 100 M$_{\odot}$.  The luminosities L$_{\nu}$ and L$_{GALEX}$ are in units of ergs sec$^{–1}$ Hz$^{–1}$, and valid over the wavelength ranges 1500–2800 $\AA$ and 1300-1800 $\AA$, respectively.

\begin{equation}
\textrm{SFR}\ (\textrm{M}_{\odot}\ \textrm{yr}^{-1}) = 1.4 \times 10^{–28}\ \textrm{L}_{\nu}  
\label{usfr}
\end{equation}

\begin{equation}
\textrm{SFR}\ (\textrm{M}_{\odot}\ \textrm{yr}^{-1}) = 1.08 \times 10^{–28}\ \textrm{L}_{GALEX}  
\label{galexsfr}
\end{equation}

Scaling relations usually adopt the ``continuous star formation" approximation, i.e., it is assumed that the SFR has remained constant over timescales that are long compared to the lifespan of the dominant UV emitting population ($<$10$^8$ yr). Therefore, it is worth noting here that if star formation has been active in a region on a timescale shorter than about 100 Myr, the cumulative UV emission of massive stars is still increasing in luminosity, and so, the UV-based SFR would consequently be underestimated. \cite{1998ARA&A..36..189K} found that a $\sim$10$^6$ yr old population in continuous burst would yield SFRs that are 57$\%$ higher than those given in eq. [\ref{usfr}]. 
\newline

\textit{Ionized gas emission}: Young, massive stars produce copious amounts of ionizing photons that ionize the surrounding gas. Only stars of masses $\gtrsim$10 M$_\odot$ and lifetimes of $<$20 Myr would produce enough photon flux to ionize the nebulae. Hence, emission line-generated SFR measures are expected to be independent of older star formation history, and so, are more sensitive to changes in SFR over short timescales ($\sim$ a few Myr) than other tracers (e.g., \citealt{1998ARA&A..36..189K, 2013seg..book..419C}). \cite{1998ARA&A..36..189K} give the conversion factors to compute the SFRs from recombination H$\alpha$ line and forbidden [OII]$\lambda3727$ doublet, employing the same Salpeter IMF ($0.1–100$ M$_\odot$) as the UV-band relations. L$_{H\alpha}$  and L$_{[O II]}$ luminosities are in ergs sec$^{-1}$.

\begin{equation}
\textrm{SFR}\ (\textrm{M}_{\odot}\ \textrm{yr}^{-1}) = 7.9 \times 10^{–42}\ \textrm{L}_{H\alpha}
\end{equation}

\begin{equation}
\textrm{SFR}\ (\textrm{M}_{\odot}\ \textrm{yr}^{-1}) = 1.4 \times 10^{–41}\ \textrm{L}_{[O II]}   
\end{equation}

\textit{mid-IR continuum}: 
An indirect SFR diagnostic, the mid-infrared emission traces the dust heated by UV-luminous, young stellar populations; for which the IR SED is more luminous and peaks at shorter wavelengths ($\sim$10$-$100 $\mu$m). \cite{2009ApJ...692..556R} derived the linear correlation between SFR and single-band 24 $\mu$m infrared luminosity at the galaxy-wide scale (eq. [\ref{irsfr}]). We use WISE 22 $\mu$m (W4) band luminosity, expressed in units of L$_\odot$, for our IR SFR estimates.

\begin{eqnarray}
\textrm{SFR}\ (\textrm{M}_{\odot}\ \textrm{yr}^{-1}) = 7.8\times 10^{–10}\ \textrm{L(24 $\mu$m, L$_\odot$),} \nonumber \\
\textrm{for}\ 6\times10^8\ \textrm{L}_\odot \leq \textrm{L(24)} \leq 1.3\times 10^{10}\ \textrm{L}_\odot \nonumber \\
\nonumber \\
= 7.8\times 10^{–10}\ \textrm{L(24 $\mu$m, L}_\odot) \nonumber \\
\times \ [7.76\times 10^{–11}\ \textrm{L(24 $\mu$m, L}_\odot)]^{0.0048}, \nonumber \\
\textrm{for}\ \textrm{L(24)}>1.3\times10^{10}\ \textrm{L}_\odot
\label{irsfr}
\end{eqnarray}

\textit{Composite calibrations}:
 \cite{2009ApJ...703.1672K} developed scaling relations from linear combinations of optical emission-line luminosities with single-band IR luminosity to produce internal attenuation-corrected SFRs. We use these to draw comparison with SFRs from other indicators not corrected for internal dust extinction. L$_{H\alpha}$ and L$_{[O II]}$ are extincted luminosities in ergs sec$^{-1}$.

\begin{equation}
\textrm{SFR}\ (\textrm{M}_{\odot}\ \textrm{yr}^{-1}) = 7.9 \times 10^{-42}\ [\textrm{L}_{H\alpha}\ +\ 0.020\times \textrm{L(24 $\mu$m)}]
\end{equation}

\begin{equation}
\textrm{SFR}\ (\textrm{M}_{\odot}\ \textrm{yr}^{-1}) = 8.1 \times 10^{-42}\ [\textrm{L}_{[O II]}\ +\ 0.029\times \textrm{L(24 $\mu$m)}]
\end{equation}

\begin{deluxetable*}{lccccccc}
\tablecaption{Estimates of star formation rates from various indicators \label{tab:sfr}}
\tabletypesize{\footnotesize}
\tablehead{
\colhead{Source} &  \colhead{SFR$_{HST/UV}$} &  \colhead{SFR$_{GALEX}$} &   \colhead{SFR$_{H\alpha}$} &  \colhead{SFR$_{[OII]}$} & 
\colhead{SFR$_{22\mu m}$} & \colhead{SFR$_{22\mu m+H\alpha}$} &  \colhead{SFR$_{22\mu m+[OII]}$}  \\
 & \colhead{(M$_\odot$ yr$^{-1}$)} &  \colhead{(M$_\odot$ yr$^{-1}$)} &  \colhead{(M$_\odot$ yr$^{-1}$)} &  \colhead{(M$_\odot$ yr$^{-1}$)} &  \colhead{(M$_\odot$ yr$^{-1}$)} &  \colhead{(M$_\odot$ yr$^{-1}$)} &  \colhead{(M$_\odot$ yr$^{-1}$)}
}
\decimalcolnumbers
\startdata
\\
1025+390 & 1.97 & 5.41 & 5.42 & 9.27 & 55.69 & 45.57 & 65.06 \\
1037+30 & 1.21 & 1.47 & 0.83 & 1.31 & 5.35 & 5.0 & 6.96 \\
1128+455 & 2.04 & 0.99 & 11.95 & 4.12 & 90.16 & 75.55 & 96.93 \\
1201+394 & 0.56 & 3.00 & \nodata & 1.95 & 54.61 & \nodata & 59.73 \\
1203+645 & 1.95 & 1.03 & 38.18 & 4.82 & 106.32 & 112.6 & 113.44  \\
1221$-$423 & 6.87 & 4.37 & 1.09 & 0.19 & 38.12 & 29.06 & 41.69  \\
\\
\enddata
\tablecomments{SFRs from observed fluxes in different bands for the 6 UV-detected CSS host galaxies. The estimates are computed from scaling relations detailed in Sec \ref{subsubsec:sfindic}. All UV, IR and emission-line luminosities used in these calculations are corrected for Galactic (foreground) extinction only. The composite SFR diagnostics in columns (7) and (8) include internal attenuation correction by definition. H$\alpha$ flux is not available for the galaxy 1201+394. Note that AGN photoionization may have contributed to the ionized gas emission and mid-IR continuum used here as star formation tracers, so these SFR estimates are to be considered upper limits. }
\end{deluxetable*}

The star forming rates estimated from the various tracers are given in Table \ref{tab:sfr}. The UV-derived SFRs are in general the lowest compared to those inferred from other indicators$-$ this suggests low contamination from scattered AGN light to the observed UV continuum. It is also possible that dust may be affecting the UV observations, although this appears unlikely given the high internal extinction estimates (Sec. \ref{subsec:photometry}).
SFRs deduced from optical emission-lines are more or less in agreement with the UV rates, with the exception of BLRG hosts 1128+455 and 1203+645. The ubiquitous high SFRs of 1128+455 in all bands are consistent with its IR-ultraluminous classification on the WISE color-color plot (Fig. \ref{fig:wise}). In the case of 1203+645, the emission-line luminosities likely have contribution from AGN-ionized regions, from evidence of nuclear photoionization in its EELRs (e.g., Shih+13; also see Sec \ref{sec:compiled}).
In general, the internal attentuation-corrected composite SFRs (Columns 7 and 8 in Table \ref{tab:sfr}) are not substantially higher than the 22 $\mu$m SFRs in all galaxies$-$ this is consistent with the low dust obscuration hypothesis (see Sec \ref{subsec:photometry}).

\subsubsection{Comparison with stellar population models} \label{subsubsec:sfmodel}

We now compare the \textit{HST}/UV photometry of the UV-detected subset of our sample with evolutionary models from the stellar population synthesis code \textsc{starburst99}\footnote{\url{https://www.stsci.edu/science/starburst99/docs/default.htm}} (\citealt{1999ApJS..123....3L, 2005ApJ...621..695V, 2010ApJS..189..309L, 2014ApJS..212...14L}) to examine star formation models and stellar ages that are consistent with a population of stars that could: (i) cause the UV continuum emitting clumpy regions in the jet vicinity, and (ii) produce sufficient ionizing photons to power the nebula. 

\textsc{starburst99} provides predictive SEDs of a young stellar population from the FUV to the NIR, with varying parameters$-$ initial mass function (IMF), mass range, metallicity and whether the starburst continuously formed stars (expressed in terms of star formation rate), as opposed to the single burst scenario (in which case a given starburst mass evolves through time). We consider both continuous and instantaneous star formation models that include stellar and nebular emission, with varying power-law index $\alpha$ for the Salpeter IMF and different mass cutoffs. 
Near-solar metallicity is chosen (Z$=$0.008), since all the scaling relations used in the previous section for SFR estimations are computed assuming solar abundance (e.g., \citealt{1998ARA&A..36..189K, 2013seg..book..419C}). Using synthetic photometry packages \textsc{synphot} \citep{2018ascl.soft11001S} and \textsc{stsynphot}\footnote{\url{https://github.com/spacetelescope/stsynphot$\_$refactor}} \citep{2020ascl.soft10003S}, each model SED is redshifted to the redshift of the target source and then convolved with the relevant \textit{HST} filter transmission curve. The redshift-corrected and bandpass-convolved model is then used to compute the predicted effective stimulus (effstim, eq. [\ref{effstim}]), i.e., integrated flux in the given filter bandpass in the desired flux units, which in our case were ergs sec$^{-1}$ cm$^{-2}$ $\AA^{-1}$ so as to be readily comparable with flux units of \textit{HST} observations.

\begin{equation} 
effstim = \frac{\int F_\lambda P_\lambda \lambda d\lambda}{\int P_\lambda \lambda d\lambda} 
\label{effstim}
\end{equation}

where $\lambda$ is the wavelength, P$_\lambda$ is the filter throughput and F$_\lambda$ is the bandpass-convolved model flux distribution. These ``artificial" fluxes synthesized from starburst models can then be directly compared with our observed, extinction-corrected photometry. 
\newline

\textit{Emission in the UV continuum}. For each source, we obtained predicted fluxes for models at ages between 1 Myr to 1 Gyr. This gave us estimates of the star formation rates or the initial starburst masses required to produce the observed UV continuum at different epochs in the evolution of a young stellar population. Tables \ref{tab:contsfr} $\&$ \ref{tab:instsfr} list the results for continuous and instantaneous bursts, respectively, for three chosen epochs-- 1 Myr, the earliest available stellar age with \textsc{starburst99} models; a reasonably young 10 Myr; and the intermediate age of 100 Myr. The selection of these epochs illustrates the general trend in the amount of star formation with each order of magnitude in age. Figure \ref{fig:sfepoch} shows the detailed variation of the predicted values with time for all the model scenarios.
\newline

On comparison of the synthetic model estimates with the extinction-corrected \textit{HST}/UV fluxes, we find that in general: 

(i) The closest fit to the observed, low SFRs (a few M$_\odot$ per year, column 2 in  Table \ref{tab:sfr}) is the model normalized to a continuous SFR of 1 M$_\odot$ yr$^{-1}$, with Salpeter IMF of slope 2.35,  upper and lower mass limits of 100 M$_\odot$ and 1 M$_\odot$, respectively, and solar abundances where Z = 0.020 (i.e., 'Model 1' in Table \ref{tab:contsfr} and Figure \ref{fig:sfepoch}). A $\sim$10$^6$ yr old continuously star-forming nebula is most consistent with the regions emitting UV continuum in the CSS hosts. The other two models considered approach the observed low SFRs at late ($>$10 Myr in most cases) epochs.

(ii) If an instantaneous burst is assumed, the intial starburst masses of 10$^7$ M$_\odot$ to 10$^8$ M$_\odot$ are required for a rapid starburst triggered between $10^6$ and $10^7$ years ago, to result in a population of hot stars producing the observed UV flux in the target sample. The variation between models is relatively small in this case.
\newline

\textit{Can the nebula be powered by hot stars?} We consider whether the observed FUV continuum is consistent with a sufficient number of hot stars that could ionize the nebula. The number of ionizing photons required to power the nebula is given by:

\begin{equation}
Q_{tot} = 2.2 \frac{L_{H\alpha}}{h\nu_\alpha}
\end{equation}

where $\nu_\alpha$ is the rest frequency of H$\alpha$ emission line and $h$ is Planck’s constant. The Case B recombination scenario is assumed \citep{2006agna.book.....O}, as is usual for most computations concerned with SFRs, for a nebula that is optically thick to ionising photons.

We compare the total number of ionizing photons derived using H$_\alpha$ luminosities with the \textsc{starburst99} predictions for ionizing photon numbers in the HI, HeI and HeII continuum (spectral range below $\sim$912 \AA) vs. age, for each stellar synthesis model.
H$_\alpha$ flux measurements for our CSS galaxies are taken from previous spectroscopic studies in literature (sources listed in Table \ref{tab:extcorr}). The 3$''$ aperture for the slit spectra include nuclear emission as well as the more extended ionized gas regions.
\newline

The star formation rates and starburst masses that would produce the estimated number of ionizing photons are included in Tables \ref{tab:contsfr} $\&$ \ref{tab:instsfr}, respectively. 
In case of continuous production of stars, our results show that the observed rates of star formation are comparable or higher than those required by our best fitting model (i.e., $\sim$10$^6$ yr old SED with Salpeter IMF slope of 2.35 and upper mass limit of 100 M$_\odot$) in most of the sources. Hot, young stars could therefore provide the bulk of the photons ionizing emission-line nebulae, in the majority of CSS galaxies in our sample. The other two continuous starburst models considered are generally inconsistent with our source sample at most epochs. Considering the instantaneous case, the masses of initial starburst required to produce the expected amount of photons lie between $\sim$10$^6$ M$_\odot$ to $10^{10}$ M$_\odot$ over the source sample. These values remain more or less constant over different models. Note that this calculation does not take into account AGN photonization for excitation of the nebular clouds, in the case of CSS sources with a HERG progenitor. So, the ``required" SFRs and starburst masses should, in general, be considered upper limits. 
\newline

The last column of Tables \ref{tab:contsfr} $\&$ \ref{tab:instsfr} indicates the ratio of SFRs (or starburst masses) needed to account for observed flux to those needed for producing ionizing photons. 
A ratio $>$1 would mean that enough ionizing photons are generated from the young stellar population to ionize the nebula. The resulting ratios are all of order unity, suggesting that FUV emission due to the young stellar population has sufficient strength to power the emission-line nebula. This is consistent with our hypothesis that the UV light is produced by young stars rather than scattered nuclear light or AGN-ionized nebulae.
\newline

\subsection{Age of starburst vs. radio source lifetimes} \label{subsubsec:ages}

Compact radio source ages have been estimated in the literature from their kinematic as well as radiative properties. From proper motion considerations, multi-epoch observations of compact CSO jets show hotspots propagating outwards (with respect to the galaxy nucleus) with velocities in the range $\sim$0.04$c$ to $0.4c$, with a median value of $\sim$0.1$c$ \citep{2021A&ARv..29....3O}. The corresponding dynamical ages evident from these velocities are of the order of $\sim$10$^2-$10$^3$ yr for CSOs (e.g., \citealt{1998A&A...336L..37O, 1998A&A...337...69O, 2002evn..conf..139P, 2012ApJ...760...77A}); while for the relatively larger CSS sources, dynamical ages of $10^4-$10$^5$ yr have been estimated (e.g., \citealt{2005A&A...441...89G}). 

Spectral ageing estimates, that take into account the synchrotron radiative lifetimes of electrons in the jet lobes, also suggest similar values. Lobe-dominated CSS sources have been found to have typical radiative ages of $\sim$10$^3-$10$^5$ yr (e.g., \citealt{1999A&A...345..769M, 2003PASA...20...19M}). 
\newline

Here, we consider the jet lobe propagation method for estimating the ages of our compact radio source sample. Assuming a typical advance velocity of $0.1c$, we roughly calculate dynamical ages for the 6 CSS radio sources in Table \ref{tab:ages}. Two of these sources, 0258+30 and 1037+30, were also part of the \cite{2005A&A...441...89G} sample. Their study estimated a $4.5\times10^4$ yr kinematic age for the CSS source in 1037+30 assuming an expansion velocity of $0.2c$. For 0258+35, the radio structure shows no hotspots to measure the jet advancement, but \cite{2005A&A...441...89G} have estimated a radiative age of $9\times10^5$ yr. These values are in agreement with our estimates.

\begin{deluxetable}{lccc}[t!]
\tablecaption{Dynamical age estimates for the UV-detected CSS sample \label{tab:ages}}
\tabletypesize{\footnotesize}
\tablehead{
 & \colhead{Source} & \colhead{T$_{dyn}$} &\\
 & & \colhead{(yr)} &
}
\startdata
\\
 & 1025+390 & 1.6 x 10$^4$ &\\
 & 1037+30 & 5.6 x 10$^4$ &\\
 & 1128+455 & 4.9 x 10$^4$ &\\
 & 1201+394 & 1.2 x 10$^5$  &\\
 & 1203+645 & 7.3 x 10$^4$  &\\
 & 1221$-$423 & 4.4 x 10$^4$ &\\
\\
\enddata
\tablecomments{T$_{dyn}=LS/v$; where $LS$ is projected linear separation at source redshift (listed in Table 1) and $v$ is the expansion velocity in the source rest frame ($\approx$ $0.1c$).}
\end{deluxetable}

\begin{figure*}[ht!]
\epsscale{1.1}
\plotone{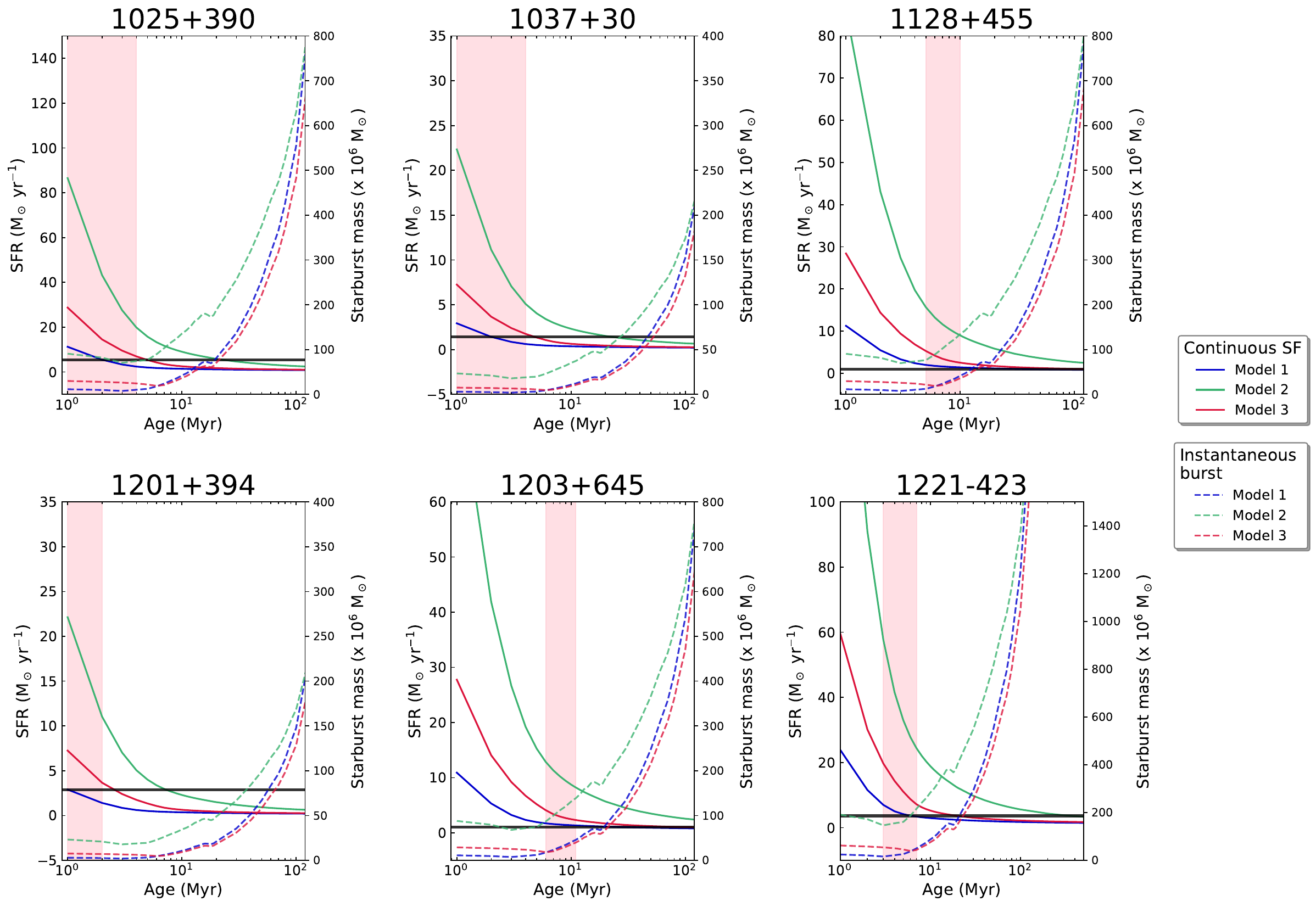}
\caption{\textsc{starburst99} population synthesis-derived parameters vs. starburst age for the 6 CSS hosts showing radio-aligned star-forming regions. The left axes indicate SFRs for continuous-starburst models (Models 1,2,3 in Table \ref{tab:contsfr}). The solid black line marks the "observed" SFR derived from $HST$/UV flux using the \cite{1998ARA&A..36..189K} calibration. The coloured region shows the starburst age estimate from the best-fit model (solid blue line). A $\sim$1$-$8 Myr old continuously star-forming nebula is most consistent with the observed UV emission in our CSS sample. On the right axes, we plot the variation of predicted initial-starburst masses required to produce the observed flux with age, in an instantaneous-starburst scenario. In this case, the tested models generally agree that a $\sim$10$^7-$10$^8$ M$_\odot$ burst triggered $\sim$1$-$10 Myr ago would produce the hot, massive stars that generate the observed UV.  \label{fig:sfepoch}}
\end{figure*}

Our analysis shows the star forming activity that would emit the observed UV continuum to be older, at $\leq$10$^6$ yr, than the mean CSS radio source aged $ \sim$10$^4$ yrs. 
For star formation induced as a result of jet propagation, the newly produced stellar population would be expected to have ages comparable to radio source expansion timescales. 
There are a few explanations for the apparent gap in our age estimations: 

Firstly, as described in Sec. \ref{subsubsec:sfindic}, UV-to-SFR scaling relations inherently produce highly underestimated star forming rates, and hence the SFR values from Table \ref{tab:sfr} we use to calibrate our models are, in fact,  lower limits. Higher SFRs correspond to much younger ($\lesssim$1 Myr) starburst ages, as evident from the population synthesis curve (Fig. \ref{fig:sfepoch}). 

Secondly, the youngest stellar population simulated by \textsc{starburst99}'s predictive modelling is 1 Myr old. This puts a constraint on star formation timescales accessible for $\sim$few M${_\odot}$/yr continuous production. For reference, massive O-types have a lifespan of $<$6 Myr. So the UV emission by a typical O-type star during its first million years on the main sequence is not taken into account by evolutionary synthesis. Due to this inherent model limitation, the possibility that the observed UV is radiated by hot, luminous stars that are at a much younger stage ($\lesssim$ 10$^5$ yrs) is not ruled out.

Thirdly, the jets may be older than estimated from their projected linear sizes. Evidence exists that compact (CSS and PS) radio sources can signify restarted radio activity (e.g., \citealt{2001AJ....121.1915O, 2005A&A...443..891S, 2009BASI...37...63S, 2010ApJ...715..172T, 2012A&A...545A..91S, 2021A&ARv..29....3O}). A known example of these "rejuvenated" scenarios is fortuitously part of our observed CSS sample (0258+35; Sec. \ref{sec:compiled}) as well, where fainter, lower-frequency radio emission from an earlier phase of activity exists distinct from newborn steep-spectrum radio lobes. It is possible that the $\sim$10$^6$ yr old stars suggested by the models were produced in star forming activity ignited by an older episode of radio emission, that may or may not have faded/cooled beyond detectable limit. Low-frequency radio observations of our source sample could help provide deeper insight in this context. Interestingly, in case of double-double radio galaxies (which are typical examples of recurring jet outbursts in the same direction, e.g., \citealt{2019MNRAS.486.5158N, 2000MNRAS.315..371S}), it has been shown that the strength of the alignment effect in optical and UV emission correlates with the linear size of radio source \citep{1996MNRAS.280L...9B, 2000MNRAS.315..381K}, and hence presumably with the age of the radio source.

Further, it is possible that the radio source is confined to the compact scales due to being frustrated by interaction with a dense host galaxy ISM (e.g., \citealt{1984IAUS..110...59V, 1984Natur.308..619W, 1991ApJ...380...66O}). This would mean that the radio source could be much older than its dynamical age estimate; the interaction with host ISM having slowed down the jet propagation (e.g., \citealt{1991ApJ...371...69D, 1999MNRAS.309..273H, 2000ApJ...534..201W}). Simulations suggest that the jets could remain frustrated for about 1 to 2 Myr in sufficiently dense environments \citep{2021A&ARv..29....3O}. In this case, the radio source will not be as young as estimated from proper motion-based arguments and thus capable of inducing star formation that results in the $\sim$few Myr old stellar population suggested by the UV emission. Observations to trace molecular gas and column densities from X-ray absorption (e.g., \citealt{2019ApJ...884..166S}) will help probe 
the likelihood of this scenario in our CSS sample.

Another, albeit less likely, possibility is that the star formation we detect could be triggered in a galaxy merger or be a hybrid of jet-induced and merger-induced starbursts. It is known that features consistent with merger interactions (companion objects and/or distorted isophotes for instance) are a common occurrence in PS/CSS sources \citep{2021A&ARv..29....3O}. 
The majority of CSS sources in our sample show evidence of a merger history, owing to their perturbed morphologies (Fig. \ref{fig:opradio}). This prompts us to consider the case where a common merger event triggered both nuclear activity and star formation in the host, or a \cite{2011A&A...528A.110F}$-$like "composite" scenario where star forming activity was triggered due to a past merger, subsequently enhanced by the AGN jet outburst. This might, in principle, explain the existence of an intermediate-age stellar population that is older than the young radio source ($\sim$10$^4-10^5$ yr). However, in a case where star formation and AGN are triggered by the same event, studies show a $\sim$10$^7-10^9$ yr time delay between the onset of the starburst and the start of radio activity (e.g., \citealt{2005MNRAS.356..480T, 2008A&A...477..491L,  2010ApJ...715..172T, 2011A&A...528A.110F}), corresponding to the timescale over which gas is transported from kpc to sub-kpc scales following a merger. The stellar population expected in this scenario would be much older than the $\sim$10$^6$ yr ages detected in our study. Hence, merger-related scenarios are less likely to have contributed in generating the young stellar populations observed in the CSS sample.

\vspace{-15mm}

\begin{deluxetable*}{cccccccccc}
\tablecaption{Continuous Star Formation \label{tab:contsfr}}
\tabletypesize{\footnotesize}
\tablehead{
\colhead{Source} & \multicolumn{3}{c}{$\dot{m}$(f$_{uv}$)} & \multicolumn{3}{c}{$\dot{m}$(Q$_{tot}$)} &  \multicolumn{3}{c}{$\dot{m}$(f$_{uv}$)/$\dot{m}$(Q$_{tot}$)} \\
 & \multicolumn{3}{c}{(M${_\odot}$ yr$^{-1}$)} & \multicolumn{3}{c}{(M${_\odot}$ yr$^{-1}$)} & \\
& \colhead{1 Myr} & \colhead{10 Myr} & \colhead{0.1 Gyr} & \colhead{1 Myr} & \colhead{10 Myr} & \colhead{0.1 Gyr} & \colhead{1 Myr} & \colhead{10 Myr} & \colhead{0.1 Gyr}
}
\startdata
Model (1) \\ 
\tableline
\\
1025+390 & 11.8 & 1.5 & 0.9 & 7.1 & 2.0 & 2.0 & 1.66 & 0.75 & 0.45 \\
1037+30 & 7.5 & 1.0 & 0.6 & 1.1 & 0.3 & 0.3 & 6.82 & 3.33 & 2.00 \\
1128+455 & 12.0 & 1.5 & 0.9 & 15.6 & 4.4 & 4.3 & 0.77 & 0.34 & 0.21 \\
1201+394 & 3.2 & 0.4 & 0.2 & \nodata & \nodata & \nodata & \nodata & \nodata & \nodata \\
1203+645 & 11.6 & 1.5 & 0.9 & 49.3 & 13.8 & 13.8 & 0.24 & 0.11 & 0.07 \\
1221$-$423 & 65.8 & 8.6 & 4.9 & 1.4 & 0.4 & 0.4 & 47.0 & 21.5 & 12.25 \\
\\
\tableline
Model (2) \\ 
\tableline
\\
1025+390 & 90.6 & 9.5 & 2.8 & 97.0 & 23.5 & 23.2 & 0.93 & 0.40 & 0.12 \\
1037+30 & 60.2 & 6.6 & 1.7 & 14.8 & 3.6 & 3.5 & 4.07 & 1.83 & 0.49 \\
1128+455 & 92.0 & 9.6 & 2.8 & 213.5 & 51.7 & 51.1 & 0.43 & 0.19 & 0.05 \\
1201+394 & 24.5 & 2.5 & 0.8 & \nodata & \nodata & \nodata & \nodata & \nodata & \nodata \\
1203+645 & 89.3 & 9.3 & 2.7 & 676.2 & 163.7 & 161.8 & 0.13 & 0.06 & 0.02 \\
1221$-$423 & 516.8 & 55.3 & 14.7 & 19.5 & 4.7 & 4.7 & 26.5 & 11.77 & 3.13 \\
\\
\tableline
Model (3) \\ 
\tableline
\\
1025+390 & 30.1 & 2.7 & 1.1 & 51.4 & 8.7 & 8.6 & 0.59 & 0.31 & 0.13 \\
1037+30 & 21.8 & 2.0 & 0.7 & 7.9 & 1.3 & 1.3 & 2.76 & 1.54 & 0.54 \\
1128+455 & 30.3 & 2.7 & 1.1 & 113.1 & 19.2 & 18.9 & 0.27 & 0.14 & 0.06 \\
1201+394 & 8.0 & 0.7 & 0.3 & \nodata & \nodata & \nodata & \nodata & \nodata & \nodata \\
1203+645 & 29.6 & 2.6 & 1.1 & 358.2 & 60.8 & 59.7 & 0.08 & 0.04 & 0.02 \\
1221$-$423 & 181.1 & 16.2 & 6.2 & 10.3 & 1.7 & 1.7 & 17.58 & 9.53 & 3.65 \\
\\
\enddata
\tablecomments{\hspace{2mm}$\dot{m}$(f$_{uv}$) and $\dot{m}$(Q$_{tot}$) are the estimated star formation rates that would produce the observed UV continuum and the required number of ionizing photons, respectively; at age epochs of $10^6$, $10^7$ and $10^8$ yrs. The \textsc{starburst99} models used for the estimates are normalized to a continuous SFR of 1 M$_{\odot}$ yr$^{-1}$, solar abundance (Z=0.008) and lower mass limits of 1 M$_{\odot}$. (1) Salpeter IMF of slope $\alpha = 2.35$ and upper mass cutoff M$_{up}=100$ M$_{\odot}$; (2) Salpeter IMF of slope $\alpha = 3.30$ and upper mass cutoff M$_{up}=100$ M$_{\odot}$; (3) Salpeter IMF of slope $\alpha = 2.35$ and upper mass cutoff M$_{up}=30$ M$_{\odot}$. For the source 1201+394, the H$\alpha$ luminosity needed for estimating Q$_{tot}$ was not available.}
\end{deluxetable*}

\begin{deluxetable*}{cccccccccc}
\tablecaption{Instantaneous Star Formation\label{tab:instsfr}}
\tabletypesize{\footnotesize}
\tablehead{
\colhead{Source} & \multicolumn{3}{c}{log $M$(f$_{uv}$)} & \multicolumn{3}{c}{log $M$(Q$_{tot}$)} &  \multicolumn{3}{c}{log $M$(f$_{uv}$)/log $M$(Q$_{tot}$)} \\
 & \multicolumn{3}{c}{(log M${_\odot}$)} & \multicolumn{3}{c}{(log M${_\odot}$)} \\
& \colhead{1 Myr} & \colhead{10 Myr} & \colhead{0.1 Gyr} & \colhead{1 Myr} & \colhead{10 Myr} & \colhead{0.1 Gyr} & \colhead{1 Myr} & \colhead{10 Myr} & \colhead{0.1 Gyr}
}
\startdata
Model (1) \\ 
\tableline
\\
1025+390 & 7.1 & 7.6 & 8.8 & 6.9 & 9.2 & 13.7 & 1.03 & 0.83 & 0.64 \\
1037+30 & 6.9 & 7.5 & 8.5 & 6.1 & 8.4 & 12.9 & 1.13 & 0.89 & 0.66 \\
1128+455 & 7.1 & 7.6 & 8.8 & 7.2 & 9.5 & 14.1 & 0.99 & 0.80 & 0.62 \\
1201+394 & 6.5 & 7.1 & 8.2 & \nodata & \nodata & \nodata & \nodata & \nodata & \nodata \\
1203+645 & 7.1 & 7.6 & 8.8 & 7.7 & 10.0 & 14.6 & 0.92 & 0.76 & 0.60 \\
1221$-$423 & 7.8 & 8.4 & 9.4 & 6.2 & 8.5 & 13.0 & 1.26 & 0.99 & 0.72 \\
\\
\tableline
Model (2) \\ 
\tableline
\\
1025+390 & 8.0 & 8.2 & 8.8 & 8.0 & 9.8 & 13.9 & 1.00 & 0.84 & 0.63 \\
1037+30 & 7.8 & 8.0 & 8.5 & 7.2 & 9.0 & 13.0 & 1.08 & 0.89 & 0.65 \\
1128+455 & 8.0 & 8.2 & 8.8 & 8.4 & 10.2 & 14.2 & 0.95 & 0.80 & 0.62 \\
1201+394 & 7.4 & 7.6 & 8.3 & \nodata & \nodata & \nodata & \nodata & \nodata & \nodata \\
1203+645 & 8.0 & 8.1 & 8.8 & 8.9 & 10.7 & 14.7 & 0.90 & 0.76 & 0.60 \\
1221$-$423 & 8.7 & 8.9 & 9.5 & 7.3 & 9.1 & 13.2 & 1.19 & 0.98 & 0.72 \\
\\
\tableline
Model (3) \\ 
\tableline
\\
1025+390 & 7.5 & 7.6 & 8.7 & 7.7 & 9.1 & 13.7 & 0.97 & 0.84 & 0.64 \\
1037+30 & 7.4 & 7.5 & 8.4 & 6.9 & 8.3 & 12.9 & 1.07 & 0.90 & 0.65 \\
1128+455 & 7.5 & 7.6 & 8.7 & 8.1 & 9.5 & 14.0 & 0.93 & 0.80 & 0.62 \\
1201+394 & 6.9 & 7.0 & 8.2 & \nodata & \nodata & \nodata & \nodata & \nodata & \nodata \\
1203+645 & 7.5 & 7.6 & 8.7 & 8.6 & 10.0 & 14.5 & 0.87 & 0.76 & 0.60 \\
1221$-$423 & 8.3 & 8.4 & 9.4 & 7.0 & 8.4 & 13.0 & 1.19 & 1.00 & 0.72 \\
\\
\enddata
\tablecomments{\hspace{2mm}$M$(f$_{uv}$) and $M$(Q$_{tot}$) are the estimated total mass of an instantaneous starburst that would produce the observed UV continuum and the required number of ionizing photons, respectively; at age epochs of $10^6$, $10^7$ and $10^8$ yrs. The \textsc{starburst99} models used for the estimates assume an instantaneous burst of star formation, solar abundance (Z=0.008) and lower mass limits of 1 M$_{\odot}$. (1) Salpeter IMF of slope $\alpha = 2.35$ and upper mass cutoff M$_{up}=100$ M$_{\odot}$; (2) Salpeter IMF of slope $\alpha = 3.30$ and upper mass cutoff M$_{up}=100$ M$_{\odot}$; (3) Salpeter IMF of slope $\alpha = 2.35$ and upper mass cutoff M$_{up}=30$ M$_{\odot}$. For the source 1201+394, the H$\alpha$ luminosity needed for estimating Q$_{tot}$ was not available.} 
\end{deluxetable*}

\section{Discussion on individual sources} \label{sec:compiled}

\textbf{0258+35:} This well-studied radio source is hosted in the early-type giant NGC 1167 with an optical spectrum typical of a LINER AGN \citep{2006PhDT........60E}. Radio data display two pairs of jet lobes -- an inner luminous, steep-spectrum pair spanning $\sim$3 kpc \citep{2005A&A...441...89G}, embedded in large-scale ($\sim$240 kpc) outer jets with extremely low surface brightness that are $\sim$110 Myr old \citep{2012A&A...545A..91S, 2018A&A...618A..45B}.  This nested jet structure and the contrast in spectral indices of the inner and outer lobes point to recurrent activity with $\sim$100 Myr quiescent phase and led to its interpretation as a restarted radio galaxy \citep{2012A&A...545A..91S}, though other evolutionary scenarios have also been suggested \citep{2018A&A...618A..45B}. \cite{2005A&A...441...89G} derived the age for the young CSS source as $\sim$0.9 Myr and cited dynamical arguments to suggest that the compact radio source will likely not grow out into an extended FRI/II structure. The sharp bend in the southern (inner) lobe indicating interaction with a dense surrounding ISM supports this source confinement scenario. This argument further agrees with the turbulent jet-ISM interaction in the heart of the galaxy, as evident in the HI absorption observations \citep{2019A&A...629A..58M}. The presence of a massive molecular gas outflow, evidently driven by the young radio source, was detected in the central region of the galaxy in recent CO(1–0) observations \citep{2022NatAs...6..488M}. At larger scales, the CO emission forms a ring of molecular gas of $\sim$10 kpc radius, which is found to be coincident with low star-forming activity along the faint spiral arms \citep{2016A&A...588A..68G}.

Our \textit{HST}/UV continuum imaging did not detect signs of a young stellar population, consistent with NGC 1167’s passive, non-star forming [NUV$-r$] and IR colors (Fig. \ref{fig:nuv-r}). This is in agreement with \cite{2006PhDT........60E} result of no young star signature in optical spectrum. However, given the clear evidence of jet-cloud interaction in the vicinity of the radio source from cold gas kinematics \citep{2022NatAs...6..488M} and X-ray emission coincident with the CO outflow region \citep{2022ApJ...938..105F}, the non-detection of star-forming UV emission is a puzzling result. Dust obscuration of UV light could be a possible explanation for this paradoxical observation. The visible band \textit{HST} image indeed shows intricate dust lanes in the nuclear region, clearly observable in the residual image (Fig. \ref{fig:galfitOP}). 

Our best-fit \textsc{galfit} model is a combination of three Sersic bulges$-$ an inner, fairly compact bulge with R$_{eff}$$\sim$5 kpc, and two larger-scale components with R$_{eff}$$\sim$16 kpc and $\sim$25 kpc; all centered within $\sim$0.5$''$ of the optical nucleus and having de Vaucouleurs elliptical ($n$=4) intensity profiles with a steep central core and extending outer wing. 
\newline

\textbf{1014+392:}
The highest redshift source in our sample, 4C 39.29 is a Type 2 quasar with powerful optical narrow-line spectrum. Classified as a LERG \citep{2013MNRAS.430.3086G}, the AGN accretion in 1014+392 is radiatively inefficient. Spatial coincidence of the $\sim$15 kpc extended optical emission-line regions with the brighter radio lobe suggests that the radio jet could have ionized the emission-line gas \citep{2006MNRAS.369.1566G}. 

Optically, the source fits cleanly with an $n$=2 elliptical model for the host galaxy with $1''$ effective radius, which translates to $\sim$6 kpc at $z$=0.536. Our \textit{HST} imaging did not detect continuum UV emission in the host. 
\newline

\textbf{1025+390:}
This elliptical galaxy is host to CSS source 4C 39.32 which has a relatively amorphous double-lobe radio morphology. The radio core identified in \cite{2006A&A...449...49R} aligns with \textit{HST} optical and UV nuclei, with the bent NW jet tail coinciding with the northern UV knot lying $\sim$5 kpc from the core (also appears as an extended arm-like structure in the optical; see Fig. \ref{fig:galfitOP}). The spatial alignment in this case strongly suggests that the clustered young stellar population in the path of the jet is due to shock-triggered star formation. \cite{2011A&A...528A.110F} found 0.06$-$0.1 Gyr old young stellar population in the host with their optical-UV SED fitting, while our modelling results show $\lesssim$5 Myr old stellar population.

Our best-fit optical \textsc{galfit} model for this source is a relatively flat bulge with R$_{eff}\sim$3.5 kpc and outer, more elliptical Sersic component of $\sim$23 kpc effective radius. Our fit suggests a nuclear point source, $\sim$4 magnitudes fainter than the bulge, is likely present.
In the bluer band, fitting shows the R$_{eff}$=4 kpc UV bright core along with a $\sim$3 kpc-wide Northern clump, best-fit with two distinct $n\approx$1 Sersic profiles. 
\newline

\textbf{1037+30:}
The relatively nearby ($z$=0.091) CSS source 4C 30.19 lies in the chaotic central galaxy of the Abell 923 cluster \citep{2005A&A...441...89G}. Marked by prominent tidal features and a disturbed-elliptical/irregular morphology \citep{2000A&AS..142..353G}, it also exhibits highly complex nuclear structure as evident from the twisted isophotes (Fig. \ref{fig:isofitOP}) and optical \textsc{galfit} residuals (Fig. \ref{fig:galfitOP}). SED fitting analysis by \cite{2015A&A...581A..33D} concluded that 1037+30’s FUV to mid-IR SED is dominated by young stellar component rather than the AGN, consistent with its clearly starburst-aligned IR colors (Fig. \ref{fig:wise}). The extended, radio-aligned continuum UV detected in our imaging is hence not likely to be highly contaminated with AGN-related radiation. 1037+30 is hence a strong candidate for jet-induced SF, considering the $\lesssim$10 Myr aged starburst population that our data suggests. This hypothesis is also supported by evidence of UV-bright kpc-scale filamentary structure in some BCGs, attributed to star formation induced by the jets and/or compression inside X-ray cavities (e.g., \citealt{2015MNRAS.451.3768T}).

Visible-band \textsc{galfit} fitting gives an $n$=3.4 Sersic bulge (effective radius $\sim$2 kpc) and a larger (R$_{eff}\sim$30 kpc) $n$=10 Sersic component with a 20 AB mag nuclear point source, as the best fit. This model offers the lowest $\chi^2_\nu$ but does not result in a clean fit, which is expected for a highly perturbed galaxy, likely undergoing interaction. In the UV band, the fitting is limited to the central region with higher surface brightness, while the more disperse extended UV regions are excluded by \textsc{galfit}. The central region is fit by a nuclear bar profile of radius $\sim$5 kpc and a 20 AB mag point source component (as in the optical fit). 
\newline

\textbf{1128+455:}
The host galaxy of this broad-line \citep{2020MNRAS.491...92L} CSS radio source shows strong spiral characteristics with a dense dust lane aligned edge-on along the major axis. \cite{2011A&A...528A.110F} found 0.06$-$0.1 Gyr old young stellar population in the host with their optical-UV SED fitting. Our results from starburst modelling indicate $\lesssim$10 Myr ages. \cite{1999A&A...345..769M} computed the synchrotron-radiation timescale for this radio source to be $1.7\times10^4$ yr. 
This suggests that the radio-aligned UV-emitting regions could have starburst contributions from past merger events or subsisting SF ignited by an earlier cycle of radio emission. Being a BLRG, this galaxy may also have contamination from nuclear activity-related emission in the UV band. 

An exponential disk with scalelength $R_{eff}$$\sim$3 kpc fits this galaxy well. The giant, galaxy-wide dust lane is clearly visible in the residual image, as is the extended filament structure not fitted by the model. Having a pure disk structure, this galaxy could be one of the rare bulgeless AGN hosts (e.g., \citealt{2009ApJ...704..439S, 2009ApJ...690..267D, 2011ApJ...742...68J}). Deeper and higher-resolution observations are needed to confirm the bulgeless nature, as the bulge might escape detection due to the dust lane and/or the (relatively) high redshift. The UV emission is best-fit with an $n$=1.7 Sersic component of $\sim$5 kpc effective radius. 
\newline

\begin{figure*}
\plotone{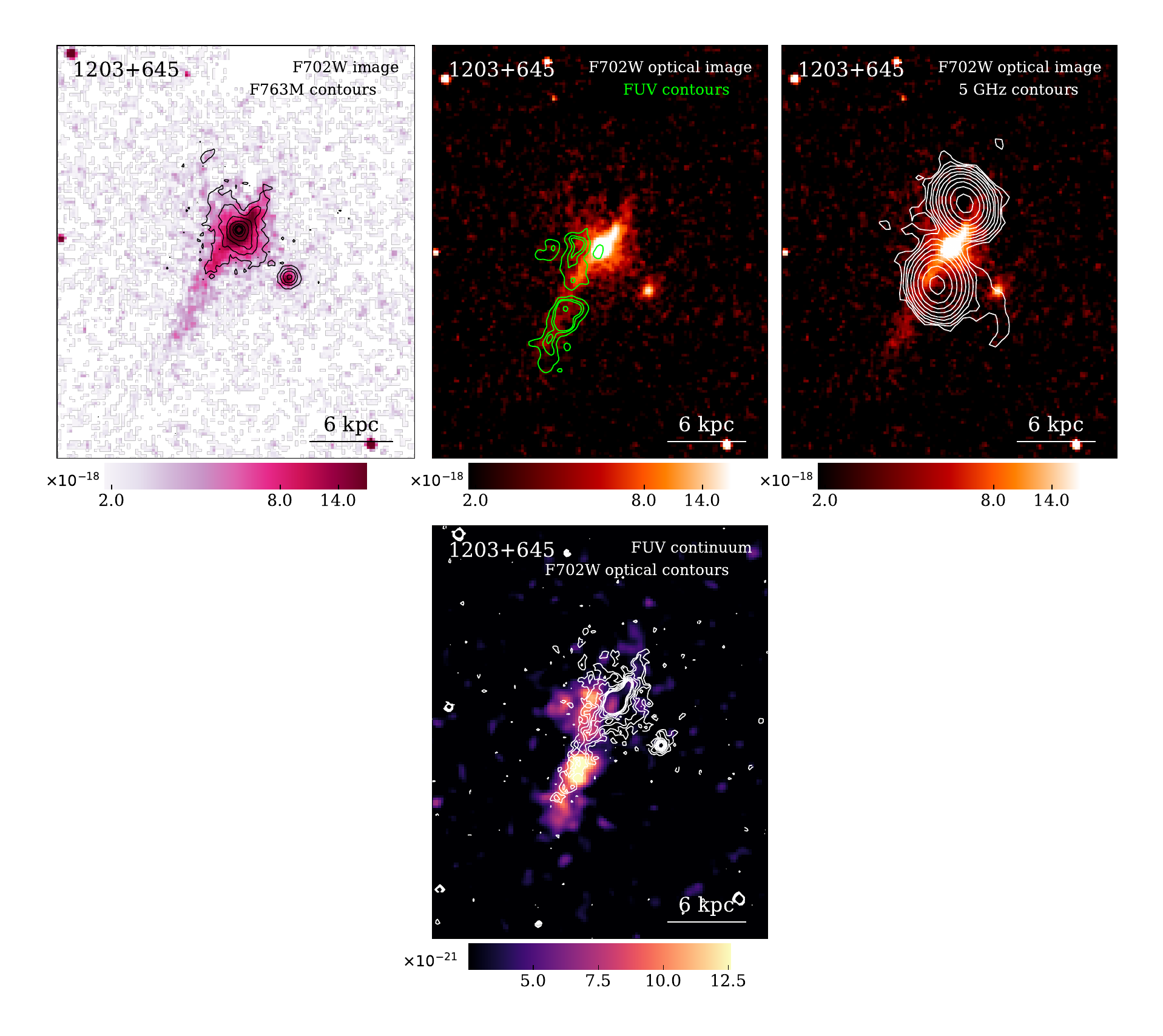}
\caption{$HST$ broad-band (F702W; \cite{1997ApJS..110..191D}) morphology in 1203+645, host to the CSS radio source 3C 268.3. The F702W filter includes [O III] line emission \citep{1999ApJ...526...27D}. All panels are rotated such that North is pointed up and East is left. Flux density is in units of ergs sec$^{-1}$ cm$^{-2}$ \AA$^{-1}$. \textit{Top: (left)} Contours from the optical line-free continuum image from this study overlaid on de Vries et al. F702W image. The $\sim$15 kpc tail-like feature extending beyond the galactic continuum. \textit{(center)} F702W image overlaid with contours of the UV-emitting regions from our data.
\textit{(right)} F702W broad-band emission with VLA 5 GHz radio contours. 
\textit{Bottom:} \textit{HST} UV-band image of 1203+645 from this work overlaid with deVries et al. F702W contours. The line-emission tail closely aligns with the extended, likely star-forming regions.
\label{fig:1203extra}}
\end{figure*}

\textbf{1201+394:}
Another one of the higher redshift sources in our sample, 1201+394 is a cluster-centric BCG \citep{2016MNRAS.460.3669Y}. Our \textit{HST} imaging revealed two other galaxies in the vicinity; we carried out surface brightness measurements for these likely cluster companions as well. The host exhibits messy, highly twisted isophotal profile in the optical. A 1.2 kpc-wide nuclear bar and an exponential disk of effective radius $\sim$6 kpc produces the cleanest residuals for the apparently bulgeless galaxy profile, though a fixed centroid and $m$=1 Fourier mode modification (amplitude a$_m$=0.35 and 10.5$^\circ$ phase angle relative to the disk component axis) needed to model this galaxy suggests a "lopsided" brightness structure \citep{2002AJ....124..266P}. 

The UV band image shows an extended clumpy region near the core, fit with an $n$=2 Sersic component of 4.6 kpc effective radius. The 1201+394 host lies on the boundary of the AGN division on the \textit{WISE} diagnostic color plot (Fig. \ref{fig:wise}), so the $\sim$10 kpc wide UV knot may include some activity-related UV emission. Considering the young $\lesssim$5 Myr old stellar population indicated by the continuous SF models and the CSS source's (dynamical) age, we suggest that the star forming activity in this BCG host could have been ignited due to gas infall and/or gas compression from possible X-ray cavities, 
and eventually enhanced by the radio source. 

We also separately fit the other two sources detected close to the target CSS host in the optical image, that were masked out for the analysis of 1201+394. The galaxy to the far NW has a 1$\sigma-$aperture magnitude of $21.58\pm0.37$ (corrected for Galactic extinction), and is best fit by an $n$=2 Sersic profile of R$_{eff}$=0.5$''$. The other neighbouring source, detected closer to the host, is best fit with an extended profile of an $n$=4 Sersic, with R$_{eff}$=3.25$''$; though the faint outer regions in the profile may have light contribution from 1201+394 in this case.
\newline

\textbf{1203+645:}
This peculiarly-shaped BLRG \citep{2008MNRAS.387..639H, 2016AJ....151..120W, 2020MNRAS.491...92L} is the extensively studied host to the most powerful radio source in our sample, 3C 268.3. Located in a cluster, this galaxy has clear signs of tidal interactions (elongated, bent arm-like structures in the N-S direction), and a neighbouring source 2$''$ to the SW, that is likely a cluster companion. The radio emission has an asymmetric double-lobed morphology with a faint, possible core detected at 5 GHz \citep{1998MNRAS.299..467L, 2004MNRAS.351..845G} about 1.5 kpc south of the northern lobe. Our \textit{HST/VLA} overlays do not take this possible core position into account, instead we align our images by matching the optical center with the mid-point of the 5 GHz radio map. (Aligning w.r.t the possible core would shift the overlaid contours 0.3$''$ towards South, further coinciding with the extended UV regions). \cite{1999A&A...345..769M} found the radiative age for the CSS source to be $3.5\times10^4$ yr. 

The best fitting model for the optical source is a $\sim$1000 pc nuclear bar along with a faint (25.3 AB mag) nuclear point source. The N-S filamentary structure was masked for the purpose of 2D modelling. The UV emission is spread out to about $2.2''$ away from the optical center of the source, hence is likely a tail of extended emission. Three Sersic components at distinct centroids, with largest effective radius of $\sim$6 kpc, are needed to fully fit the UV regions. The low Sersic indices (0.1$-$1) indicate a rather flat core and sharply decreasing intensity outwards. This is consistent with clumpy, star-forming regions.

The bright source to the SW of the CSS host in the optical image is likely a cluster companion. It shows a 1$\sigma$ magnitude of $22.94\pm0.53$ (corrected for Galactic extinction) and best fits with a nuclear bar profile ($\alpha$=2, $\beta$= 2.1, $\gamma$=0.8) of effective radius $\sim$0.6 kpc.

\begin{figure}[t!]
\plotone{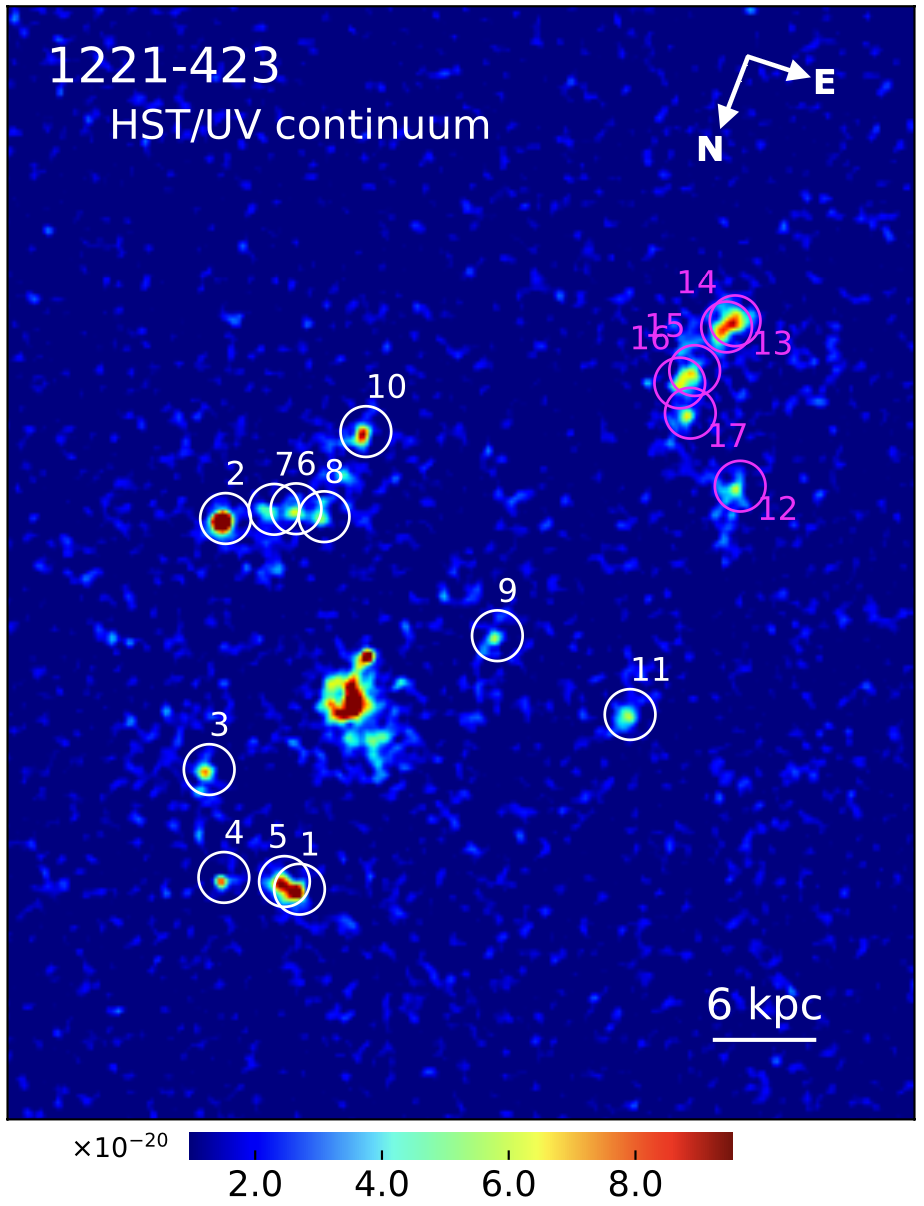}
\caption{\textsc{SExtractor} identification of individual UV  clumps, farther away from the circumnuclear region, in the merging system of 1221-423. A detection was defined as minimum five pixels of the source brighter than $3\sigma$ flux threshold. The circles represent the Kron apertures used for photometric measurements of the likely star-forming regions in the spiral arms of the CSS-host parent galaxy (white) and the Southern companion (magenta). Flux density is in units of ergs sec$^{-1}$ cm$^{-2}$ \AA$^{-1}$. \label{fig:1221extra}}
\end{figure}

\begin{deluxetable}{ccL}[t!]
\tablecaption{\textsc{SExtractor} photometry of the outlying UV clumps in 1221-423 system \label{tab:hii}}
\tabletypesize{\footnotesize}
\tablehead{
\colhead{Region \#} & \colhead{Aperture radius } & \colhead{Kron magnitude} \\
 & \colhead{($''$)} & \colhead{(mag)}
}
\startdata
    1     &        0.30        & 22.37\pm0.77 \\
    2     &       0.26        & 21.81\pm0.59 \\
    3     &       0.35        & 23.17\pm1.11 \\
    4     &       0.28        & 23.98\pm1.61 \\
    5     &       0.36        & 22.24\pm0.72 \\
    6     &       0.34        & 23.28\pm1.16 \\
    7     &       0.38        & 23.25\pm1.15 \\
    8     &       0.34        & 23.55\pm1.32 \\
    9     &       0.34        & 23.80\pm1.48  \\
    10    &       0.34        & 22.87\pm0.96 \\
    11    &       0.34        & 23.22\pm1.13 \\
    12    &       0.38        & 23.02\pm1.04 \\
    13    &       0.39        & 21.50\pm0.51  \\
    14    &        0.40        & 21.86\pm0.60  \\
    15    &       0.39        & 23.06\pm1.05 \\
    16    &       0.37        & 23.55\pm1.32 \\
    17    &       0.39        & 23.27\pm1.16 \\
\enddata
\tablecomments{Projected spatial scale at source redshift is 2.923 kpc/$''$.}
\end{deluxetable}

The galaxy 1203+645 has previously been imaged with \textit{HST} in the broad-band \citep{1997ApJS..110..191D, 1999ApJ...526...27D} and narrow-band \citep{2000AJ....120.2284A, 2008ApJS..175..423P} optical filters. Figure \ref{fig:1203extra} compares the de Vries et al. broad-band image with our optical and UV band data. 
The bright [O III]-line emitting tail \citep{1999ApJ...526...27D} extending well beyond the galactic continuum in our optical image (overlaid contours) is an interesting feature. The line-emission tail closely aligns with the $\sim$15 kpc extended UV continuum regions as well as the radio source. This strongly suggests that while the 
peculiar NW-SE filaments (seen in our line-free continuum image) might have been a result of galactic interaction in the cluster environment, the expanding jet is likely driving shocks that ionize the line-emitting gas and cause starburst activity in the extended tail structure. Narrow emission-line regions in this galaxy reveal dense, radio source-aligned ionized gas clouds in the ISM. Spatially-resolved kinematic analyses \citep{2005A&A...436..493L, 2009AN....330..226H, 2013ApJ...772..138S} have suggested that a combination of AGN photoionization and shock-ionization might be responsible for emission-line regions. \cite{2016MNRAS.455.2242R} have agreed that shocks were triggered by jet–cloud interaction and must be taken into account to explain the spatial behaviour of emission lines. HI-absorption has been detected in the northern lobe of 3C 268.3 \citep{2006A&A...447..481L} and found to be consistent with the HI being produced in emission-line clouds in vicinity of the radio source, further supporting the jet-ISM interaction hypothesis. The findings support our argument that jet-induced SF might be the cause of the UV-bright tail. 

The \textit{WISE} colors for this galaxy show AGN-dominant IR emission, which would explain the high 22$\mu$m luminosity (heavy contribuiton from AGN-heated dust) and much higher SFR estimates compared to other bands. \cite{2016AJ....151..120W} analyzed the optical-to-infrared SED for 1203+645 and estimated broadband far-IR (8–1000 $\mu$m) SFR of 17.4 M$_\odot$ yr$^{-1}$. 
\newline

\textbf{1221$-$423:}
Part of an interacting galaxy pair, this face-on spiral hosts a young CSS radio source emerging from a LINER nucleus in a gas-rich environment. The irregular radio structure is evidence of a vigorous jet-ISM interaction that has caused the southern jet to be bent in a 180$^\circ$ turn. Strong extended line emission has been observed in this radio galaxy with spatial structure suggestive of both AGN-ionized regions and star formation signature \citep{2013MNRAS.431.3269A}. Three distinct stellar populations have been found in an earlier study by \cite{2005MNRAS.356..515J}: the old ($\sim$15 Gyr) population in outer boundary, intermediate-age ($\sim$300 Myr) population around the core and along the tidal tail with southern companion galaxy and a young star population ($\sim$10 Myr) near the nucleus and blue knots. The more recent starburst episodes were attributed to tidal interactions with the companion galaxy; the same event also been suggested to have triggered the radio source, after a $\sim$100 Myr interval. The radiative age of the CSS source has been estimated to be 10$^5$ yrs \citep{2003PASA...20....1S}, which led \cite{2005MNRAS.356..515J} to conclude that there must be substantial time delay between the most recent star formation and the birth of the radio source. However, considering the striking spatial correlation of star forming regions with jet structure, we suggest that the gas infall from the ongoing galactic interaction proposed to have triggered the AGN, might also have ignited star formation near the nuclear region which was then enhanced by the expanding jet. This could explain the southern filament extending out to about $\sim$1 kpc from the nucleus and the UV knot cospatial with the southern radio peak (Fig. \ref{fig:uvradio}). Observing a low index of the 4000\AA$ $ break in the nuclear region, \cite{2005MNRAS.356..515J} confirmed the presence of recent star formation. The AGN contribution to the blue excess likely to be less than half the total UV light \citep{2005MNRAS.356..515J}. Further, optical continuum modelling by \cite{2010MNRAS.407..721J} found evidence of $\sim$5 Myr-old stellar population in the nuclear region of the host, consistent with our results (Fig. \ref{fig:sfepoch}). 
\newline

In the R- and V- band imaging of \cite{2005MNRAS.356..515J}, 1221$-$423 host was fit with a de Vaucouleurs bulge of half-light radius of 42 kpc plus an exponential disc. Our best-fit visible band \textsc{galfit} model is a combination of a flat, sharply truncated Sersic bulge ($n$=0.3) and a R$_s$$\sim$6 kpc exponential disk component oriented at 42 degrees w.r.t. the bulge and brighter by 3 orders of magnitude. The inner 1 kpc also shows another disky component, fainter than the other two components. The nuclear UV emission is best-fit with an $n=2.5$ Sersic component of $\sim$3 kpc effective radius. Another UV knot about a kpc from the nucleus coincides with the southern radio lobe and so is likely jet-induced. 

The UV clumps scattered along the spiral arms (Fig. \ref{fig:opuv}) are likely HII-emitting star-forming nebulae. We extracted photometry for each individual knot (shown in Figure  \ref{fig:1221extra}), using \textsc{SExtractor} source detection, with a $3\sigma$ threshold. The cicumnuclear region was left out due to its spatial association with the AGN jet.
Table \ref{tab:hii} lists the aperture size and observed magnitude of each knot.
\newline

\textbf{1445+410:} The host galaxy of a $\sim$26 kpc steep-spectrum radio source. Although it lies in the region of high star-formation activity on the \textit{WISE} color plot (W2$-$W3 $>1.6$; Fig. \ref{fig:wise}), our imaging observations did not detect UV continuum in this galaxy. Recent study by \cite{2022MNRAS.511..214N} found a low star formation rate of 0.05 M$_\odot$/yr over the last 50 Myr in the host, using stellar population synthesis with SDSS/DR12 spectral data. 

Best-fit \textsc{galfit} model for this galaxy is a $n$=10 Sersic ellipse of $\sim$70 kpc half-light radius, along with an bar structure of radius $\sim$1.6 kpc offset from the optical centroid by 0.2$''$.  The fit also suggests a faint nuclear point source component with integrated magnitude of 23.2 (AB mag).

\section{Summary $\&$ Conclusions} \label{sec:concl}

We obtained and analyzed sub-arcsecond resolution \textit{HST} imaging for 7 CSS radio galaxies against 2 larger radio galaxies as control. The radio galaxies with $>$20 kpc-sized radio sources do not show star-forming UV emission. We use the visible band observations  and archival IR data for these galaxies to examine their stellar continuum morphology and optical-IR spectral energy distribution. For the galaxies where UV continuum emission is detected, the UV data were compared with the optical and radio maps. We also modelled the UV continuum with synthesized spectra of stellar populations of different ages. Our key findings are summarized below:

\begin{itemize}

\item About half of the sample has large-scale perturbed morphological features$-$ large scale tidal tails in 1037+30, faint extended arm-like structure in B1128+455, extended x-shaped filaments in 1203+645, the 1221-423 system$-$ that hint at probable merger history or ongoing interactions. Two CSS hosts in the sample (1201+394 and 1203+645) are located in galaxy clusters and show close neighbouring galaxies that are likely gravitationally-interacting cluster companions. 

The presence of strong spiral and/or disk-like structure in 3 out of 7 CSS galaxies in the sample (0258+25, 1128+455 and 1221-423) is an interesting result, since PS/CSS sources are generally hosted in large ellipticals (see \cite{2021A&ARv..29....3O} for a review) with very few known late-type hosts. Observations with larger samples of compact radio galaxies could test for this effect.

\item Surface brightness modelling of the line-free optical continuum shows that almost all of the host galaxies in the sample exhibit isophotal twists with highly varying ellipticity in their radial profiles and heavy 2D-model residuals in the core, which suggests mildly (e.g., 1025+390) to highly complex (e.g., 1037+30) nuclear structure. 4 galaxies may be rare bulgeless AGN hosts; deeper observations are needed to confirm this because of their higher redshifts ($0.3<z<0.6$). 

The best-fit models for 4 out of 7 CSS radio galaxies in the sample show a faint nuclear point source component. This suggests that these are possibly home to low-luminosity AGN (1025+390, 1037+30, 1445+410) or a broad-line AGN that is heavily obscured (1203+645).

\item Near-UV continuum maps reveal spatially resolved, clumpy emission in 6 out of 7 CSS hosts. The UV emission clearly comes from extended star-forming regions beyond the nucleus. While direct UV light from the AGN will not contribute significantly to the extended emission, some fraction of the observed UV excess may arise from continuum emission from ionized gas nebulae and/or scattered AGN radiation. Follow-up IFU spectroscopic observations are needed to spatially resolve nebular emission in the extended UV regions, while polarized-UV imaging can help constrain the scattered UV contamination. Our extinction estimates show that heavy line-of-sight dust obscuration in the host  galaxies is unlikely.

\item Continuum UV emission is aligned along the kpc-scale steep-spectrum radio lobes in 5 CSS galaxies (plus possible alignment in 1201+394). Three CSS galaxies with aligned UV light (1025+390, 1037+30 and 1221-423) are strong candidates for jet-induced star formation; while others (1128+455 and 1203+645) may have added contribution from radiative outflows in the jet-aligned regions, or need deeper UV observations (1201+394). In general, these detections expand the sample size for suspected footprint of positive radio-source feedback; 3C 303.1 has been the only known CSS candidate for radio source-triggered SF so far \citep{2008A&A...477..491L, 2009AN....330..261O}.

\item Low ($\sim$ few to a few tens of M$_\odot$/yr) rates of star formation are observed in the UV-bright CSS hosts, though these are lower limits by a likely significant margin. The \textit{WISE} 22$\mu$m-derived SFRs are likely heavily contaminated by nuclear activity (i.e., dust heated by the AGN) and hence, not reliable.

\item Young stellar population produced in a recent ($\sim$1$-$8 Myr old) continuous starburst is most consistent with the observed UV continuum in the CSS galaxies.
But given the limitations of the synthesis modelling and SFR calibrations, we do not rule out $<$1 Myr old stellar population. The dynamical radio source ages of the sample range from 0.01$-$0.1 Myr. This suggests the starbursts could have been induced by the current young radio source, or in a past episode of jet activity. 
It is also possible that the radio source might be confined to a small size by a dense ISM environment, in which case it could easily trigger the starbursts that produce the $\sim$1$-$10 Myr old stellar population. X-ray observations will help probe the jet frustration scenario.

Although most CSS hosts show disturbed morphologies typical of galactic interactions, observed star formation is unlikely to have been triggered due to a merger event, since our estimates show starbursts too young for merger-driven gas infall timescale. 
Assuming an instantaneous single burst, an initial $10^7-10^8$ M$_\odot$ starburst triggered $\sim$1$-$10 Myr ago would likely produce enough hot massive stars to generate the observed UV.

\item The observed UV flux is consistent with the emission-line nebula being ionized by hot, massive stars. The young stellar population would produce a significant fraction of the ionizing photons required to power the nebula, although other sources (e.g., photoionization by the AGN) may also contribute in certain regions of the nebula.    

\end{itemize}

Our results show that compact, young radio galaxies hold the key to understanding jet-ISM interplay on sub-galactic scales. 
Recent simulation studies of radio-mode feedback have found that most of the star formation occurs in the first few Myr (e.g., \citealt{2021MNRAS.506..488B, 2018MNRAS.479.5544M}), in agreement with our observations. Detection of radio-aligned star-forming regions, even in low- to moderate-power CSS hosts, is a promising diagnostic of the mechanically-driven positive AGN feedback that stimulates galaxy growth on short timescales. The presence of jet-induced star formation confirms a salient prediction of the radio-jet feedback model in AGN host galaxies. Further investigation with integral-field observations of the ionized gas kinematics as well as measurements of cold molecular gas distribution in the CSS hosts, will shed more light on the extent and efficiency of the positive feedback mechanism.
\newline

All Python codes $\&$ Jupyter notebooks created for the analyses in this paper are available in publicly accessible repositories at \url{https://github.com/chetnaduggal}

\begin{acknowledgments}
C.D. thanks Dr. Biny Sebastian for performing radio imaging for some of the archival VLA data, and Dr. Sravani Vaddi for sharing her PyRAF photometry scripts. C.D., C.O., and S.B. acknowledge support from the Natural Sciences and Engineering Research Council (NSERC) of Canada. This research has made use of NASA’s Astrophysics Data System Bibliographic Services. This research has made use of the NASA/IPAC Extragalactic Database (NED), as well as the NASA/IPAC Infrared Science Archive (IRSA), which are funded by the National Aeronautics and Space Administration and operated by the California Institute of Technology. 
\end{acknowledgments}

%

\facilities{HST, VLA, SDSS, 2MASS, WISE, GALEX, PanSTARRS.}


\software{Ned Wright's Cosmological Calculator \citep{2006PASP..118.1711W}, SAOImage ds9 \citep{2003ASPC..295..489J}, NumPy \citep{harris2020array}, Astropy (\citealt{2022ApJ...935..167A, 2018AJ....156..123A, 2013A&A...558A..33A}), Matplotlib \citep{Hunter:2007}, IPython \citep{2007CSE.....9c..21P}, 
\textsc{extinction} \citep{2016zndo....804967B}, \textsc{reproject} \citep{2020ascl.soft11023R}, EllipSect \citep{2020zndo...4033448A}, Drizzlepac \citep{2021drzp}, IRAF \citep{1999ascl.soft11002N}, Tiny Tim \citep{2011SPIE.8127E..0JK}, \textsc{galfit} (\citealt{2002AJ....124..266P, 2010AJ....139.2097P}), \textsc{stsynphot} \citep{2020ascl.soft10003S}, \textsc{starburst99} \citep{1999ApJS..123....3L}, AIPS \citep{1996ASPC..101...37V}, \textsc{SExtractor} \citep{1996A&AS..117..393B}.}



\appendix

\section{atlas of images and galaxy fitting results}

The following pages contain the images and plots showing the results of \textsc{ellipse} photometric analysis (referenced in Sec. \ref{subsec:photometry}) and \textsc{galfit} morphological decomposition (referenced in Sec. \ref{subsubsec:galfit}), in the optical and UV bands.

For all galaxies, the layout contains one row with three columns:

\textbf{Left column:} (Clockwise from left) Target image, 2D composite \textsc{galfit} model and residual image ($=$ data $-$ model). The white solid line shows the image scale. 

\textbf{Middle column:} 1D surface brightness profile fit rendered from \textsc{galfit} using the \textsc{ellipsect} software. The solid red line marks the galaxy light profile, dashed lines trace the individual component profiles and the solid blue line shows the sum of all components comprising the final model. 

\textbf{Right column:} Galaxy radial profiles as a function of the semi-major axis r$^{1/4}$ (arcsec) from isophote fitting of the \textit{HST} images with \textsc{iraf/ellipse}. (From top to bottom) Surface brightness $\mu$ (mag arcsec$^{-2}$), ellipticity $\epsilon$ and major axis position-angle PA (in degrees; from North to East). Where not visible, the error bars are smaller than the plot markers.

\subsection{Visible band}

1D and 2D surface brightness modelling results for \textit{HST/V} images for the 9 target radio galaxies. Blue regions mark the masked-out objects.

\begin{figure}[h!]
\figurenum{11}
\gridline{\fig{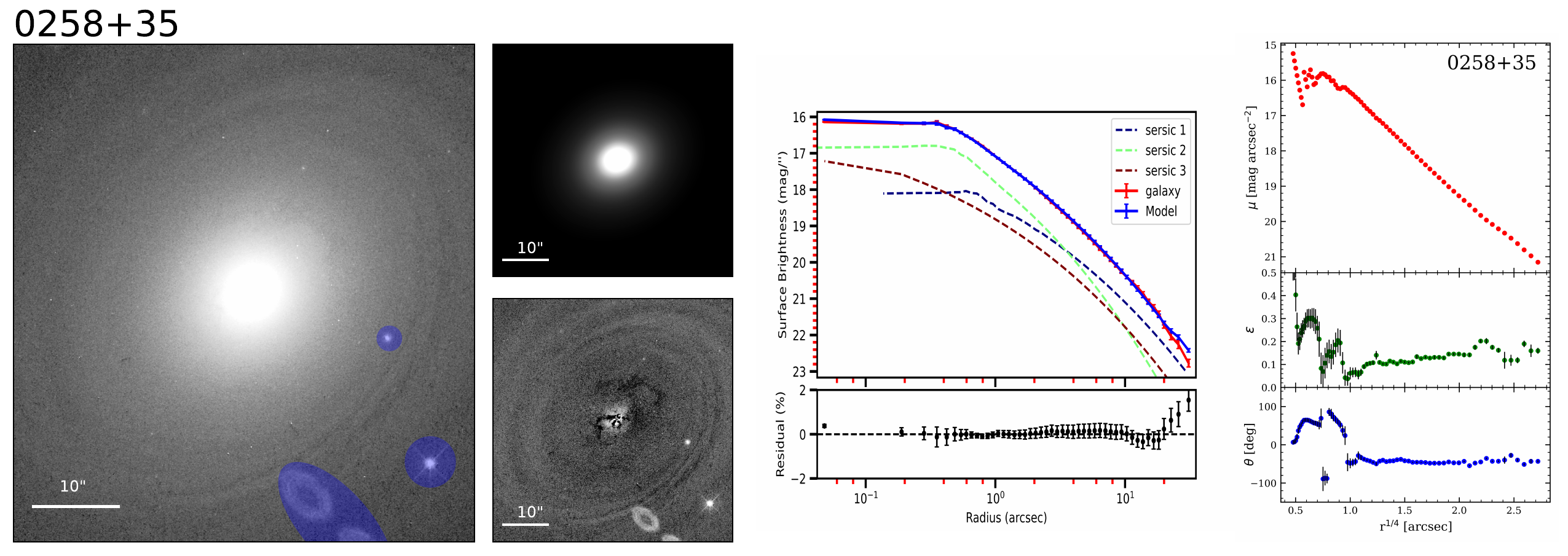}{\textwidth}{}}
\gridline{\fig{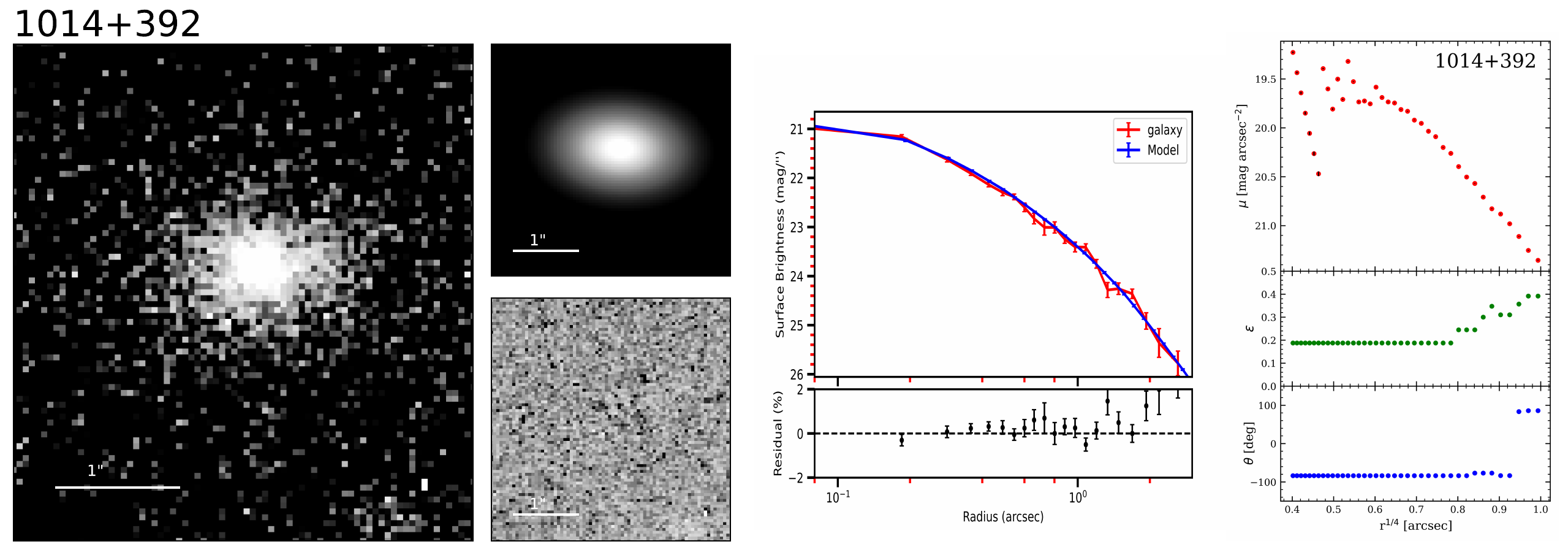}{\textwidth}{}}
\gridline{\fig{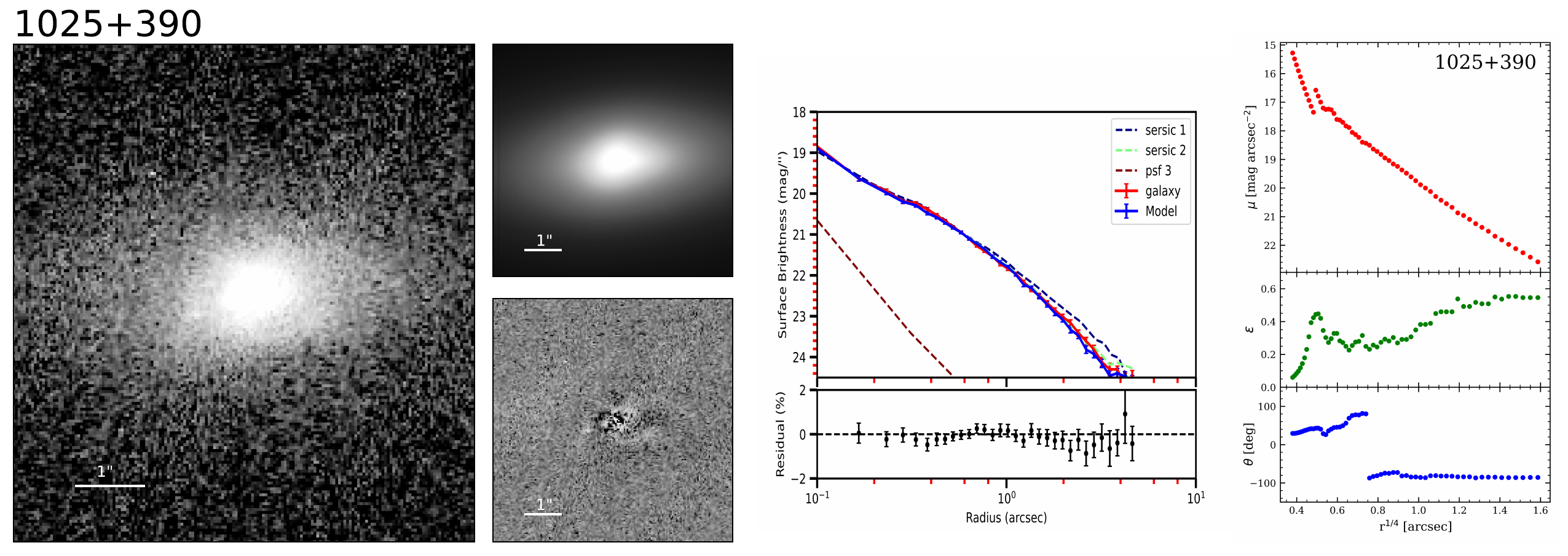}{\textwidth}{}}
\caption{Optical photometric analysis and structural decompositions.}
\end{figure}

\begin{figure*}[h!]
\figurenum{11}
\gridline{\fig{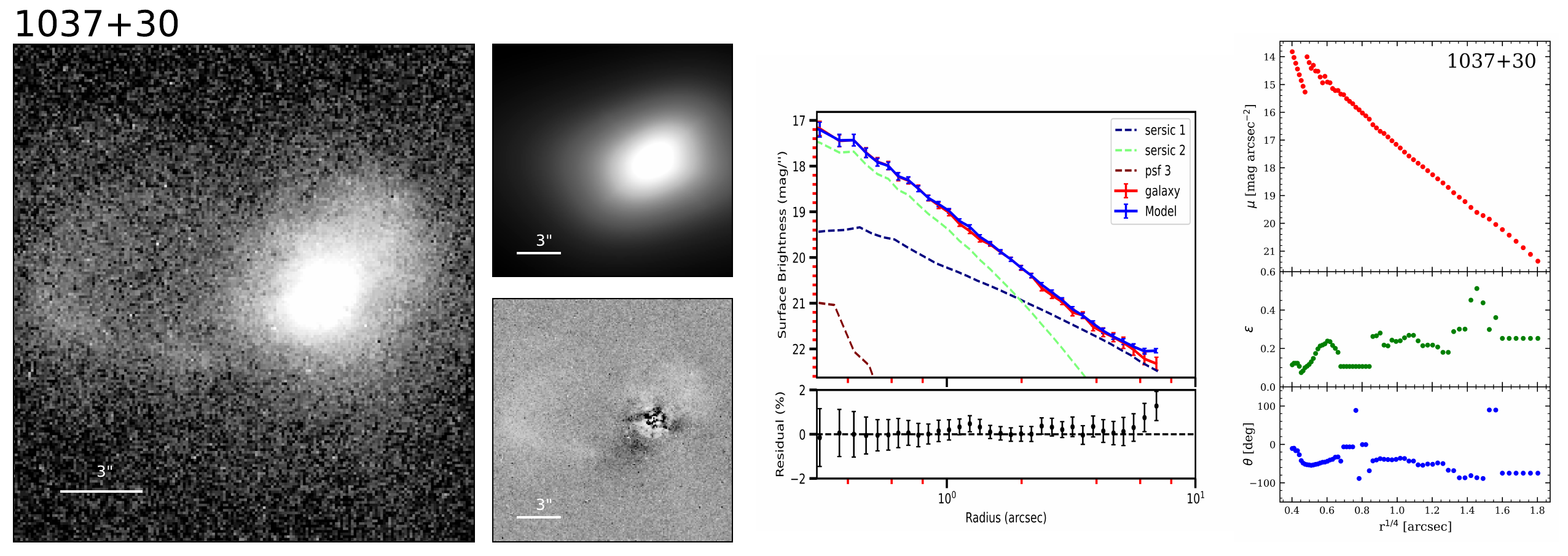}{\textwidth}{}}
\gridline{\fig{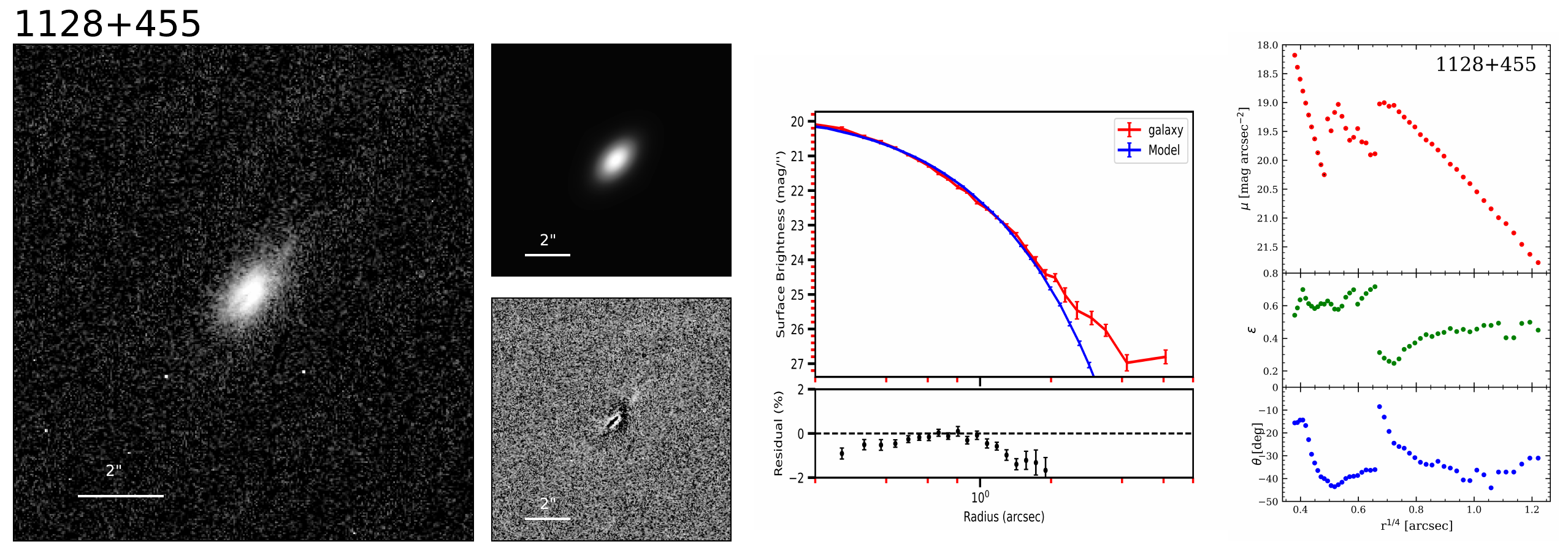}{\textwidth}{}}
\gridline{\fig{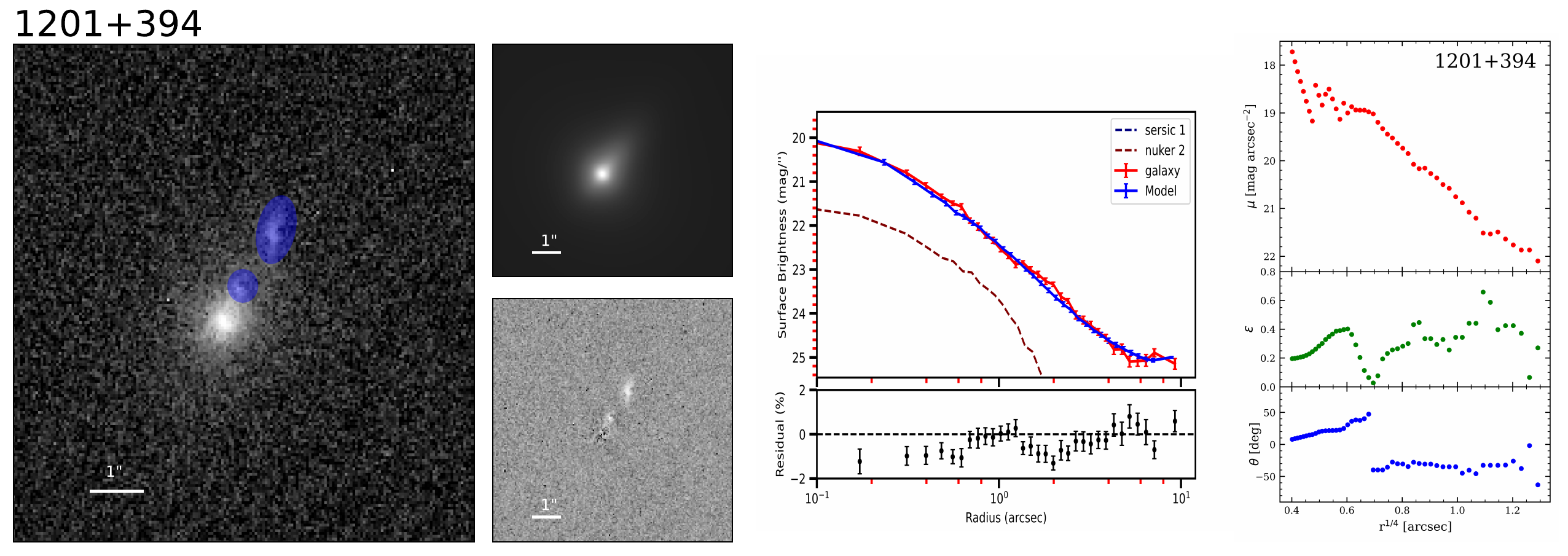}{\textwidth}{}}
\caption{$continued$}
\end{figure*}

\begin{figure*}[h!]
\figurenum{11}
\gridline{\fig{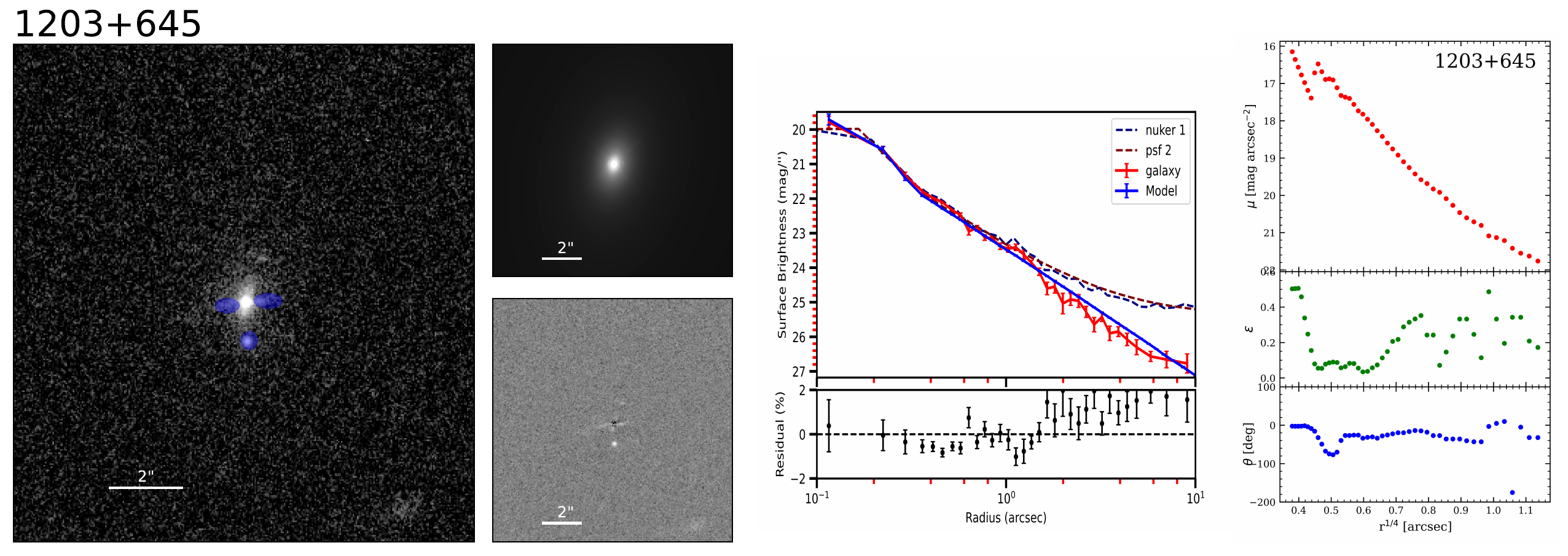}{\textwidth}{}}
\gridline{\fig{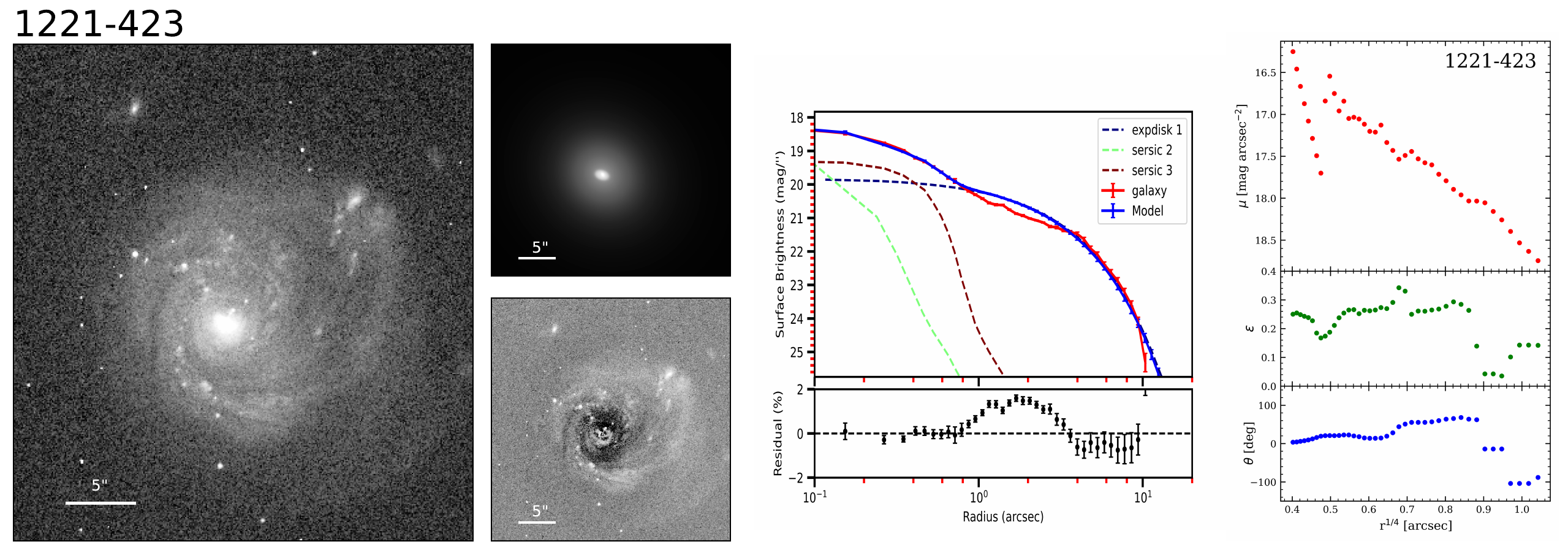}{\textwidth}{}}
\gridline{\fig{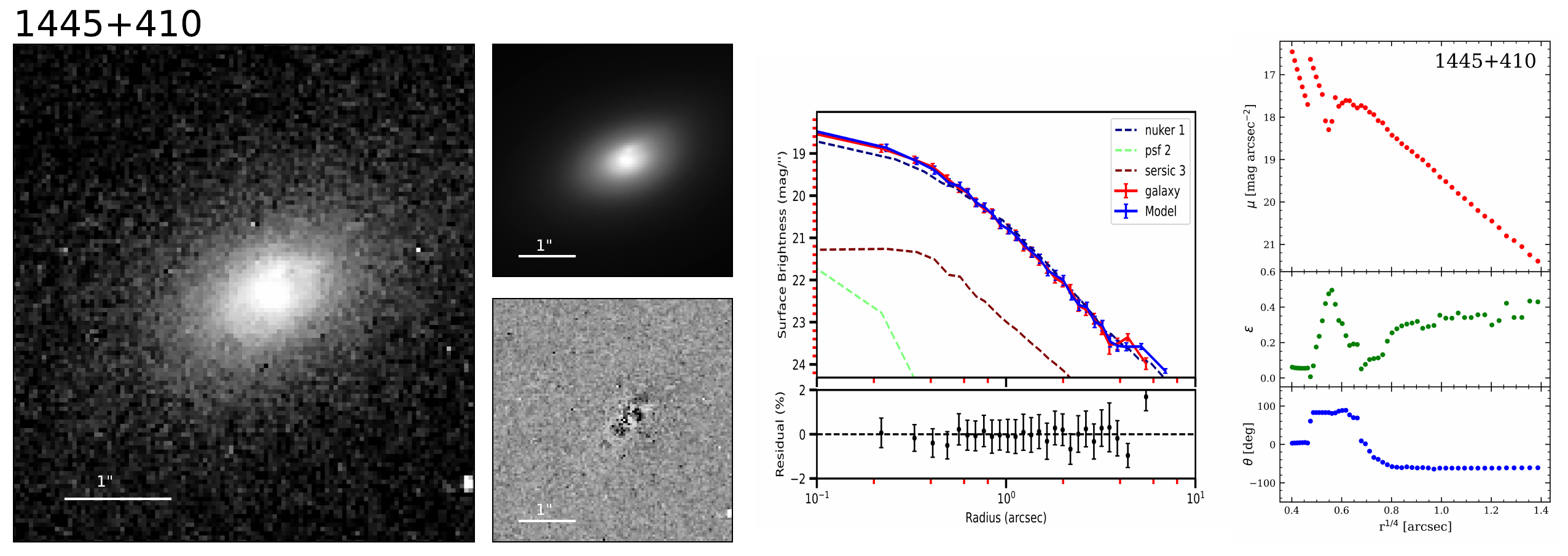}{\textwidth}{}}
\caption{$continued$}
\end{figure*}

\subsection{UV band}

1D and 2D surface brightness modelling results for \textit{HST/UV} images for the 6 UV-detected radio galaxies.

\begin{figure*}[h!]
\figurenum{12}
\gridline{\fig{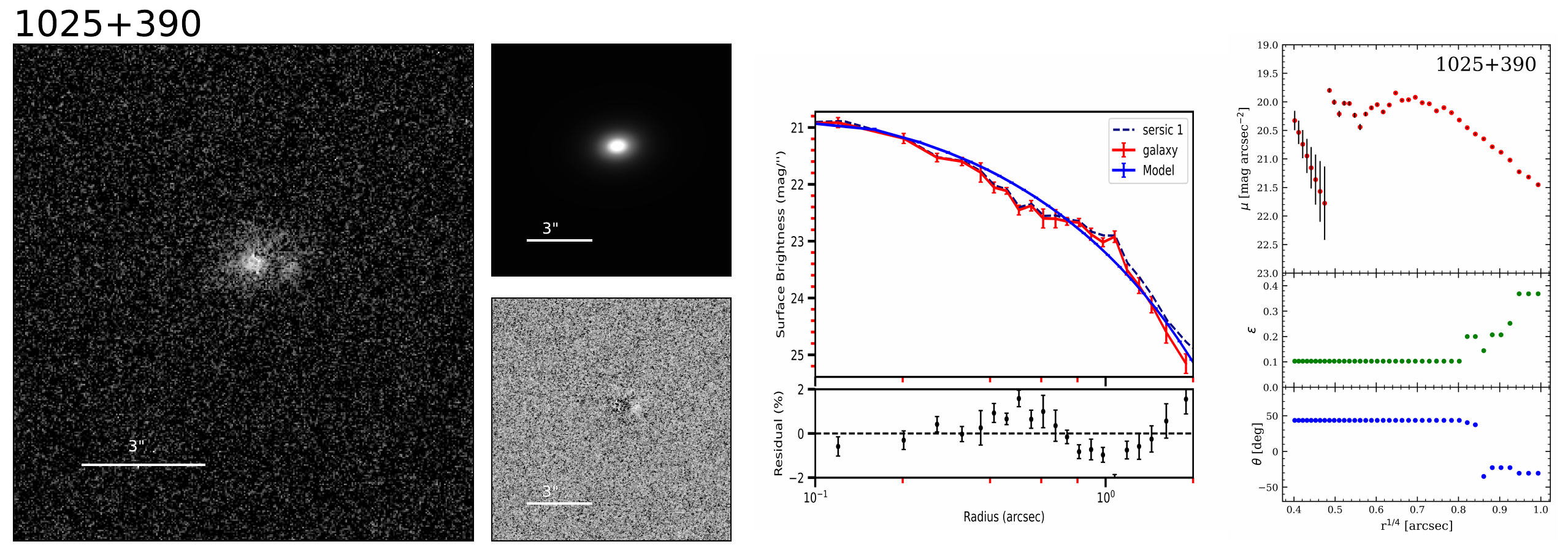}{\textwidth}{}}
\gridline{\fig{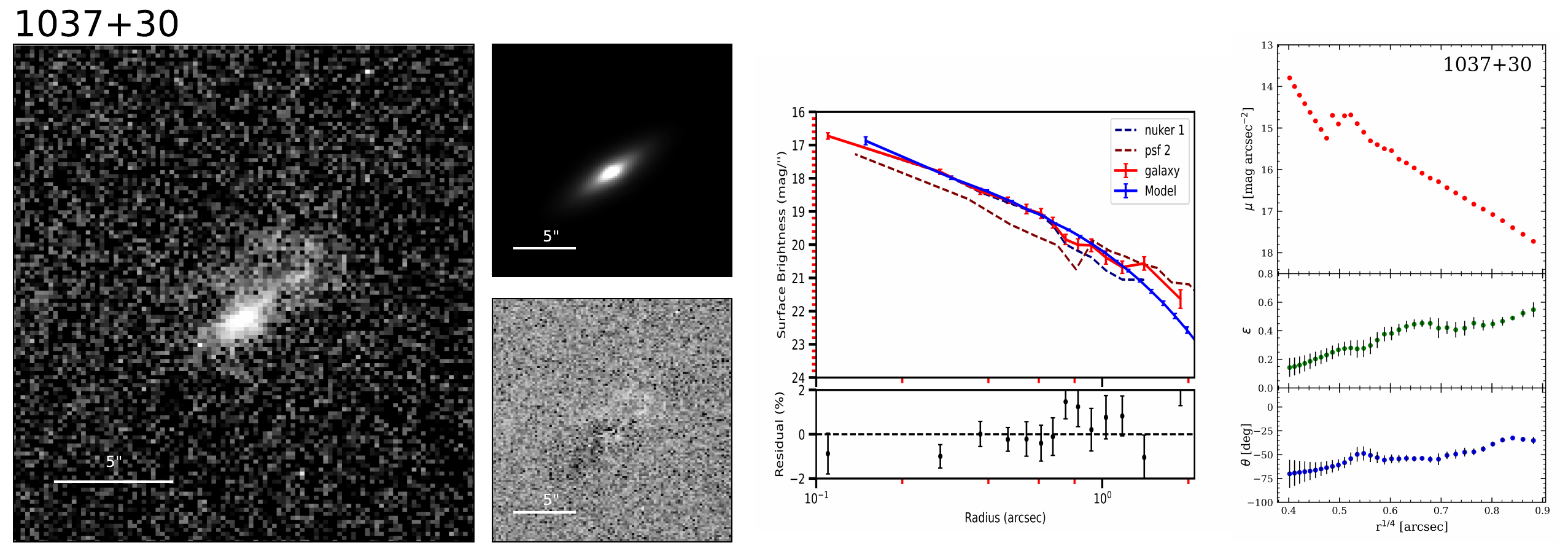}{\textwidth}{}}
\caption{UV photometric analysis and structural decompositions.}
\end{figure*}

\begin{figure*}[h!]
\figurenum{12}
\gridline{\fig{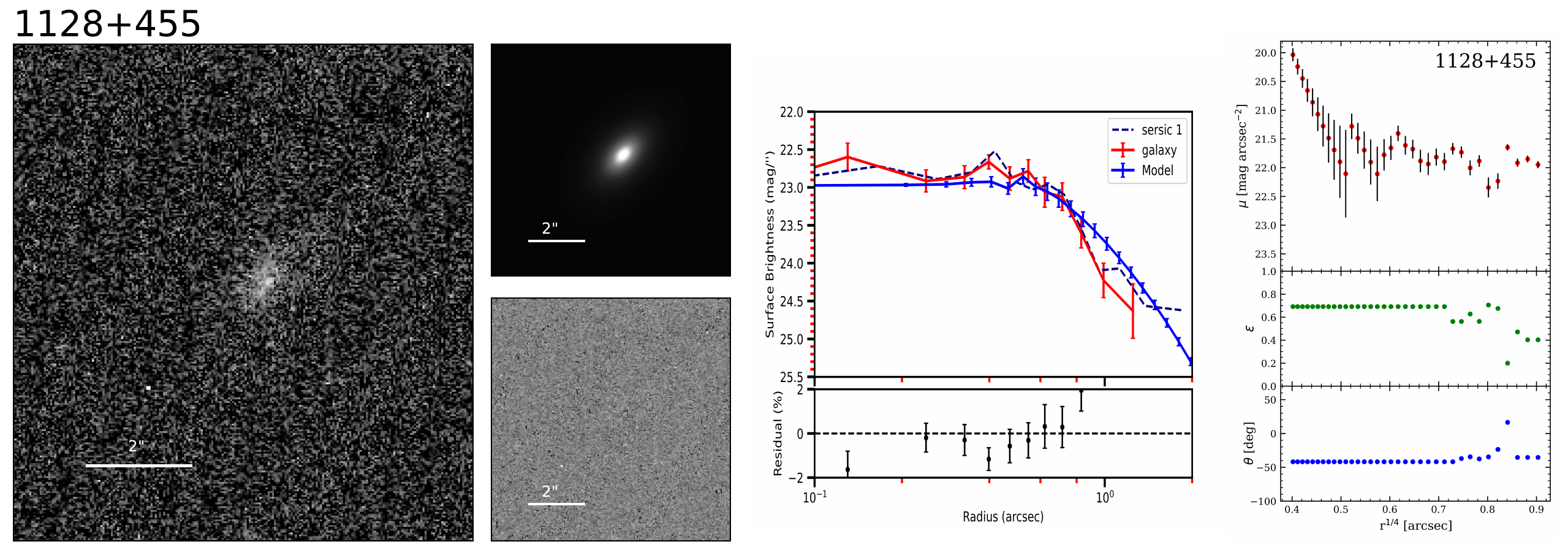}{\textwidth}{}}
\gridline{\fig{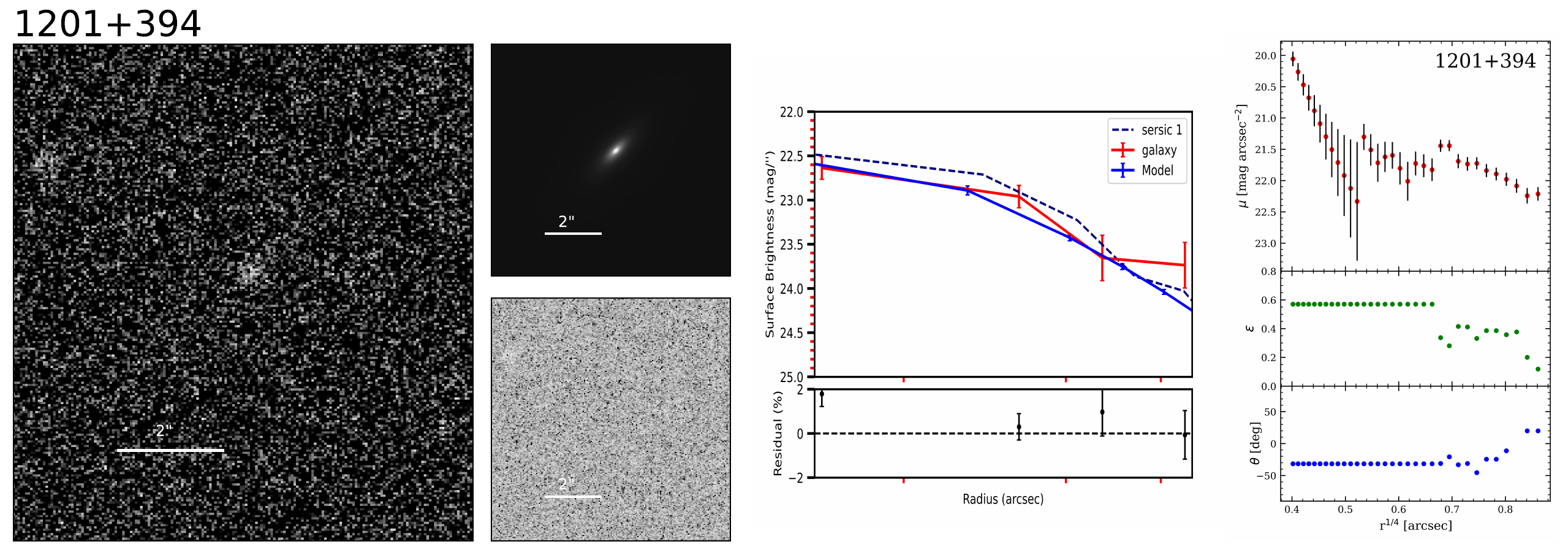}{\textwidth}{}}
\gridline{\fig{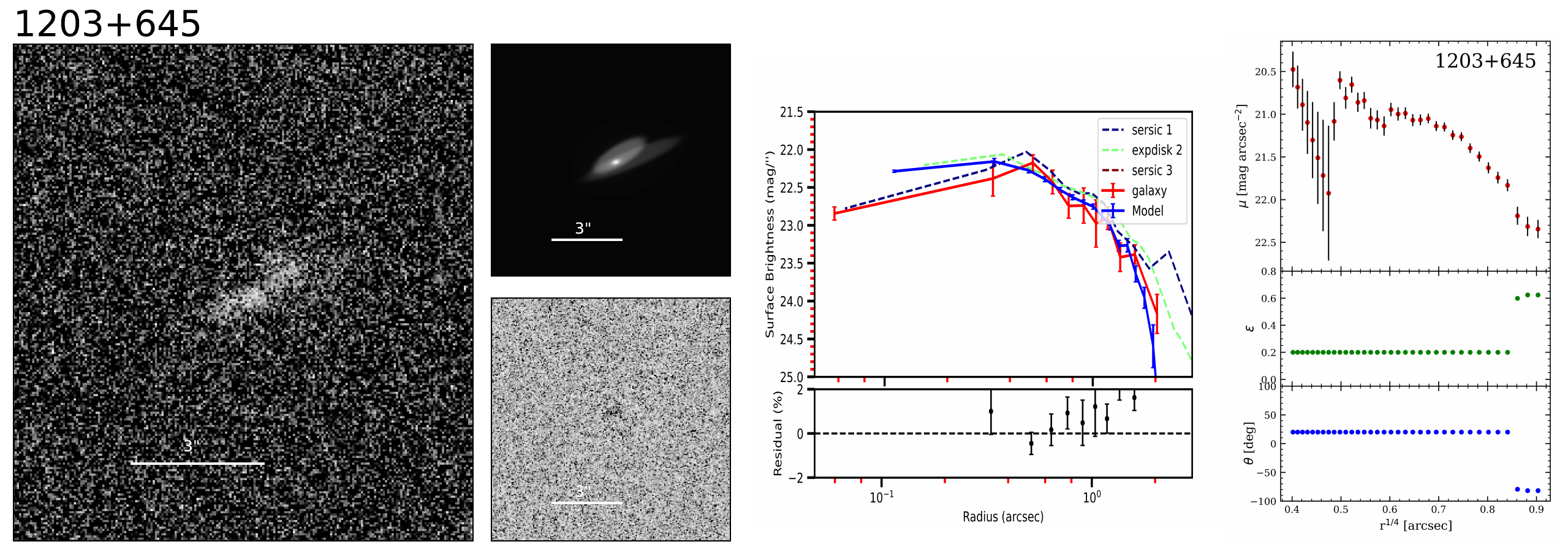}{\textwidth}{}}
\caption{$continued$}
\end{figure*}

\begin{figure*}[h!]
\figurenum{12}
\gridline{\fig{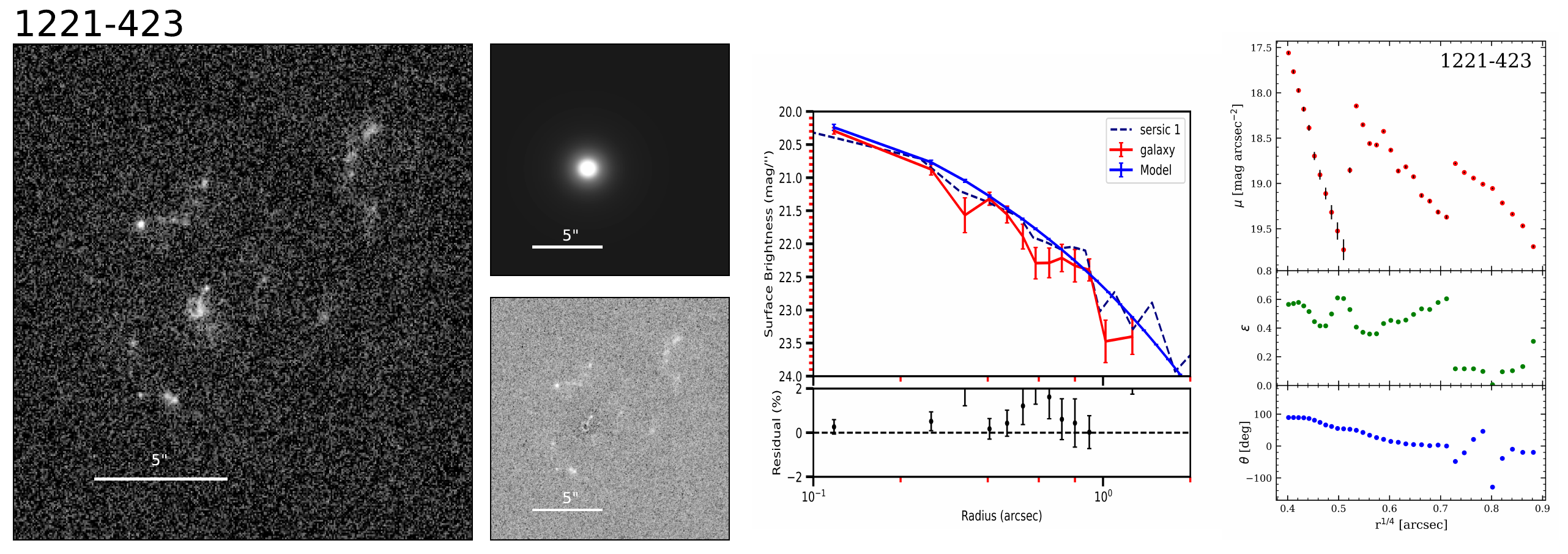}{\textwidth}{}}
\caption{$continued$}
\end{figure*}




\bibliography{ref_list}{}
\bibliographystyle{aasjournal}



\end{document}